%% file: ms.tex
\documentclass[10pt,preprint]{aastex}
\slugcomment{{\sc Accepted to ApJ:} September 20, 2010} 
\usepackage{natbib}
\usepackage{lscape}
\tighten

\newcommand{\lsun}{\mbox{$L_\odot$}}
\newcommand{\flum}{\mbox{$f_{lum}$}}
\newcommand{\ldisk}{\mbox{$L_{disk}/L_*$}}

\newcommand{\lstar}{\mbox{$L_\star$}}

\newcommand{\msun}{\mbox{$M_\odot$}}

\newcommand{\mic}{\mbox{$\mu$m}}
\newcommand{\app}{\mbox{$\sim$ }}

\newcommand{\ha}{\mbox{$H\alpha$}}
\newcommand{\ewha}{\mbox{EW($H\alpha$)}}
\newcommand{\fwha}{\mbox{FW.1H($H\alpha$)}}
\newcommand{\dg}{\mbox{$^\circ$}}
\makeatletter
\newcommand{\Rmnum}[1]{\expandafter\@slowromancap\romannumeral #1@}
\makeatother

\begin{document}
%\section {Observation}
%\include{table1}
\title {The {\bf {\it Spitzer}} c2d Survey of Weak-Line T Tauri Stars. III. 
The Transition from Primordial Disks to Debris Disks.
}
\author {Zahed Wahhaj\altaffilmark{1}, 
Lucas Cieza\altaffilmark{1},  
Karl R. Stapelfeldt\altaffilmark{2},   
Deborah L. Padgett\altaffilmark{3},  
David W. Koerner\altaffilmark{4},
April Case\altaffilmark{4},  
James R. Keller\altaffilmark{5},  
Bruno Mer{\'i}n\altaffilmark{6}, 
Neal J. Evans, II\altaffilmark{7},  
Paul Harvey\altaffilmark{7},
Anneila Sargent\altaffilmark{8},   
Ewine F. van Dishoeck\altaffilmark{9},  
Lori Allen\altaffilmark{10},  
Geoff Blake\altaffilmark{12},  
Tim Brooke\altaffilmark{3},   
Nicholas Chapman\altaffilmark{2},   
Lee Mundy\altaffilmark{13},   
Philip C. Myers\altaffilmark{11}   
}

\altaffiltext{1} {Institute for Astronomy, University of Hawaii, Honolulu, HI 96814}
\altaffiltext{2} {Jet Propulsion Laboratory, California Institute of Technology, MS 183-900, 4800 Oak Grove Drive, Pasadena, CA 91109} 
\altaffiltext{3} {Spitzer Science Center, California Institute of Technology, Mail Code 220-6, Pasadena, CA 91125} 
\altaffiltext{4} {Northern Arizona University,  Building 19, Rm. 209, Flagstaff, AZ 86011-6010}
\altaffiltext{5} {Department of Physics, 301 Weniger Hall, Oregon State University, Corvallis, OR 97331-6507}
\altaffiltext{6} {Herschel Science Centre, European Space Astronomy Centre (ESA), P.O. Box 78, 28691 Villanueva de la Ca\~nada (Madrid), Spain}
\altaffiltext{7} {Astronomy Department, University of Texas, 1 University Station C1400, Austin, TX 78712}
\altaffiltext{8} {Division of Physics, Mathematics, and Astronomy, California Institute of Technology, MS 105-24, Pasadena, CA 91125.} 
\altaffiltext{9} {Leiden Observatory, Postbus 9513, 2300 R.A. Leiden, Netherlands.}
\altaffiltext{10} {National Optical Astronomy Observatory, 950 North Cherry Ave., Tucson, AZ 85719}
\altaffiltext{11} {Smithsonian Astrophysical Observatory, Harvard-Smithsonian Center for Astrophysics, 60 Garden Street, MS 42, Cambridge, MA 02138} 
\altaffiltext{12} {Division of Geological \& Planetary Sciences, California Institute of Technology, MS 150-21, Pasadena, CA 91125} 
\altaffiltext{13} {Astronomy Department, University of Maryland, College Park, MD 20742}
\altaffiltext{14} {Department of Astronomy, University of Maryland, College Park, MD 20742.}

\begin{abstract}
We present 3.6 to 70~\mic\ {\it Spitzer} photometry of 154 weak-line 
T Tauri stars (WTTS) in the Chamaeleon, Lupus, Ophiuchus and Taurus  
star formation regions, all of which are within 200~pc of the Sun. For a
comparative study, we also include 33 classical T Tauri stars (CTTS) which
are located in the same star forming regions.  
{\it Spitzer} sensitivities allow us to robustly detect the photosphere in the IRAC bands (3.6 to 8~\mic) and the 24~\mic\ MIPS band. 
In the 70~\mic\ MIPS band, we are able to detect dust emission brighter than roughly 40 times the photosphere. 
These observations represent the most sensitive WTTS survey in the mid to far infrared to date, 
and reveal the frequency of outer disks (r = 3-50 AU) around WTTS.
The 70~\mic\ photometry for half the c2d WTTS sample (the on-cloud 
objects), which were not included in the earlier papers in 
this series, \citet{2006ApJ...645.1283P} and \citet{2007ApJ...667..308C}, are presented here for the first time.
We find a disk frequency of 19\% for on-cloud WTTS,  but just 5\% for off-cloud WTTS, 
similar to the value reported in the earlier works. 
WTTS exhibit spectral energy distributions 
(SEDs) that are quite diverse, spanning the range from optically thick
to optically thin disks. Most disks become more tenuous than $L_{disk}/L_* = 2\times 10^{-3}$ in 2 Myr, 
and more tenuous than $L_{disk}/L_* = 5\times 10^{-4}$ in 4 Myr.

\end{abstract}

\section {Introduction}
Observational and theoretical studies are converging on the consensus
that planet formation occurs in circumstellar disks within a few million years (Myr) of the 
central star's core formation. Beyond an age of 10~Myr, the primordial disks of dust 
and gas are no longer observable. The evolution of the disks during this short period 
when they are observable contains indispensable information about the 
universal prospects for planet formation. 
The nearest star forming 
regions are over 120~pc away, and they harbor many hundreds of young stars.
However, it is difficult to ascertain which stars are young cloud members
and which are old field stars. 
Surveys for H$\alpha$ emission identified very young stars (age \app few Myr) 
which are actively accreting disks. Young stars were also identified 
by high X-ray activity and strong lithium absorption (in late type stars).  
Weak H$\alpha$ emission T Tauri stars (WTTS) are also thought to 
be young but their age distribution is wider than that of CTTS. Their modern definition on the basis 
of spectral properties, which is explained below, was chosen to select non-accreting young stars.  They are found in very 
young clusters and also in older off-cloud regions \citep{2007ApJ...667..308C,2006ApJ...645.1283P}.
Thus, their evolutionary status is quite uncertain. In this paper, we will address this question in terms 
of the spectral energy distributions (SEDs) of WTTS disks, and thus try to understand where WTTS fit in 
the larger disk evolution picture.

Current theories are trying to explain what drives the dispersal of the circumstellar gas, what halts gas accretion,  
in what order the different parts of the dust disk disappears, and how all these processes affect planet formation. The 
mechanisms proposed  as pathways for disk dissipation include viscous accretion \citep{1998ApJ...495..385H, 2005A&A...442..703H}, 
grain growth \citep{2005A&A...434..971D}, 
photo-evaporation \citep{2001MNRAS.328..485C, 2006MNRAS.369..229A} 
and dust sweeping by companions \citep{1979MNRAS.186..799L, 1994ApJ...421..651A}, among others. 
The general strategy for sorting out the relative importance of these mechanisms has been
to observe large numbers of systems with different multiplicities, SEDs, accretion rates and ages 
and to try to find correlations and trends between these measures. 
Early on, it was found that some WTTS had passive disks 
that could be detected in the submillimeter \citep{1990AJ.....99..924B}.  
\citet{1995ApJ...439..288O} showed that WTTS had discernibly 
smaller millimeter wave emission than CTTS. 
The scheme for the identification of accreting disks based on \ewha\ went through some refinement 
over the years.
\citet{1998AJ....115..351M} adjusted the CTTS/WTTS
classification by \ewha\ according to spectral type. \citet{2003ApJ...582.1109W} suggested 
a further adjustment of this classification and also showed that 
requiring full widths at 10\%  height $>$ 270 km/s 
of the \ha\ line was a much preferred way of identifying accretion disk stars (CTTS). 
Further refinements and higher resolution spectra helped robustly identify 
stars which were weak accretors but not WTTS 
\citep{2003AJ....126.2997B, 2005A&A...443..541G,2006AJ....132.2135S}. 
Following these refinements, \citet{2005ApJ...631.1134A} found that only 15\% of WTTS had submillimeter emission
while 91\% of CTTS were detected in the submillimeter. This signaled that most WTTS had lost the small dust grains 
and pebbles in their outer disks (r $>$ 50 AU) where most of the dust mass in primordial disks is usually found.

Before {\it Spitzer}, disk surveys performed at wavelengths longward of the L-band 
provided an incomplete picture, as cool dust disks could remain undetected in 
regions beyond a few tens of AUs from the stars.
\citet{2001ApJ...553L.153H} reported on L band excess rates in clusters and 
clouds from 0.3 to 30~Myr, showing a well behaved decline with cluster age. 
Excess rates were found to be 85\% at 0.3~Myr \citep{1996PhDT.........3M}, 52\% at 
3.2~Myr \citep{2000ApJ...540..255P}, and
only 3\% at 30~Myr \citep{1985SvA....29..499B}. 
With Spitzer's dramatic improvement in sensitivity in the mid and far infrared, 
the dust out to several 10's of AU can be detected in the nearby star-forming regions.
There have been several {\it Spitzer} studies for disks around young stars,
and they can be categorized as:
disk surveys in clusters/associations, volume-limited debris disk
surveys (often age selected) and star forming region surveys. 
A plethora of {\it Spitzer} studies on clusters and moving groups
have been conducted recently.
\citet{2007AAS...211.8918G} recently reported on an IRAC and MIPS survey of NGC 1333 (d=300~pc) 
showing the extremely high excess rate of 83\% . 
An IRAC study of 2-3~Myr old IC 348 cluster in the Perseus cloud 
yielded an excess rate of 50\%, while the WTTS excess rate was 
found to be 36\% \citep{2006AJ....131.1574L,2007AJ....134..411M}. 
Surveys of several middle aged groups
like the NGC 2362 (5~Myr), $\eta$ Chamaeleontis association (5-9~Myr),  Upper Sco-Cen OB association (5-20~Myr),
NGC 2547 (30~Myr) have yielded excess rates from 10 to 40\% \citep{2007AJ....133.2072D,2005ApJ...634L.113M,2005ApJ...623..493C,2006ApJ...651L..49C,2006ApJ...652..472H,
2007ApJ...659..599C,2007ApJ...670..516G}. 
Excess rates for older clusters like IC 2391 (50~Myr) and the Pleiades (100~Myr)
are found  to be around 25\% \citep{2006ApJ...649.1028G, 2005AJ....130.1834S, 2007ApJ...654..580S}. But these are 
due to second generation debris disks.
Despite the higher sensitivities attainable for the nearby associations like Beta Pictoris Moving Group (12~Myr)
and the TW Hydra association (8~Myr), their excess rates are found to be no higher than about 35\%
\citep{2008ApJ...681.1484R, 2005ApJ...631.1170L}.
In studies of young (1~Myr) molecular clouds,  where membership has not been established for every member by spectroscopic 
indicators, only the number of stars with excess can be estimated.
In Spitzer studies of the excess populations in Ophiuchus 
\citep{2008ApJ...672.1013P} and Lupus \citep{2007ApJ...667..288C}, it was found that
10-30\% were Class III, 40-60\% Class II, and the rest were less evolved. 
\citet{2008ApJ...676..427A} presented a
Spitzer study of Chamaeleon II, where they found that the cloud has
similar star forming efficiencies to Taurus and Lupus, but a disk
fraction that is much higher (70\%-80\%) than other star forming regions.

The debris disk surveys generally attest to the fact that second generation dust disks are common in older systems (age $>$ 1 Gyr).
In   {\it Spitzer} studies of A stars, ranging 5-850~Myr in age, the mean excess rate is \app 30\%. The maximum excess is found to be an inverse function
of age, with a lifetimes of 150~Myr and 400~Myr at 24~\mic\ and 70 \mic , respectively \citep{2005ApJ...620.1010R, 2006ApJ...653..675S}. 
Much lower excess rates (10-20\%) have resulted from other main sequence
surveys \citep{2006ApJ...639.1138S, 2007prpl.conf..573M,2008ApJ...674.1086T}.
A surprising discovery was the high rates of 
infrared excess around main sequence binary stars. 
The observed rates were 9\% at 24~\mic , 40\% at 70~\mic , and almost 60\% when
only considering very close binaries (separation $<$ 3 AU) \citet{2007ApJ...658.1289T}. 

One of the major goals of the {\it Spitzer} Cores to Disks (c2d) Legacy project \citep{2003PASP..115..965E} was to 
characterize the dust disks around WTTS and find their place in the planet formation
picture. Thus, 187 WTTS/CTTS in the nearest star forming regions, 
Chamaeleon, Lupus, Ophiuchus and Taurus were observed with IRAC at 3.6, 4.5, 5.8 and 
8.0~\mic\ and with MIPS at 24 and 70~\mic . There have already been two publications 
on the preliminary data from these observations. We discuss these works in the next section. 

\section {Observations}

\subsection {Sample Selection}
The sample for this survey is well described in \citet{2006ApJ...645.1283P}, where
photometry for about half the c2d WTTS were presented. 
Basically, we attempted 
to select young stars with weak H$\alpha$ emission for which the photosphere 
could be robustly detected at 24~\mic . This meant we could only choose objects 
from the closer (D$<$200~pc) star forming clouds, Chamaeleon, Lupus, Ophiuchus 
and Taurus. Young stars had been identified in these clouds by ROSAT detections 
of strong X-ray activity and by spectroscopic detections of H$\alpha$ lines 
with small equivalent widths (EW)  (generally less than 10\AA) 
\citep{2000A&A...359..181W,1997A&A...328..187C,1999MNRAS.307..909W,1998AJ....115..351M}.
A further selection condition was that the objects exhibit 
lithium absorption stronger than a Pleiades star of the same spectral type.
Thus, our sample should on average be younger than the Pleiades (100~Myr). 
Of course lithium is only used as a reliable indicator of youth for spectral types
later than G.  Almost all of our stars are of spectral types K and M. 

As mentioned earlier, preliminary results from these observations have already been 
presented in two earlier works.
\citet{2006ApJ...645.1283P} studied the 83 c2d WTTS that had both pointed IRAC and MIPS observations.
Only a few of these deep pointed observations were inside the c2d scan maps of the five
large clouds \citep{2003PASP..115..965E}. 
The rest were off-cloud but still less than 6 degrees away from the cloud boundaries in 
projected separation. 
These deeper off-cloud observations were made when the scan 
maps did not provide adequate sensitivity with respect to 
the target's photosphere. An excess rate of 6\% was found for these objects. 
\citet{2007ApJ...667..308C} collected all the known WTTS that were located within the
cloud maps, adding many targets in Perseus that were not in the original c2d
WTTS sample. They found that \app 20\% of the WTTS covered by the cloud maps had excess emission
from dust at 24~\mic . Seventy micron photometry was not presented at
that time. Here we present 3.6-70~\mic\ data on the entire c2d WTTS sample, 
both on-cloud and off-cloud, thus 
revealing the nature of dust disks around WTTS out to several tens of AU from the star. 
For half our sample we are sensitive to  disks as tenuous as the debris disk $\beta$ 
Pictoris ($L_{disk}/L_* = 2\times 10^{-3}$), while for the other half we are only sensitive to brighter disks.    

Since the publication of the first two papers in this series, we have refined the 
CTTS/WTTS classification criteria according to \citet{2003ApJ...582.1109W}, who required 
 a full width at 10\% of \ha\ line height 
(FW.1H(\ha)) of greater than 270 km/s (in high resolution spectra) to classify a star as CTTS. In the absence of high  
resolution spectra, we use the \ewha\ to classify the objects.   
According to \citet{2003ApJ...582.1109W},  
we classify a star as CTT when \ewha\ $> 3$ \AA\ for spectral types earlier than K0, 
when \ewha\ $> 10$ \AA\ for K7-M2.5, when \ewha\ $> 20$ \AA\ for M2.5-M5 and when \ewha\ $>40$ for later types. 
The idea is that for an M star, stellar activity alone can produce a line width  
of 10~\AA\ \citep{1998AJ....115..351M}, whereas for earlier types the line is often 
saturated. We had available reduced high resolution optical spectra 
(KPNO 4m, Echelle Spectra, R \app 42000, \citealt{2003AAS...203.0502K}, \citealt{2004MST..........3M}) for 161 of our 187 targets. 
From these, we were able to measure the \fwha\ 
to a precision of roughly 30 km/s. 
The measured line widths are presented in Table~\ref{stprops}. Some of the targets which would have been classified as WTTS
in the earlier papers in this series are now classified as CTTS, and thus old results 
are not directly comparable. Sixteen stars, which were originally classified as WTTS objects, are now reclassified as CTTS. They are  
UX~Tau, FX~Tau, ZZ~Tau, V710~Tau, V807~Tau, V836~Tau, Sz 41, RX~J1150.4-7704, RX~J1518.9-4050, Sz~65, Sz~96, RX~J1608.5-3847, RX~J1608.6-3922, ROX~16, SR~9 and ROX~39.
Furthermore, two objects were reclassified as WTTS from CTTS and they are RX~J1149.8-7850 and RX~J1612.3-1909.

\subsection {{\it Spitzer} Photometry}

In the 50 {\it Spitzer} hours allotted to this program (PID, 173) we observed 
154 WTTS objects and 33 CTTS. 
One-third of our WTTS were in the denser clouds regions (with A$_v >$ 3), one-third
were between 0 and 3 degrees from the cloud edge (A$_v =$~3 isoline), and one-third were 
between 3 and 6 degrees from the cloud edge. Given the velocity 
dispersion of these cloud associated sources, stars originating in the clouds should travel no farther 
than 6 degrees in 10~Myr \citep{1991AJ....101.1050H}. We obtained {\it Spitzer} photometry with its 
{\it Infrared Array Camera} (IRAC) \citep{2004ApJS..154...10F} and {\it Multiband Imaging Photometer} (MIPS) 
\citep{2004ApJS..154...25R} instruments. The details of the observing parameters are 
well-described in \citet{2006ApJ...645.1283P}, so we will just provide the basic 
template here. Two 12 second exposures, plus one short 0.6 exposure 
in ``HDR'' mode were taken with IRAC. These were adequate to detect the 
photosphere with $S/N>50$ in the four IRAC bands centered on 3.6, 4.5, 
5.8 and 8.0~\mic . At 24~\mic , we attempted to detect the photosphere, 
with $S/N>20$ which resulted in integration times between 42-420 seconds.
At 70~\mic , the photospheres are too faint to detect, and so we used the 
same exposure time of 360 seconds for all objects. This allows us to detect 
disks as faint as \flum\ (or \ldisk )~$= 10^{-3}$ for the brighter half 
of our sample, and disks as faint as $f_{lum} = 10^{-2}$ for the 
other half. Photometry were obtained from the IRAC and MIPS maps 
using the c2d pipeline, as described in 
\citet{2006ApJ...644..307H}, \citet{2005ApJ...628..283Y} and the c2d data 
delivery document. 

At 70~\mic , aperture photometry was conducted 
on the filtered post-BCD (Basic Calibrated Data) images obtained from the Infrared Science Archive ({\it http://irsa.ipac.caltech.edu/}).
The images are super-sampled so that the pixels are of size 4$''$ instead of the original 10$''$.
The aperture centers were fixed using the astrometry information in the FITS file headers.
We used an aperture of radius 16$''$ and a sky 
annulus with inner and outer radii of 18 and 39$''$, for which an aperture correction factor of
2.07 is recommended (MIPS data handbook, updated October 2009). The {\it IDL} routine {\it APER.PRO}
was used to perform the photometry and estimate the uncertainties which are 
obtained from the pixel-to-pixel noise in the sky annulus. Detections with 
signal-to-noise (SN) above 3 are initially considered real, but subsequently checked for shape,
confusion with nearby sources and nearby nebulosity. Three-sigma upper limits,
for sources not detected at 70~\mic, are given in Table~\ref{stphot}. The absolute 
calibration uncertainty for 70~\mic\ is 15\% \citep{2007PASP..119.1019G}. This error is added in quadrature 
to the photometric uncertainty. We note that the 70 \mic\ fluxes presented in this
paper are larger by a factor of \app 2 than those reported in \citet{2006ApJ...645.1283P}, as the 
earlier photometry was done before aperture radii and correction factors were properly calibrated
and standardized by the {\it Spitzer Science Center (http://ssc.spitzer.caltech.edu/mips/calib/)}.

Because of the significantly lower imaging resolution
at 70~\mic , it is much more difficult to decide which detections are bona~fide,
compared to the other bands. This is especially true when the signal-to-noise
of the detections are below 10, or when there is neighboring nebulosity. 
We consider any detected 70~\mic\ emission to be associated with the
target star with high confidence when the separation of the center of emission and the star 
as seen in the IRAC bands is less than FWHM$_{70}$/SN$_{70}$. Thus, for a 70~\mic\ detection 
with a signal-to-noise of 3, the separation between the emission centers must be less
than $16''/3 \sim 5.3''$. When the separation is larger than this limit, but less than 
the nominal FWHM at 70~\mic\ (16$''$), we check in the IRAC images to see if the 
70~\mic\ emission center is better matched by any other source. If there is no better match, and the surrounding 
nebulosity at 24~\mic\ and 70~\mic\ can be ruled out as sources of confusion, then 
the emission is still assumed to be coming from the target star.  Where the 70~\mic\ detections
suffer from confusion, it is possible that the 24~\mic\ photometry, as given in the 
{\it IRSA} catalogs are also contaminated. For these targets, the 24~\mic\ aperture photometry was repeated (replacing the {\it IRSA} catalog values) according to 
the MIPS data handbook recommendations for an aperture radius of 7$''$. This resulted in new 
photometry for the WTTS sources RX~J1607.2-3839, RX~J1608.3-3843 and RX~J1609.7-3854.

\begin{figure}[ht]
  \centerline{
   \vbox {
     \hbox {
       \includegraphics[height=3.3cm]{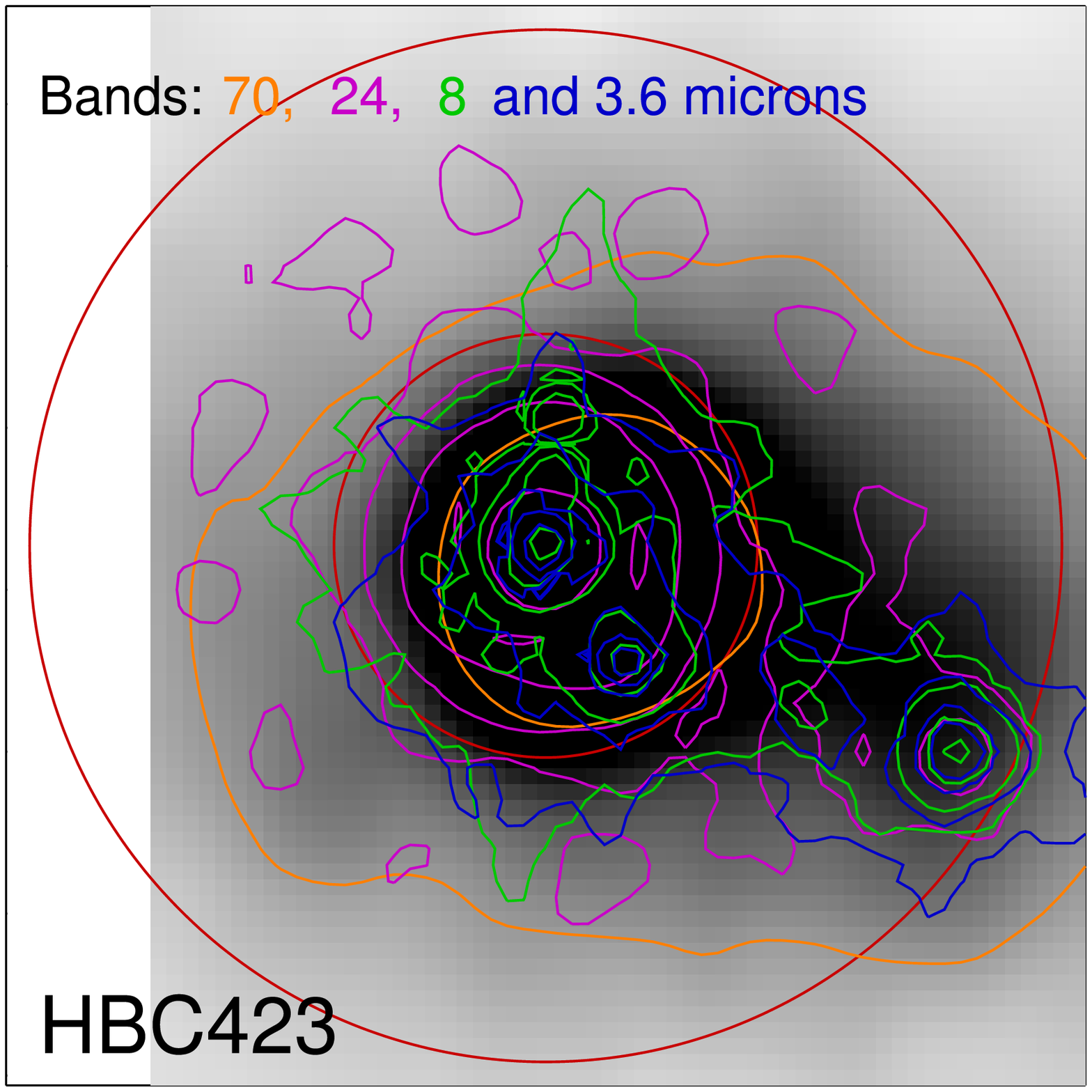}
       \includegraphics[height=3.3cm]{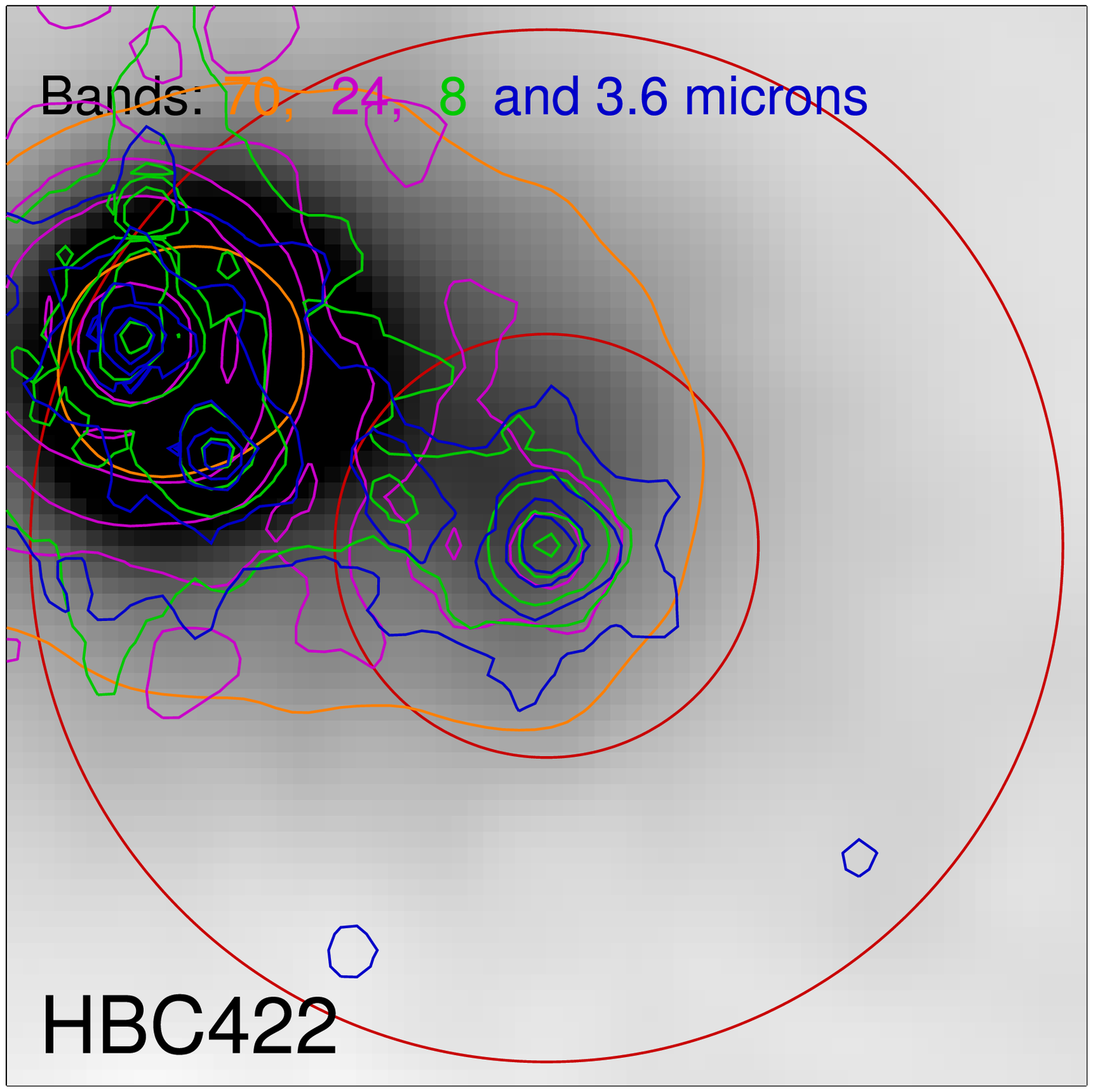}
       \includegraphics[height=3.3cm]{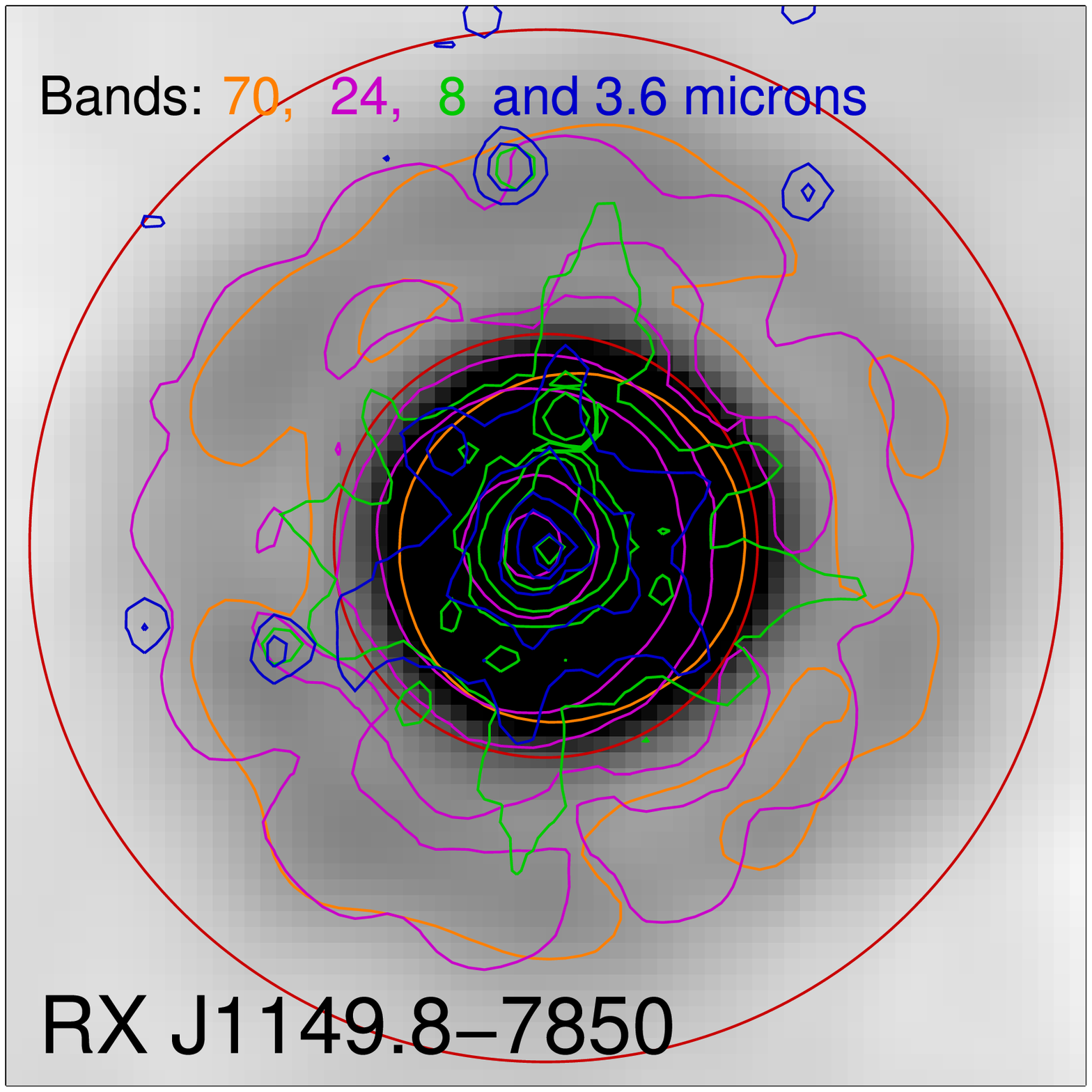} 
       \includegraphics[height=3.3cm]{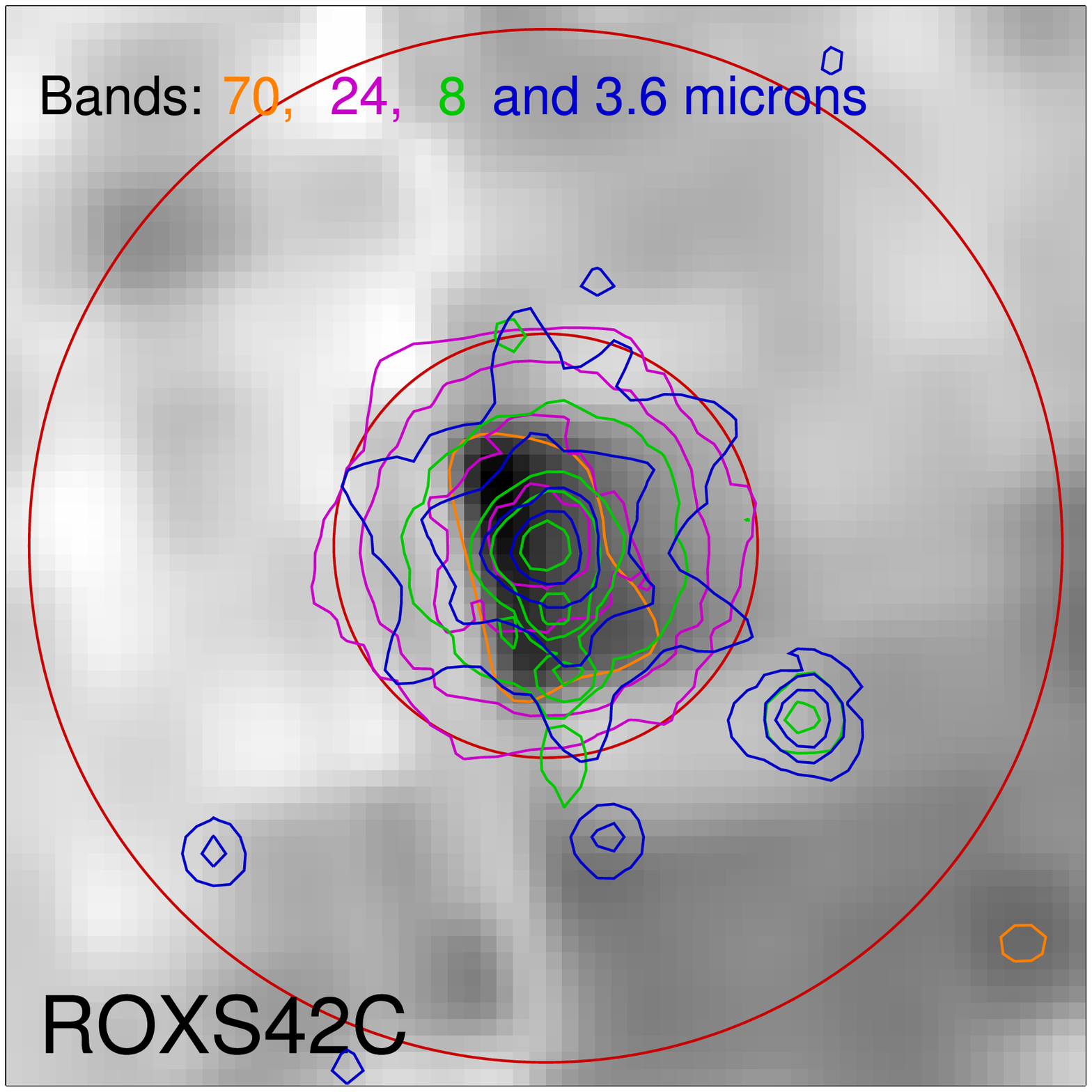} 
     }
     \hbox {
       \includegraphics[height=3.3cm]{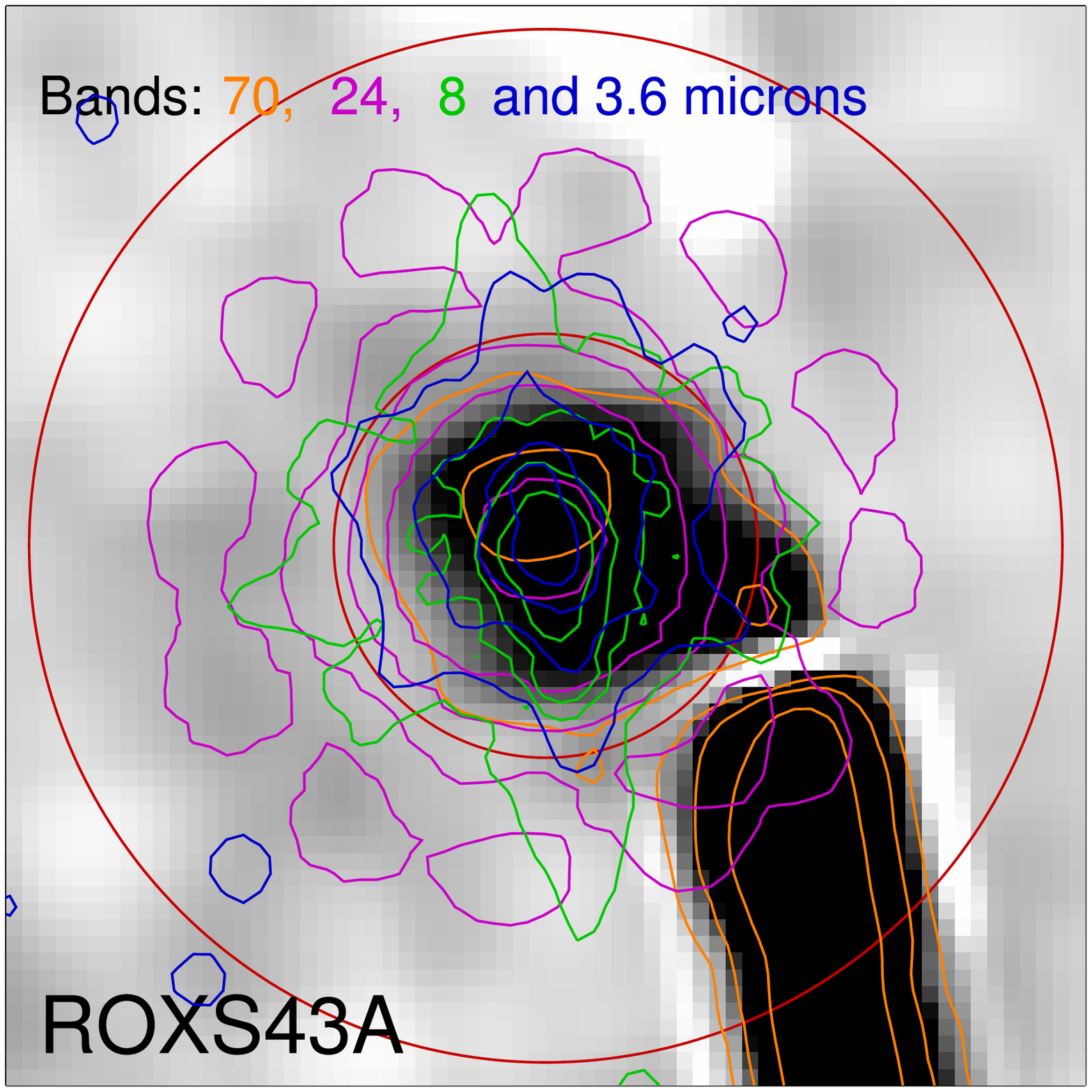} 
       \includegraphics[height=3.3cm]{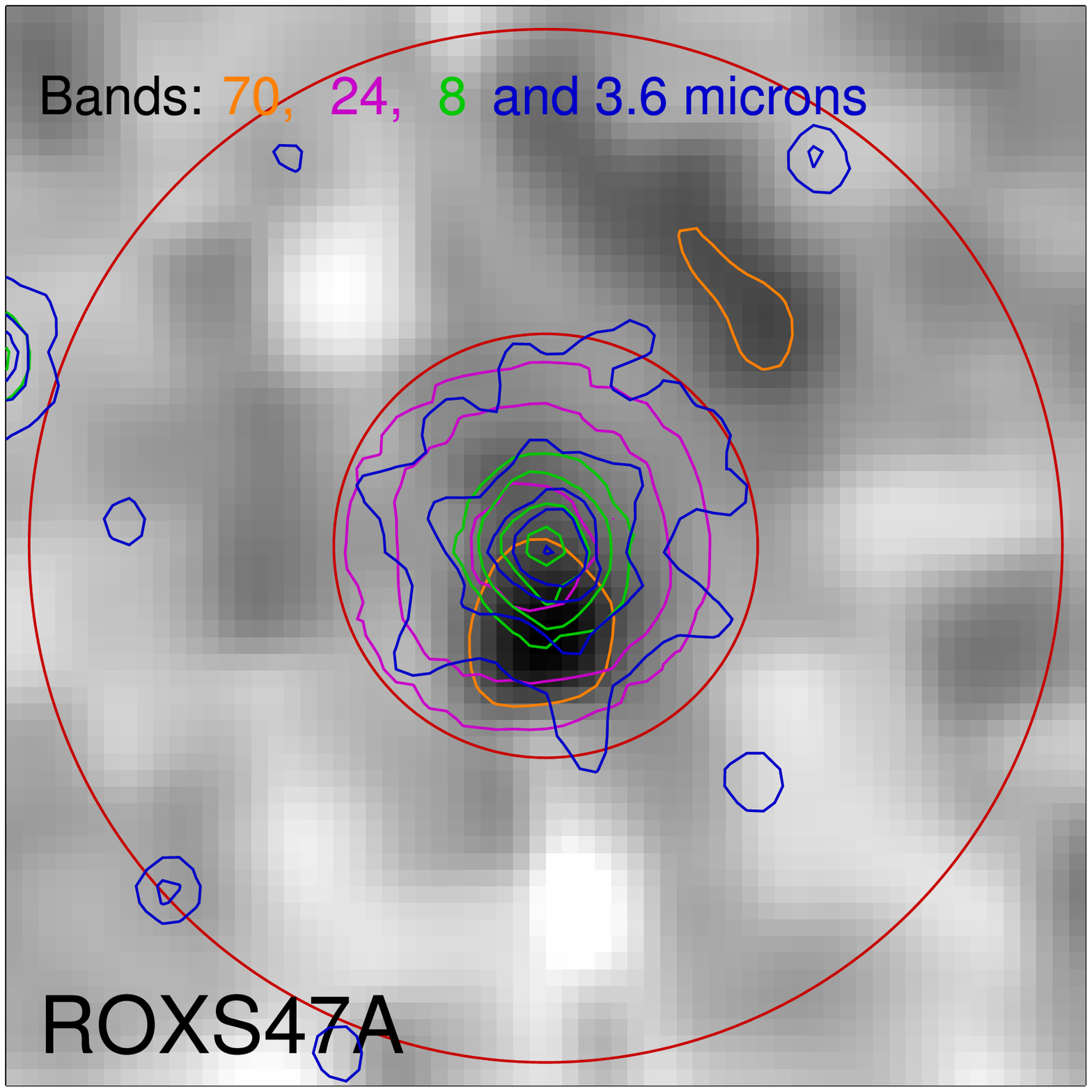} 
       \includegraphics[height=3.3cm]{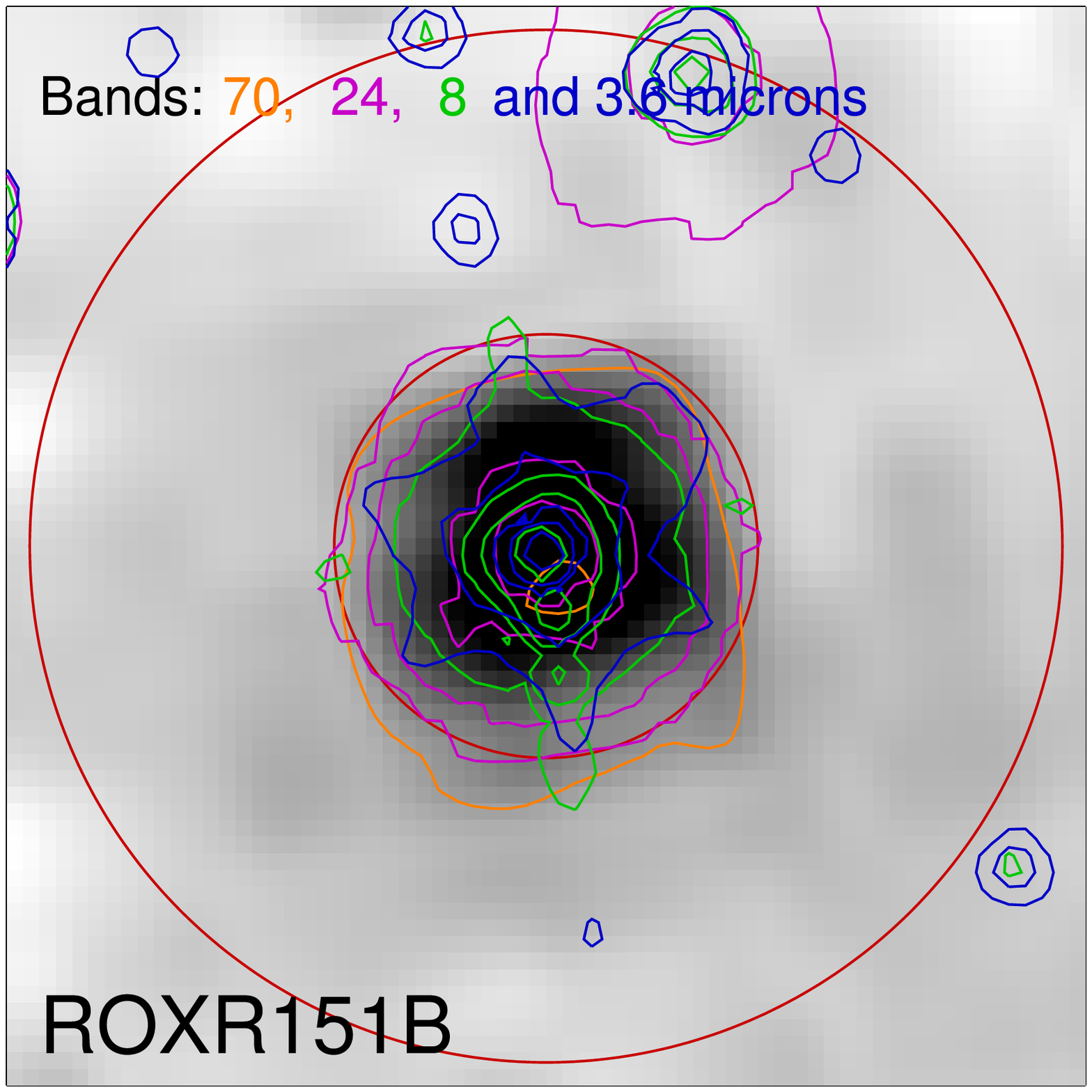}
       \includegraphics[height=3.3cm]{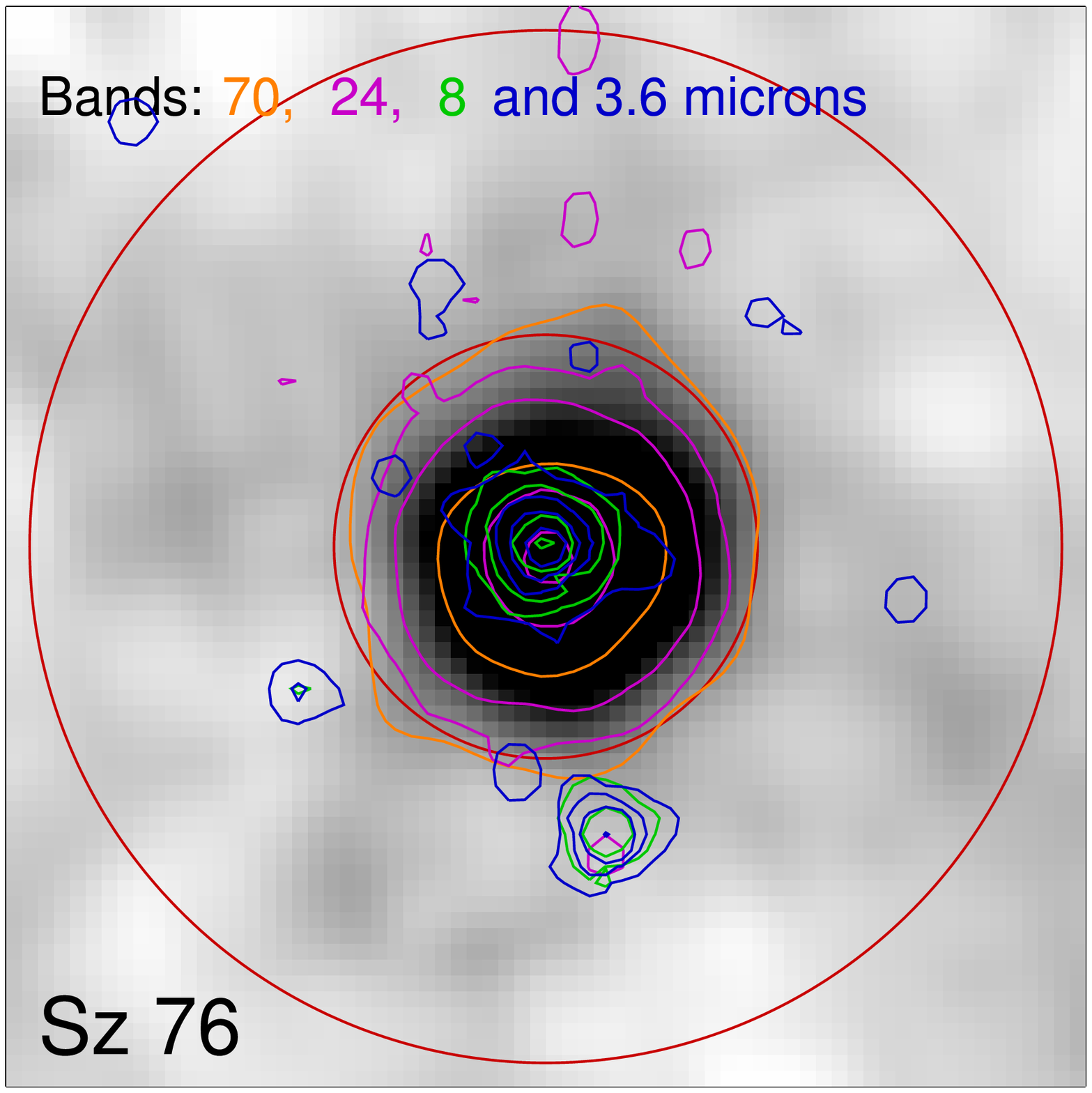}
     }
   }
 }
 \caption{ 
   70 \mic\ images of the detected WTTS targets. North is up and East is left. Contour map overlays from the IRAC1, IRAC4, MIPS-24~\mic\  and MIPS-70~\mic\  bands are shown 
in blue, green, pink and orange respectively. The contour levels shown are at 3, 15, 75 and 375 times the background RMS. 
The red circles represent the 16$''$ (radius) object aperture and the 39$''$ outer sky aperture used for photometry. The 18$''$ 
inner sky aperture is not shown.}
\label{wttsreal}
\end{figure}

In Figure~\ref{wttsreal}, we show all the WTTS 70~\mic\ 
detections which we consider to be bona~fide.   Overlaid on the figure
are contours from {\it Spitzer} maps in the 3.6, 8 and 24~\mic\ bands. 
When a nearby source was detected close
to the target aperture, it was subtracted using a custom IDL routine. 
This routine takes the centers of nearby sources and removes the median
flux in annuli around this center. Adjacent sources were removed in the 
case of HBC~423, HBC~422, DI~TAU, HV~TAU and RX~J0445.8+1556 (see
Figures~\ref{wttsreal}).  A bar-like artifact also had 
to be removed in the case of ROXS~43A.  In some cases, 
the spurious detection is entirely due to a known source close to the target.
We discuss these cases below. 
The flux detected in the DoAr 21 aperture,
probably originates from the bright HII region FG Oph 17 which reaches within 4$''$ of the target.
In the {\it Spitzer} 8 and 24 \mic\ images, we also see the surrounding nebulosity getting stronger 
at longer wavelengths. The 70 \mic\ emission is also highly irregular in shape and it is
difficult to claim with confidence that it results from disk emission. \citet{2009ApJ...703..252J} 
detected PAH emission in an irregular distribution over hundreds of AU from the star, but 
could not ascertain if there was any emission coming from a circumstellar disk. 
In the case of HV~Tau, the 70 \mic\ emission is actually coming from HV~Tau~C \citep{2003ApJ...589..410S}, which is an 
edge-on disk that lies 4$''$ to NE of the target (the AB component).
There is a very bright YSO, 2MASS J16272146-2441430, contaminating the field
and causing a spurious detection for ROX~21. 
The RX~J0445.8+1556 detection was because of contamination from the nearby star, GSC 01267-00433.
For RX~J0842.4-8345, the 70~\mic\ flux clearly originates from a separate nearby source
which appears clearly resolved at 8~\mic .
In the case of RX~J1129.2-7546, there is confusion from a known nearby source 8$''$ to the East
(2M~J11291470-7546256, \citet{2008ApJ...675.1375L}), which appears at 3.6~\mic\ and begins to 
dominate at 24~\mic . 
False detection of RX J1301.0-7654 was also caused by the nearby YSO, 2MASS J13005323-7654151.
Near WA~Oph1, a faint resolved 24~\mic\ source appears 8$''$ to the North, and although its location does not
match well with the 70~\mic\ emission, it is a better match to the emission than our target. A similar situation 
arises in the case of RX~J1612.0-1906A. For RX~J1621.2-2342a, RX~J1623.8-2341, ROX16 and SR9, it
is clear that there is too much nearby nebulosity to be confident that 70~\mic\ detections
are associated with circumstellar disks. In some cases, the nebulosity is already seen at 8 and 24~\mic\ and 
gets stronger at longer wavelengths.
Photometry at 70~\mic\ was not obtained for NTTS~043230+1746 and RX J1159.7-7601,
as the c2d Spitzer observations had the wrong pointings for these objects.

\section {Results}

\subsection{Estimating Infrared Excess}

Targets with redder colors than main-sequence stars imply emission due to 
warm dust. The warmer the dust, the shorter the wavelength at which 
the excess is observed. As in the previous c2d papers on WTTS 
\citep{2006ApJ...645.1283P}, we look for color excess using the 2MASS 
and {\it Spitzer} bands. First, we look at $K-$[24] colors, since the MIPS-24~\mic\  
band is our most sensitive band to typical 
circumstellar dust emission.  

The stellar $K$ band magnitudes have to be corrected for extinction before 
we calculate $K-$[24] colors since many of our targets are embedded in their
parent star forming clouds. We use 
$A_V=5.88 E(J-K)$ to estimate the extinction for our sources, where
$E(J-K)$ is the excess in $J-K$ color with respect to the expected stellar
photosphere. The main sequence $J-K$ colors are from \citet{1995ApJS..101..117K}. 
Extinction corrections were made according to \citet{2005ApJ...619..931I}, 
who based their empirical law 
on IRAC data from the GLIMPSE ({\it Galactic Legacy Infrared Mid-Plane Survey Extraordinaire}) project. 
The law is recommended for 2MASS and 
IRAC bands and relates $A_{\lambda}$ to $A_{K}$ as shown in 
Table~\ref{extinc}.
The derived extinctions, $A_V$ are displayed in Table~\ref{stprops}. 
Extinction in the $K$ band is below 0.2 magnitudes for most our objects, but 
some have very high extinctions. Thus, corrections to the $K-$[24] colors 
were applied, even though adding extinction correction  
introduces noise into the $K-$[24] color estimates, thus reducing  
our sensitivity to photospheric excess. 
%There seem to be quite a few high extinction sources in our sample. 
There are 49 sources with $A_V > 2$ in our sample.
We note that other than 
foreground cloud material, extinction could also result from
occultation by an optically thick circumstellar disk.
 
While the photospheric 
$K-$[24] color for A to G dwarfs is almost zero, it can be up to 1 magnitude 
for an M dwarf, thus it is important to subtract the intrinsic  photospheric 
colors before determining excess. The estimated $K-$[24] colors for the 
photosphere were taken from \citet{2007ApJ...667..527G}.  The photospheric colors 
$K-L$, $K-M$ and $K-N$ were also available from \citet{1995ApJS..101..117K}, and these were used 
for color corrections in the IRAC bands.
Now, even though the exposure times were enough to robustly detect the
photosphere at 24~\mic , 6 of our 154 WTTS are close to the confusion limit. 
To estimate the excess in the 24~\mic\ band, we select
the objects brighter than 0.5 mJy at 24~\mic , a flux level at which we 
are assured negligible spurious detections and better than 95\% completeness
\citep{2004ApJS..154...70P}. We end up with 148 WTTS and 33 CTTS
with reliable $K-$[24] colors. The rejected objects were 
RX~J0439.4+3332a, RX~J1123.2-7924, RX~J1614.2-1938, RX~J1614.4-1857a and RX~J1615.1-1851.

\begin{figure}[ht]
  \centerline{
    \includegraphics[height=8cm]{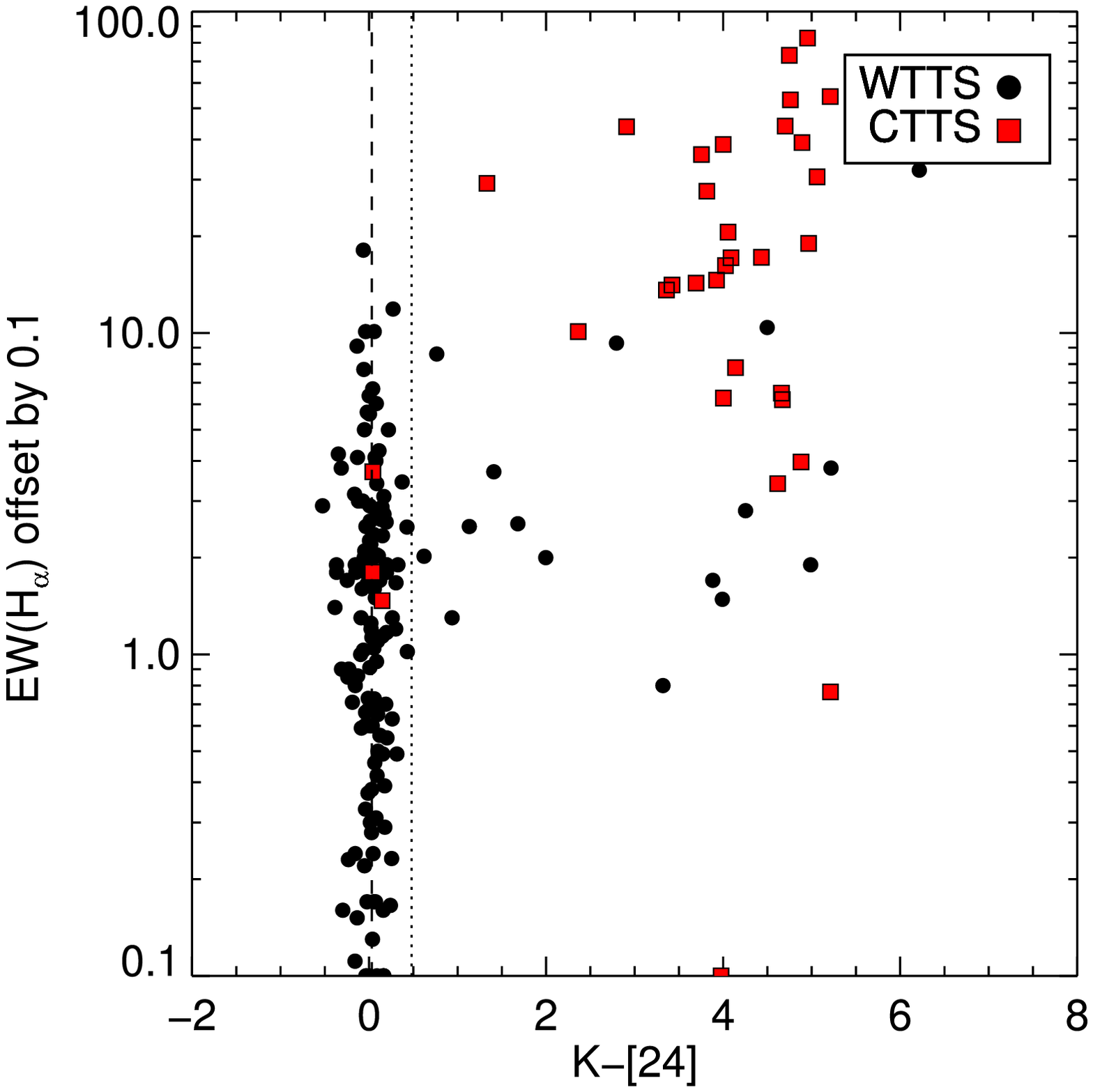}        
    \includegraphics[height=8cm]{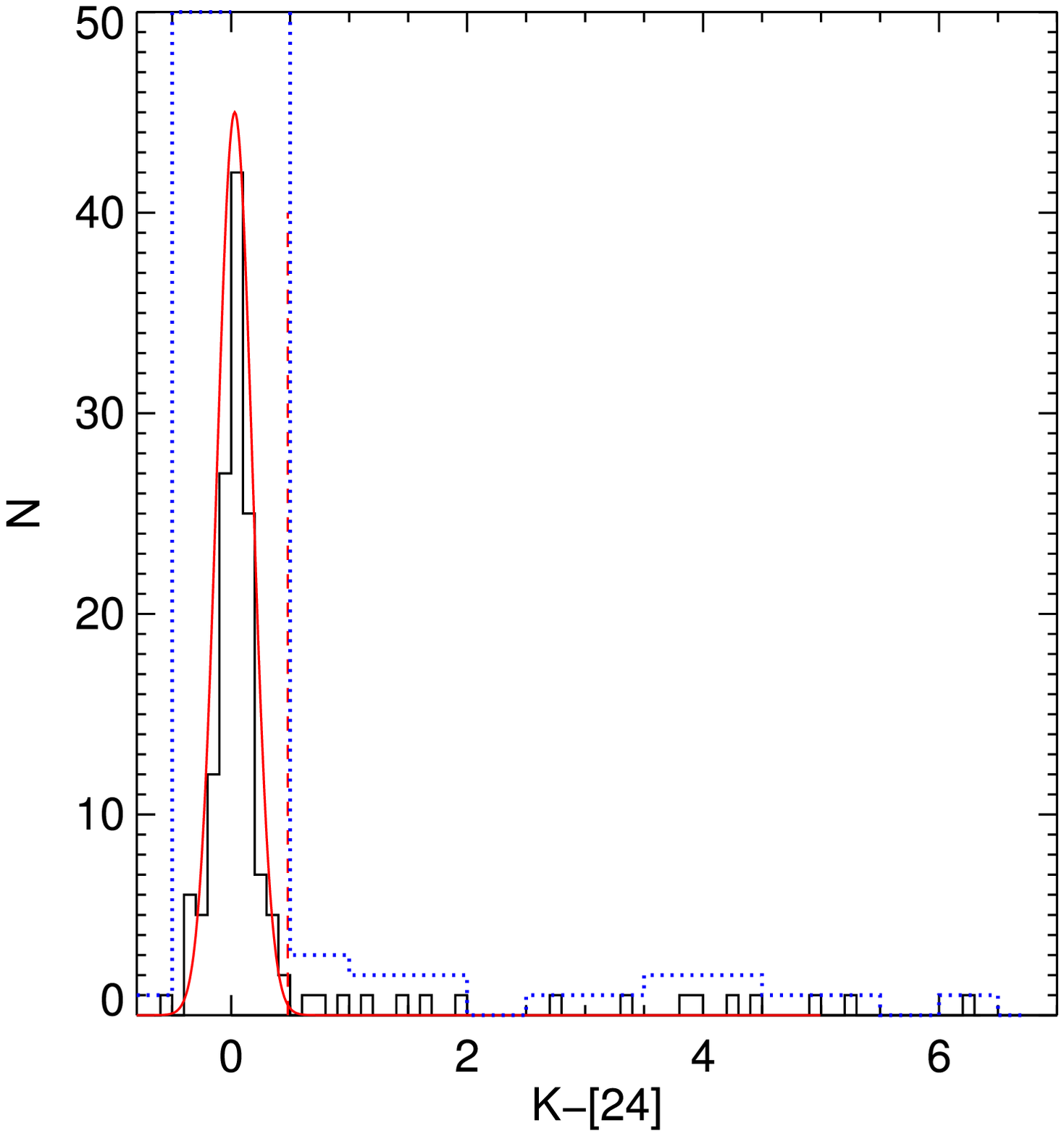}        
  }
  \caption{ (Left) EW(H$\alpha$) vs the excess $K-$[24] color for our sample of 
    181 stars. 148 WTTS are shown as circles. 33 CTTS are shown 
    are squares. The \ewha\ come from low resolution spectra, but the CTTS classification is adjusted 
    using \fwha\ estimated from high resolution spectra.
    The dashed-line shows the median color for the objects 
    without excess. The dotted-line is the 3$\sigma$ marker for excess 
    identification. A negligible offset has been added to the
    EW(H$\alpha$) values to allow easier plotting. 
    (Right) A histogram of $K-$[24] excess for just the 
    WTTS. The solid red curve is a Gaussian with mean at $K-$[24] = 0.02
    mags, and 1-$\sigma$ dispersion of 0.15 mags. The blue
    dotted-line histogram is just a coarser binning of the same distribution. (Caution: It extends beyond
    the plot boundary at the peak of the distribution).
  }
  \label{k-24}
\end{figure}

After subtracting the photospheric colors, we get a 
robust median (rejecting outliers) of 0.03 magnitudes for the $K-$[24] colors, 
and a robust standard deviation of 0.15 mags.
Throughout this paper, we identify the sources with excess color by looking at the color
distribution itself. This method has been demonstrated in earlier {\it Spitzer} papers 
\citep{2006ApJ...653..675S,2006ApJ...645.1283P,2007ApJ...667..308C}. 
Most of our sources exhibit bare photospheres and should have
zero excess colors. This group manifests itself as a Gaussian distribution with zero mean and 
some dispersion. The sources that lie 3$\sigma$ away from the mean 
are the ones we identify as excess sources (see Figure~\ref{k-24}; right). 
This criteria yields 16 (11$\pm$3\%) objects with a 3 sigma detection 
($K-$[24] $>$ 0.45 mags) of excess at 24~\mic. 

We plot the $H\alpha$ equivalent widths versus the excess 
$K-$[24] colors in Figure~\ref{k-24}.  It is clear from Figure~\ref{k-24}, that the WTT excess objects do not
all lie near some  WTTS/CTTS boundary, and the vast majority of these objects
have \ewha\ much lower than 10\AA. Almost all (30 out of 33) CTTS show 24~\mic\ excess. 
The exceptions are RX J1150.4-7704, RX J1518.9-4050 and ROX 39. The first two   
stars would have been WTTS according to the low resolution \ewha , but are 
classified CTTS because of large \ha\ widths seen in high resolution spectra. 
ROX 39 is a K5 star with \ewha\ $=3.6$ \AA\ which is near the WTTS boundary.
Thus, they are relatively weak accretors. 

\begin{figure}[ht]
  \centerline{
    \includegraphics[width=8cm]{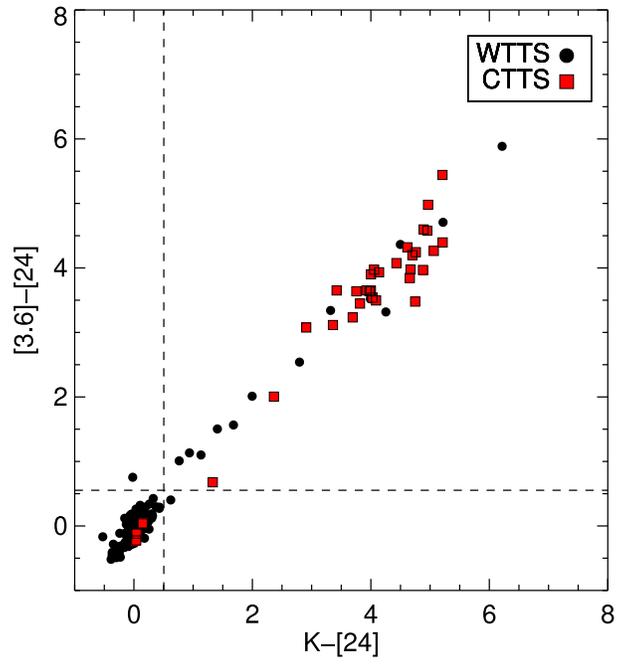}        
   }
  \caption{ (Left) Excess [3.6]$-$[24] vs $K-$[24] colors for our sample. 
    WTTS are shown as filled circles. CTTS are shown 
    as squares. The dashed-lines are the 3$\sigma$ markers for excess 
    identification. 
  }
  \label{i124}
\end{figure}

In Figure~\ref{i124}, we plot the [3.6]$-$[24] color excess against 
the $K-$[24] color excess (or EX($K-$[24]) ) for the sources that had detections in the 
involved bands. We note a high degree of correlation between the 
two excesses and infer that these two measures are consistent 
with each other. All of the 148 WTTS with reliable $K-$[24] colors also have
IRAC1 photometry. Of these, 16 (11$\pm$3\%) had excess detection above 
the 3$\sigma$ level, the same as our estimate from $K-$[24] colors. 
However, one of the candidate sources from the $K-$[24] excess objects,
is not found to be a [3.6]$-$[24] excess source, and vice versa. 
This discrepancy is likely due to the fact that there is 
excess in the IRAC1 band itself. It is easy to see from Figure~\ref{i124},
that the WTTS color excesses have a much wider 
distribution than the CTTS excesses.  Counting only the excess objects, the fraction of
CTTS with EX($K-$[24])$<$2.5 mags is only 6.7\% (2/30). 
However, 44$\pm$16\% (7/16) WTTS excess sources fall in
this range. Thus, roughly half of the excess WTTS fall between
CTTS and diskless stars, in terms of 24 \mic\ excess. 

\subsubsection{Disk ``Turn-On'' Wavelengths and Disk Holes}

\begin{figure}[ht]
  \centerline{
    \vbox {
      \hbox {
        \includegraphics[height=8cm, width=8cm]{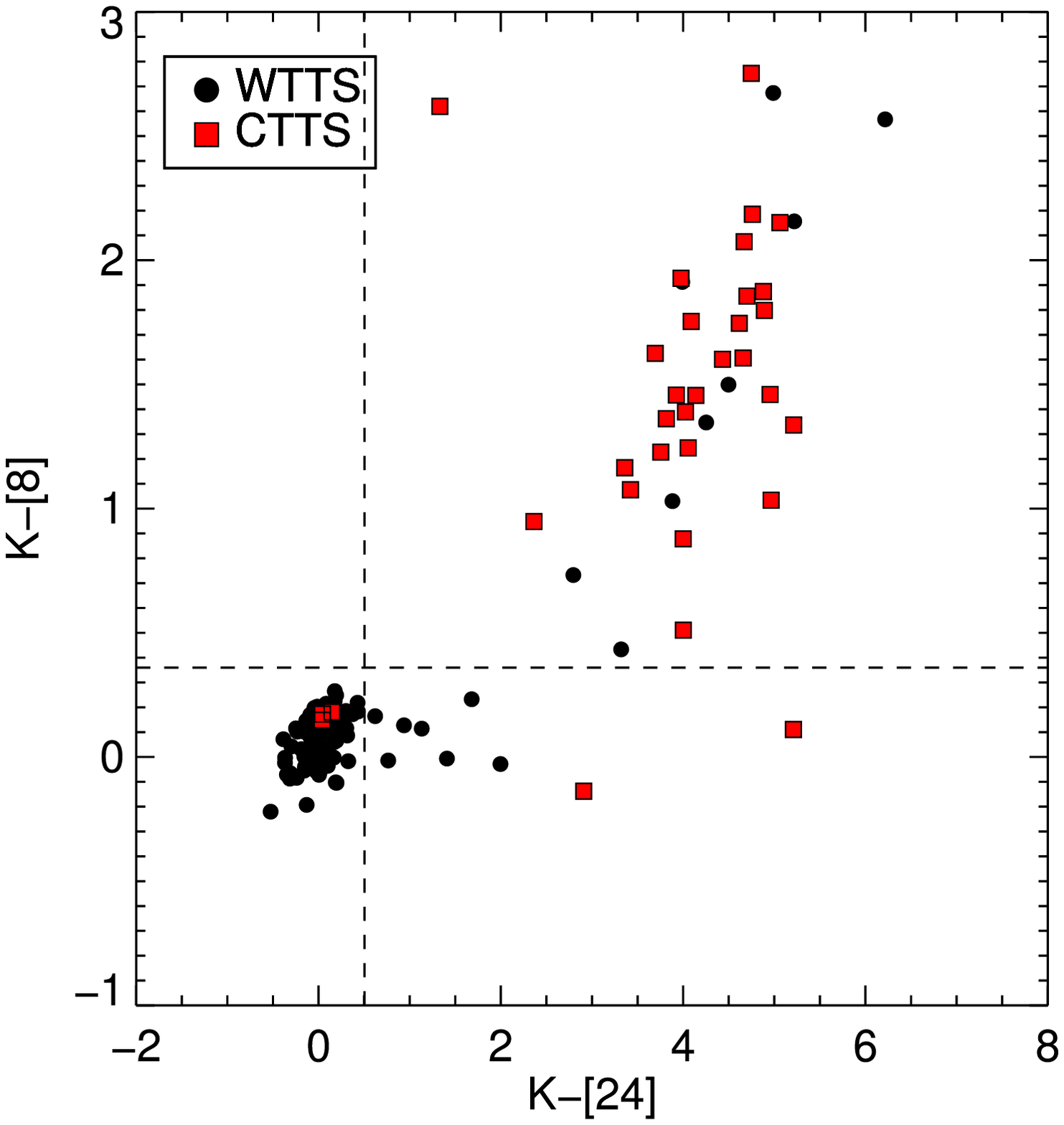}        
        \includegraphics[height=8cm, width=8cm]{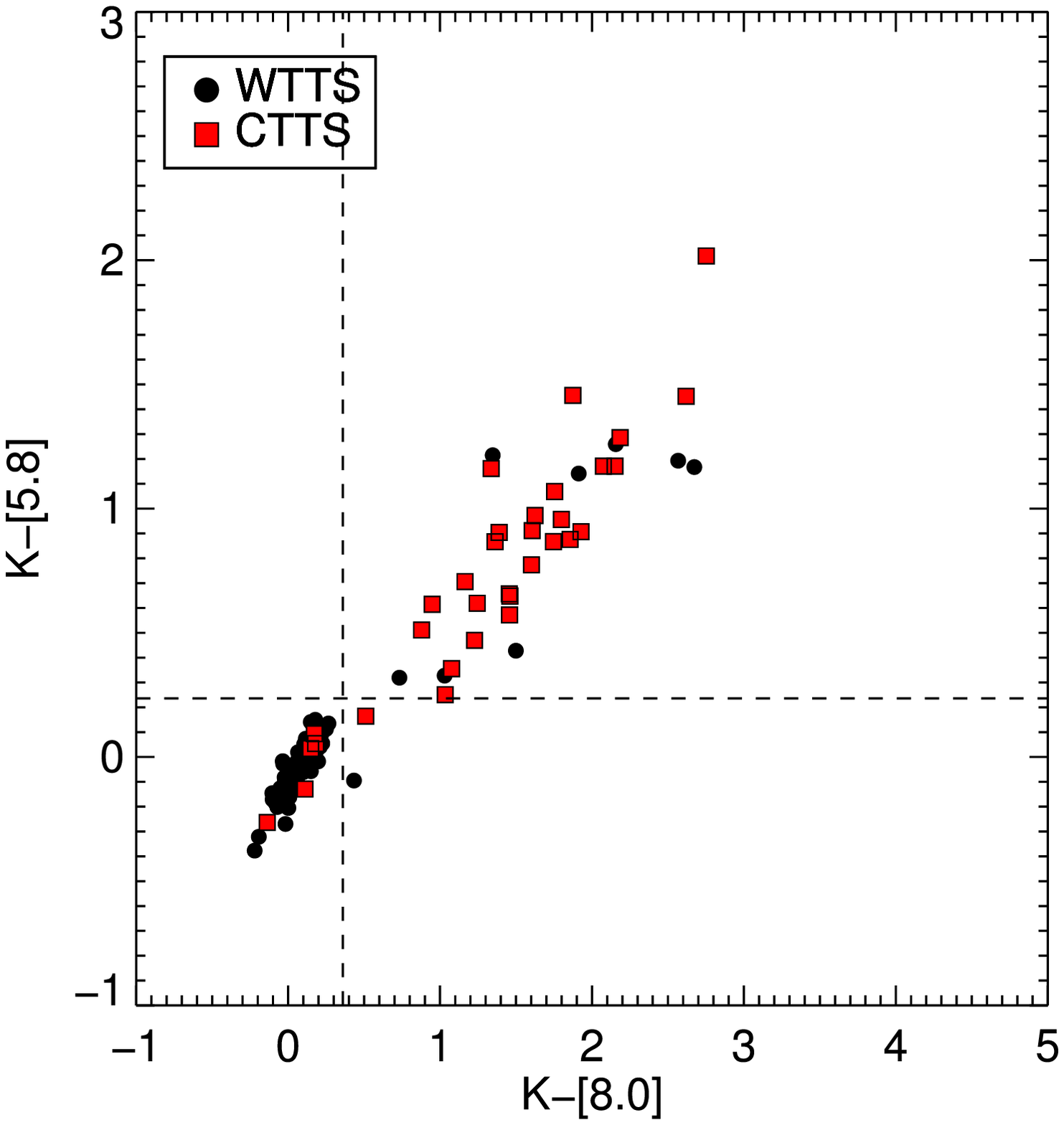}
      }
      \hbox {
        \includegraphics[height=8cm, width=8cm]{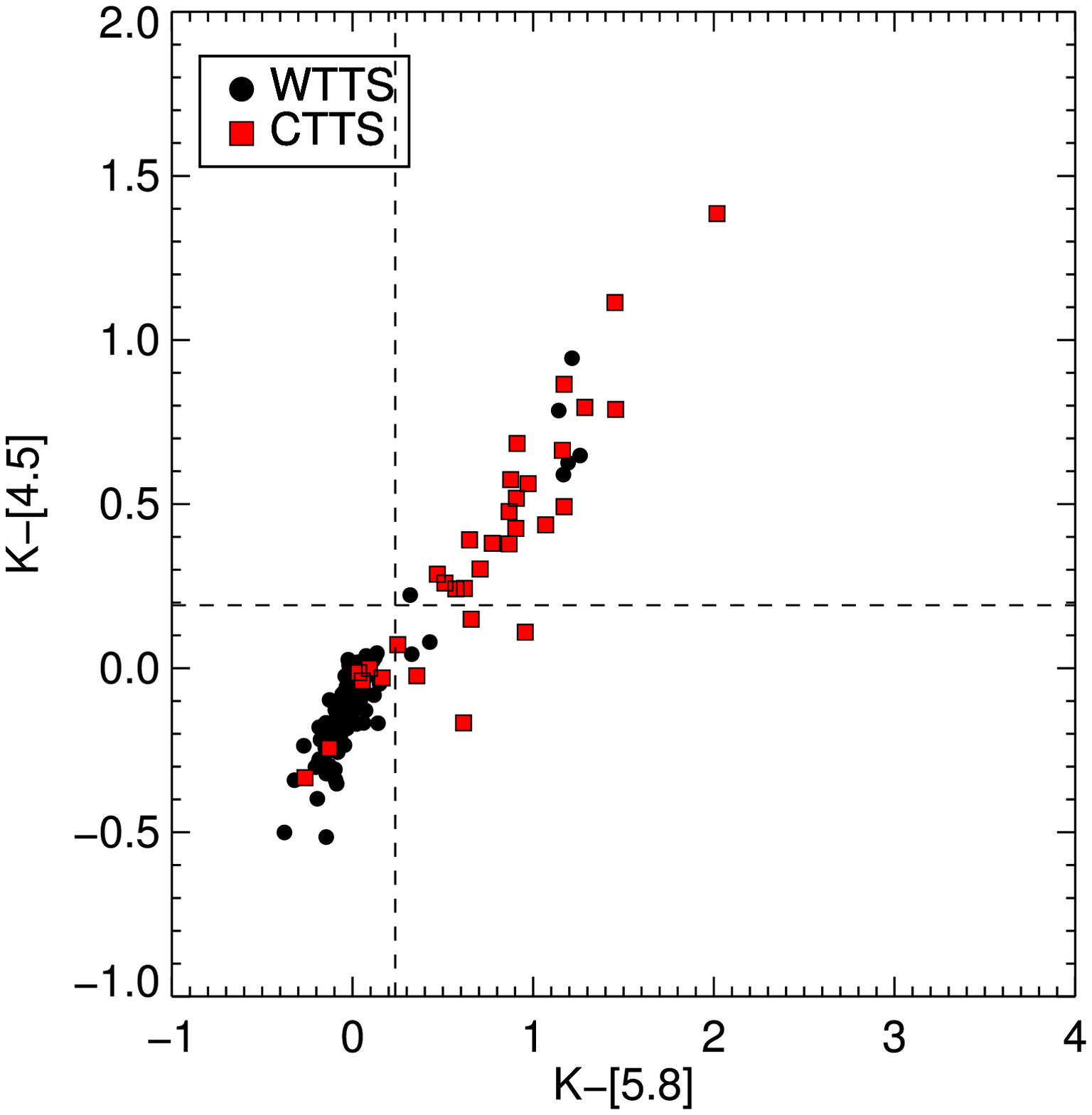}        
        \includegraphics[height=8cm, width=8cm]{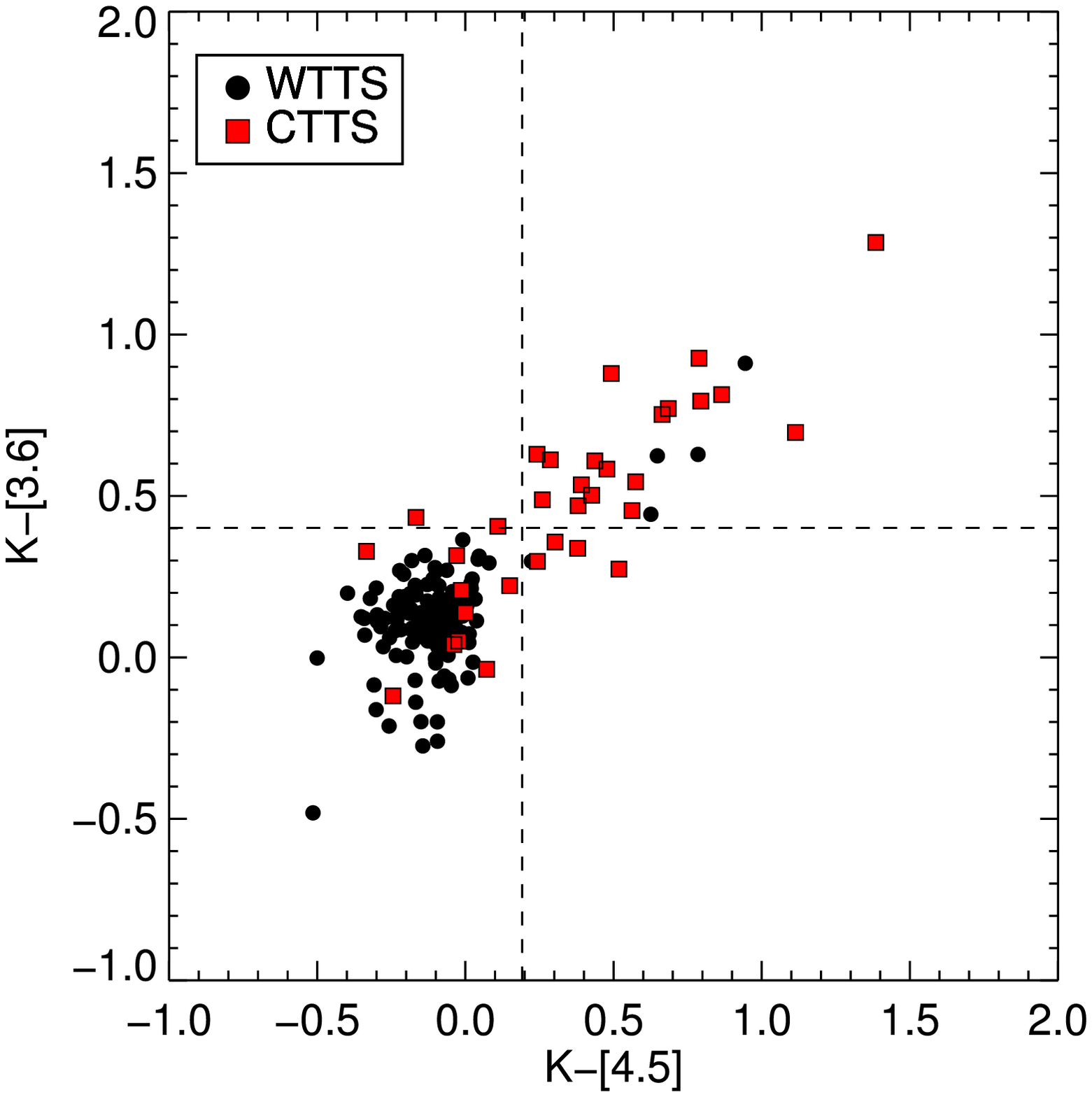}
      }
    }        
  }
  \caption{ 
    WTTS and CTTS symbols are the same as before. We plot the excess $K-$[X] color in adjacent bands against 
    each other, revealing the shortest wavelength band at which the disk emission first ``turns on''. 
    Here [X] represents the wavelength of the {\it Spitzer} band. The dashed-lines represent 
    the 3$\sigma$ markers for excess identification. 
  }
  \label{kxvskx}
\end{figure}

Excess in a particular band indicates the presence of circumstellar dust with temperatures in a certain
range. Consideration of the frequency form of Wien's displacement law, suggests that these 
temperatures are roughly 40~K, 120~K, 350~K, 480~K, 620~K, 780~K for the 70, 24, 8, 5.8, 4.5 and 
3.6~\mic\ bands respectively. For the median stellar luminosity of our sample, 0.5~\lsun , we estimate that
these temperatures are reached by circumstellar dust at orbital radii of 
100, 10, 1.3, 0.7, 0.4 and 0.25~AU respectively. When infrared excess in a particular band 
is accompanied by the lack of excess at shorter wavelengths, an 
inner cleared region or hole in the circumstellar disk is indicated.  We plot
the excess in adjacent {\it Spitzer} bands against each other, in Figure~\ref{kxvskx}, in order to identify
the disks with inner holes. We also try to identify the shortest wavelength at which the disk emission 
first ``turns on'', i.e., appears above detection limits, as we trace the excess from short to long wavelengths.  
In the same sense, the disks ``turn off''  as we go from 
long to short wavelengths, when their excess emission becomes too small to be detectable.

In the top-right panel of Figure~\ref{kxvskx}, we plot the excess $K-$[8.0] colors 
against the $K-$[24] colors of our sample.
There are 7 WTTS with 24~\mic\ excess that lack 8~\mic\ excess. Thus, 10 disks ``turn  on'' at 
24~\mic , another one turns on at 8~\mic , 2 more turn on at 5.8~\mic\ and 2 more turn on at 4.5~\mic ,
leaving 4 WTTS which were already turned on at 3.6~\mic\ and possibly also have excess in the $K$ band. 
We designate these objects according to their disk ``turn-on'' wavelengths. 
Thus, we have 7 {\bf T24} (turns on at 24~\mic ) objects, 5 {\bf TIRAC} objects (turns on at IRAC wavelengths), and 4 {\bf TNIR} 
objects (already show excess at IRAC1 and could have $K$ band excess). We will use these designations to describe
the spectral energy distributions (SEDs) of the objects throughout the paper, and later connect these to plausible physical 
interpretations of the evolutionary states of the disks. CTTS disks also turn off at IRAC wavelengths, but in fewer numbers than WTTS. Only one-third of CTTS disks turn off in one of the observed bands, 
whereas nearly 80\% of WTTS disks are observed to turn off. 

\begin{figure}[ht]
  \centerline{
    \includegraphics[width=8cm]{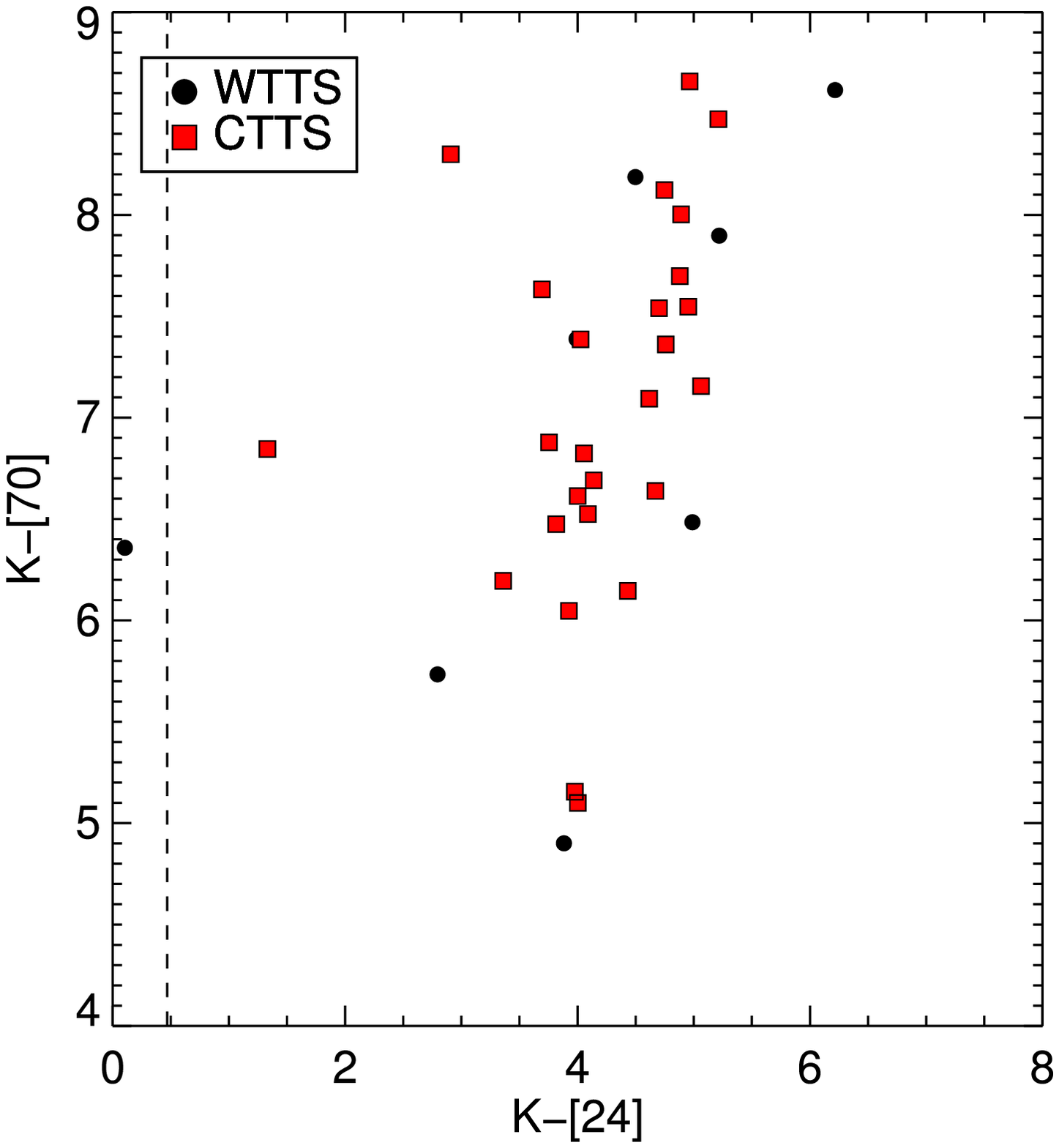}        
  }
  \caption{ $K-$[70] excess color vs $K-$[24] excess color for our
    sample. The dashed-line is the 3$\sigma$ marker for 24 \mic\ excess
    identification. 
  }
  \label{k70k24}
\end{figure}

In Figure~\ref{k70k24}, we compare infrared excesses at 24 and 70~\mic. 
%At 70~\mic , we are unable to detect the photosphere for any 
%of our objects because of the sensitivity limit of our survey.
We see that the 70~\mic\ excess has to be 
roughly 40 times (i.e. 4 magnitudes) greater than the photosphere to be detected.  
The 3$\sigma$ 70~\mic\ sensitivity for our survey 
is roughly 10 mJy (see Table~\ref{stphot}). We detect 8 WTTS and 27 CTTS at 70~\mic\ and all of these are 
detections of excess at least 3.5 magnitudes brighter than the
photosphere. 
Our 70~\mic\ excess detection rate for WTTS is 5$\pm$2\%, while for CTTS it is 82$\pm$16\%. 
We find only 1 WTTS with 70~\mic\ excess for which 
we detect no 24~\mic\ excess. This hardly changes the overall excess rate for WTTS
11$\pm$3\% (17 of 148).  This WTTS is HBC 422, a companion to another WTTS, about 35$''$ away, HBC 423 
which itself has a 70 \mic\ excess. If the excess is truly associated with the star, 
then this indicates cool dust in orbit relatively far from the star (\app 50 AU).
To keep track of the SED designations we call this a  {\bf T70} object since its excess turns on 
at 70~\mic\ .  The disk is very likely optically thin. 
None of the CTT SEDs show evidence of such large inner holes. 

In summary, while \app 89\% of WTTS are diskless,  the rest have a rich variety of SEDs, indicating 
a wide range of evolutionary states. If circumstellar disks
progressively clear from the inside out, the SED types ordered from least to most evolved would be 
{\bf TNIR, TIRAC, T24} and {\bf T70}. We designate diskless objects as {\bf RJ} indicating that their SEDs have a Rayleigh-Jeans slope longward of the $K$ band.

\subsection{Properties of On-cloud and Off-cloud Sources}

The discrepancy in the WTTS excess fractions found in the earlier
papers in this series (\app 20\% in \citet{2007ApJ...667..308C} and \app 6\%
in \citet{2006ApJ...645.1283P}) strongly suggested that the distance from the
cloud may be an important factor affecting the excess rate. It is
likely that the WTTS with larger separations from the clouds include 
a much older population of stars than the ones close to the clouds. 
To investigate this connection, we attempted to find the projected 
separations of our sample from their parent clouds
in a systematic manner. We defined the cloud edge as the $A_V=3$ contour line.
However, extinction maps produced in the same fashion for all the clouds were not available, 
especially not for regions a few degrees away from clouds. Thus, we decided to use the 
dust temperature derived all-sky extinction maps created by Schlegel and Finkbeiner (1998). 
These maps were created from COBE/DIRBE and IRAS/ISSA maps and have the calibration 
quality of the former with the spatial resolution of the latter. The maps show a great deal of filamentary 
structure and trace the H~\Rmnum{1} maps well, but are less reliable in regions where H~\Rmnum{1} is saturated. 
In comparison to more ideally produced extinction maps, as in \citet{1999A&A...345..965C}, we find that 
the agreement of the contour lines are at the 10$'$ level. We should
note that, while in cold regions these maps may give reliable
extinctions, where there is heating by nearby hot stars as in the case of
Ophiuchus, the maps become suspect. From the analysis that follows, however, 
we will see that we have sufficient accuracy to obtain meaningful results.   

\begin{figure}[ht]
  \centerline{
    \includegraphics[width=8cm, height=8cm]{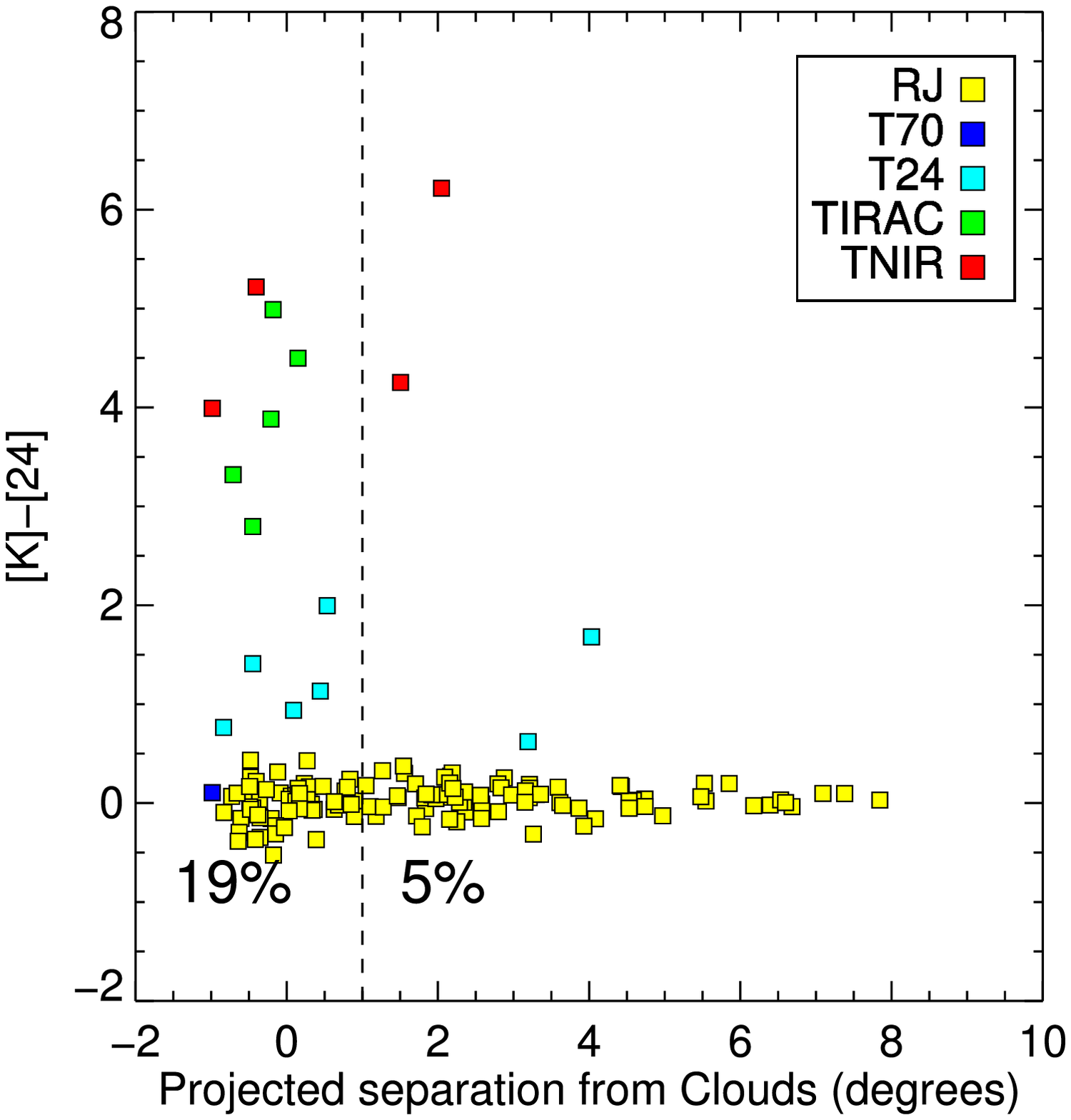}        
    \includegraphics[width=8cm, height=8cm]{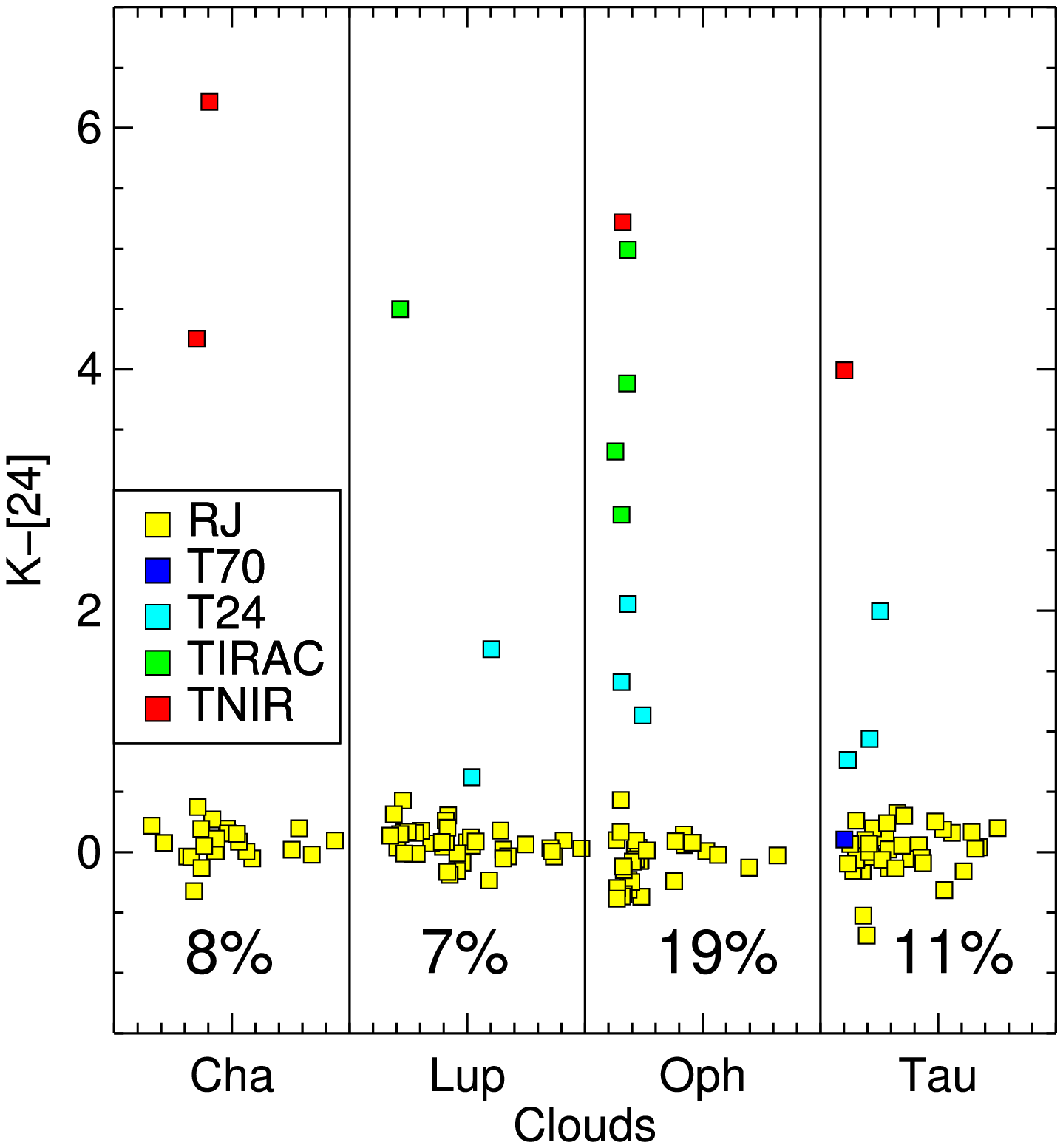}        
  }
  \caption{ (Left) Excess $K-$[24] color versus distance from cloud edge in degrees. 
    Objects are colored according to their disk ``turn-on'' wavelengths. {\bf RJ} indicates diskless objects.
   The dashed line is the 1$^o$ from cloud edge demarcation. The excess fractions on either side of this 
    boundary is given at the bottom of the plot. 
    (Right) The excess $K-$[24] color of the WTTS shown for each cloud 
   separately. A small offset proportional to the projected separation from the parent cloud 
   has been added in the horizontal direction for ease of viewing. 
 }
  \label{dist}
\end{figure}

Plotting the projected distances ($r_c$) of the WTTS estimated from these extinction maps against their 
$K-$[24] excess colors, we find that of the 70 WTTS within 1\dg\ of their parent cloud, 
13 (19$\pm$5\%) have disks (Figure~\ref{dist}). Regions within 1\dg\ of the cloud edge will be called ``on-cloud'' hereafter.
The excess rate for the 78 WTTS that lie farther away is 5$\pm$3\%, i.e., 
4 WTTS disks are clearly off-cloud.  Given the significant discrepancy between on-cloud and off-cloud excess rates, it seems 
likely that the off-cloud WTTS are a physically different group of objects, perhaps older. 
Moreover, we should remember that the 70 WTTS which are projected within 1\dg\ 
of the clouds still have some foreground off-cloud objects, so the actual on-cloud excess rate
could be much higher than we estimate here. 
%From Figure~\ref{seds}, we see that the longer the disk ``turn-on'' wavelength, the smaller is the
%$K-$[24] excess color. Thus, as the disk inner holes grow, the disks also become more tenuous.

In the right panel of Figure~\ref{dist}, we display the $K-$[24] excess colors of each   
of our targeted WTTS clouds separately. 
We find excess   
fractions of 8$\pm$6\%, 7$\pm$4\%, 19$\pm$7\% and 11$\pm$5\% for the Chamaeleon, Lupus, Ophiuchus and Taurus   
regions respectively.  In terms of the distribution of the different kinds of WTTS disks, all the clouds are quite similar, although
we should note it is difficult to draw distinctions from so few detections.
Only Ophiuchus has a notably higher WTTS disk fraction, while the rest of the clouds have
virtually indistinguishable  fractions given the large uncertainties.
An offset proportional to the projected separation 
of WTTS from their parent clouds has been added to the x-axis values in Figure~\ref{dist} (right). 
We note that the off-cloud WTTS objects with excess are   
two {\bf TNIR} disks near Chamaeleon and two {\bf T24} disks near Lupus.   
 
The WTTS with excess  in 
Taurus and Ophiuchus are all within 1 \dg\ of their 
cloud edge, while all the Chamaeleon excess 
sources lie off-cloud (Figure~\ref{dist}). 
In the literature the mean ages for Taurus and Ophiuchus
are often found to be younger than the other two clouds \citep{2007ApJ...657..511A}. 
  
\begin{figure}[ht]
  \centerline{
    \includegraphics[height=10cm]{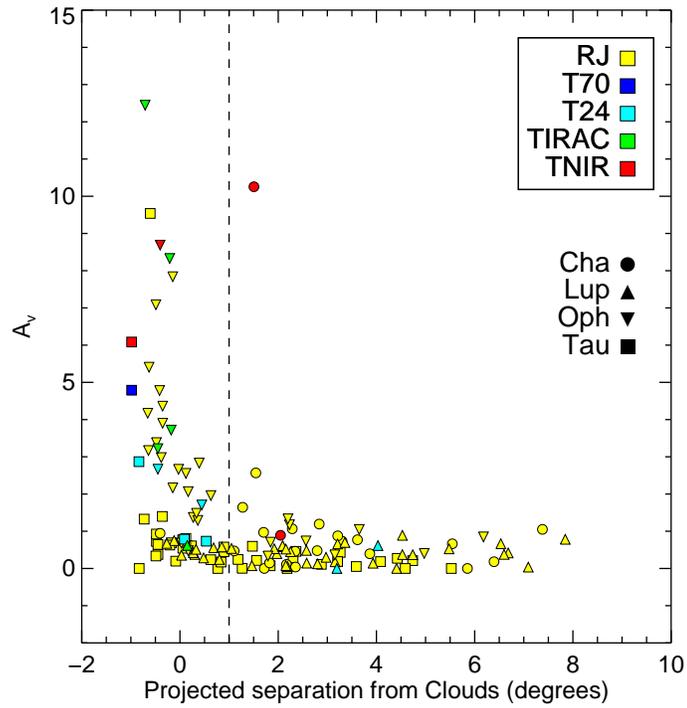}        
  }
  \caption{$A_V$ derived from $J-K$ color plotted against projected distance from cloud edge. Negative projected 
distances indicate objects inside the clouds. Objects are colored
according to SED types and are given symbols according to parent cloud region. As expected almost all high 
extinction objects are on-cloud, while almost all off-cloud objects have low extinctions.
}
  \label{AvsDist}
\end{figure}

In Figure~\ref{AvsDist}, the $A_V$s derived from $J-K$ colors are plotted against the 
projected distances from the respective clouds. Apart from one clear outlier (T Cha, an optically thick disk) with 
an $A_V\sim~10$, roughly 1.5 degrees from the cloud edge, we see that all 
high $A_V$ objects lie within 1 degree of the cloud edge and the highest $A_V$ sources 
lie well within. T Cha is thought to have a close to edge-on disk \citep{1993A&A...272..225A,2007ApJ...664L.107B} 
and so in this case the high extinction is probably caused by 
disk occultation. Almost all the other off cloud sources have low extinctions. 
This suggests that both $A_V$ and distance from cloud edge has been 
derived with high fidelity and, moreover, the high $A_V$ sources are very likely embedded in
the clouds, and thus real cloud members. The excess rate for WTTS with $A_V > 1$ 
is 11 out of 41 (27$\pm$8\%). Objects with less than zero projected cloud 
separations (these are within the cloud boundaries), have extinctions spanning the 
entire range starting from zero. Thus, these objects constitute both foreground and 
cloud-embedded objects. 

\begin{figure}[ht]
  \centerline{
    \includegraphics[height=10cm]{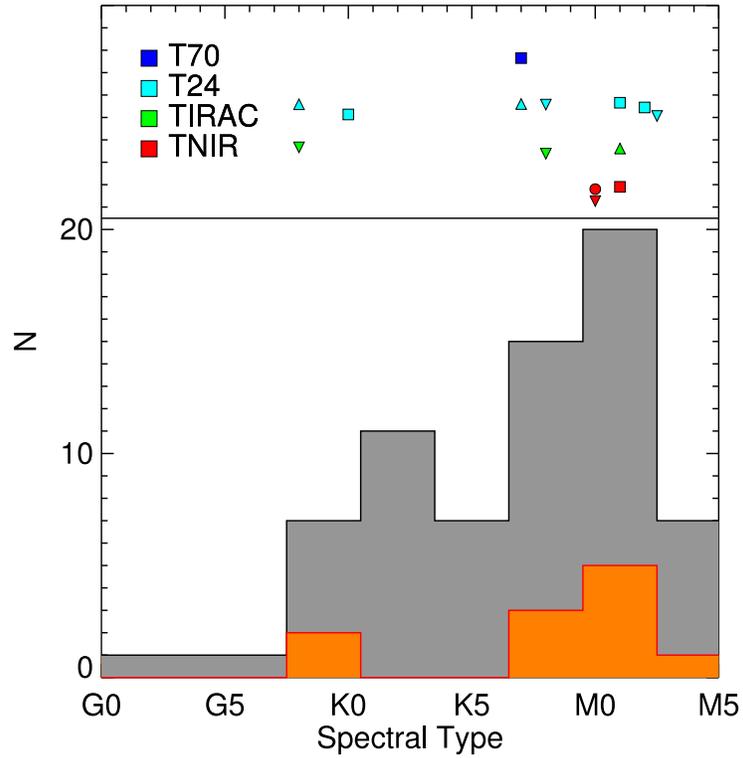}        
  }
  \caption{Histogram of the spectral types of WTTS with $r_c < 1$\dg\ are shown in gray. 
    The histogram of the subset with disks are shown in orange. A very small minority of WTTS with spectral types earlier than 
    G0 are not shown. Also overlaid on this plot are the symbols representing the types of the individual WTTS disks. 
    The shapes of the symbols represent different clouds, same as in Figure.~\ref{AvsDist}.
    The excess rate for WTTS between G0 and K5 is 2/20 (10$\pm$7\%), while the excess rate for 
    WTTS between K6 and M5 is 9/46 (20$\pm$7\%).
    The significance of the difference in these rates is only about 1$\sigma$.
  }
  \label{spthist}
\end{figure}

Since the on-cloud WTTS are a younger population with much lower contamination by older objects, 
we investigate the excess rate of these objects with respect to spectral type. In Figure~\ref{spthist},
we plot the histogram of the spectral types of WTTS with $r_c < 1$\dg\ with and without disks. We leave 
out a few very early type objects, because of very low number statistics at that end. The excess rate for 
spectral types G0 through K5 was 2 of 20 (10$\pm$7\%), while the excess rate for spectral types K6 through
were M5 was 9 of 46 (19$\pm$7\%). The later spectral types seem to have a higher excess rate, but the result is only 1$\sigma$
significant. However, we see from the disk type designations of the WTTS in Figure~\ref{spthist} that the less evolved SEDs
seems to be concentrated in the later spectral types. 

\section {Discussion}
\subsection{Disk Fractional Luminosities}
The fractional luminosities of disks ($f_{lum}=L_{disk}/L_*$) provide a 
model independent measure of the dust contents of the circumstellar environments.
It is also a convenient way to characterize the rich variety of SEDs found around WTTS
by a single number, which is expected to have some connection to their evolutionary status. 
We compute the fractional luminosities in the following way. We consider only the 
photometry and upper limits in the IRAC and MIPS bands, since excesses 
were only computable past the $K$ band. The appropriate photospheric 
fluxes are subtracted from the {\it Spitzer} photometry of each star, assuming that the 
$K$ band fluxes come solely from the photosphere. The flux densities are integrated 
according to Simpson's rule from the $K$ band through to 70~\mic . 
The stellar flux density is similarly integrated from 0.01~\mic\ to 100~\mic\ assuming 
a blackbody spectrum with the star's temperature, normalized to the $K$ band flux. Thus, 
\flum\ is computed by taking the ratio of these two integrals. 

However, most of the time, we
have only a 70~\mic\ upper limit and/or no 24~\mic\ detection of excess. Thus, there are
three cases which we treat separately. For case 1, when we have detections of excess in both
MIPS bands, we compute the minimum $f_{lum}$. Past 70~\mic , we assume a modified
blackbody (T=40 K) where the flux density behaves as $f_{\nu} \sim\ {\lambda}^{-3}$, instead of  $f_{\nu} \sim\ {\lambda}^{-2}$ 
(Rayleigh-Jeans slope). For case 2, when we have a 24~\mic\ excess detection but only a 70~\mic\ upper limit,
we use a modified blackbody (T=150~K) past 24~\mic . In a few cases where the modified blackbody
extrapolates to fluxes above the 70~\mic\ upper limit, the upper limit is adopted as the flux
estimate and used as the starting extrapolation point for the modified blackbody. Thus for case 2, we 
are also computing minimum $f_{lum}$s. For case 3, we have no detections of excess in any of the bands,
thus we can only compute an upper limit to the  $f_{lum}$s. This is done by adopting a modified blackbody,
which passes through both 24 and 70~\mic\ upper limits. 
Our estimates for the disk $f_{lum}$s are given in Table~\ref{stprops}.  It seems that the median sensitivity limit for
the WTTS sample is $f_{lum} < 5\times10^{-4}$. We also note that we are not sensitive to disks cooler than 40~K. 
Most of the detected disks have $f_{lum}$s ranging form  $5\times10^{-2}$ to $5\times10^{-4}$, although there are 
disks which are as tenuous as $f_{lum} \sim\ 1\times10^{-5}$. 

The most robust debris disks have
$f_{lum} < 1\times10^{-2}$ (HR~4796A, $\beta$ Pictoris). We have 9 WTTS with smaller $f_{lum}$s,
making them the youngest debris disk candidates. However, we do not know whether these
excess emissions results from remnants of primordial disks or second generation disks 
created by collisional grinding of planetesimals. Only two of these 9 debris disk candidates
lie beyond 1\dg\ of the cloud edge. Moreover, the robust mean age (explained in the next section) 
of the 9 WTTS is 1.5$\pm$0.4~Myr, with a spread of 1.2~Myr, which makes them as young as the on-cloud
WTTS population. Seven of these are {\bf T24} objects, one  is a {\bf T70} and the other one is a {\bf TIRAC} object. Thus, as expected, lower $f_{lum}$s correlate very well with long ``turn-on'' wavelengths.

%%% Edited this far [Move physical interpretation of SED here ?]

\subsection{Age Analysis}

\begin{figure}[ht]
  \centerline{
    \includegraphics[height=10cm]{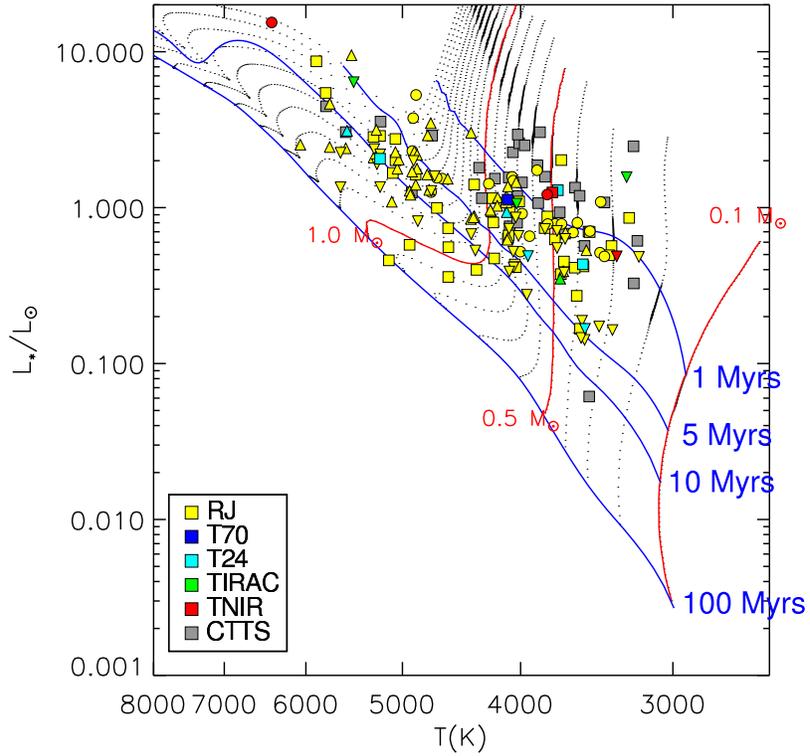}        
  }
  \caption{ Hertzsprung-Russell (HR) diagram for our sample of stars with \citet{2000A&A...358..593S} evolutionary tracks. The solid blue lines are isochrones for 1, 5, 10 and 100~Myr. The 100~Myr line basically coincides with the main sequence track (not shown here). The solid red lines represent mass tracks for 0.1, 0.5 and 1.0\msun, as labeled on the figure.}
  \label{hrd}
\end{figure}

It is interesting to compare the  
ages of the WTTS to the evolutionary states inferred from their disk SEDs.
Age analysis is always a difficult task since evolutionary tracks  
are often extremely close together on the HR diagram.  According to the 
stellar evolutionary tracks of \citet{2000A&A...358..593S}, all the mass tracks 
are in a very narrow temperature range at 1~Myr (3000$-$4500~K) (see Figure~\ref{hrd}). 
For our sample, we have an additional difficulty in estimating ages. 
The distances to the individual objects, especially 
the ones located outside the clouds, are not known precisely 
and should be assumed to be uncertain by \app 20\%. This 
introduces a 40\% uncertainty into the intrinsic luminosity of the 
object, dominating any calibration or bolometric correction uncertainty in the age determination.
Moreover, there is the uncertainty in temperature which is assumed to 
be a spectral subtype. However, these problems are somewhat mitigated 
by the fact that, for low mass stars, luminosities can decrease by a factor of 
6 as the star ages from 1 to 10~Myr. Moreover, when analyzing the properties of an 
ensemble, the median ages of the various classes of objects are 
less disturbed than the individual ages due 
to the present uncertainties. Large variations in the mean ages 
of two classes of objects may still be discernible.

As shown by \citet{2007ApJ...667..308C}, even though absolute ages can vary significantly
depending on which evolutionary models are used, the relative ages between objects are quite 
reliable. Our age estimates, all of which come from the \citet{2000A&A...358..593S} models, should only
be used to compare the relative ages of groups of objects.

We assume that each star in our sample is at the distance of their 
parent cloud which we take from the literature: 180~pc for Chamaeleon II,
200~pc for Lupus, 125~pc for Ophiuchus and 145~pc for 
Taurus\citep{1989A&A...216...44D,1997A&A...327.1194W,1998MNRAS.301L..39W,2008hsf2.book..295C}. 
These distances are used to obtain the absolute $K$ magnitudes from the extinction corrected 
apparent $K$ magnitudes. The intrinsic luminosities 
were computed from bolometric corrections applied to the absolute $K$ magnitudes.
Bolometric corrections to the $K$ magnitudes were derived from a table  
presented in Kenyon~\&~Hartmann~(1995).  
The intrinsic luminosities and the stellar spectral types are then used to estimate the stellar ages, 
using the \citet{2000A&A...358..593S} evolutionary models. 
 In Figure~\ref{hrd}, we place our objects on an HR diagram. The error 
bars in the intrinsic luminosities and the temperatures discussed 
above are not shown in the figure. Plotting these large errors 
bars on an already complex plot would only serve to muddle the subtle
separations between different groups of objects.

In the following analysis, we quote the robust means of the ages
of various groups of stars. This quantity is measured by taking the
median and the robust sigma of a group of objects. The robust sigma
is calculated using the {\it IDL} routine, {\it robust\_sigma.pro}. All values
that lie 3$\sigma$ (robust) away from the median are rejected. The mean of 
the rest of the values is then calculated along with the uncertainty in this mean.
The percentage of rejected values is never more than 20\%.

The first thing to notice in Figure~\ref{hrd}, is that most CTTS are 
younger than 1~Myr (61\% or 20/33). All of them are younger than 10~Myr, 
with just one exception. 
For the WTTS, only 16\% (22/148) are younger than 1~Myr.
For the WTTS younger than 1~Myr, 41$\pm$14\%  (9/22)  
have disks, while only 6$\pm$2\%  (8/126)  WTTS older than 1~Myr have disks.
We can also look at the robust mean ages of CTTS, WTTS with disks and WTTS without disks which are 
0.8$\pm$0.1, 1.3$\pm$0.3 and 3.7$\pm$0.3~Myr respectively.

There is a strong trend in overall excess fraction with age.  
In the three age bins, t $<$ 1~Myr, 5 $>$ t $>$ 1~Myr, 
and t $>$ 5~Myr, the excess fractions were 45$\pm$14\%, 8$\pm$3\% and 4$\pm$3\% of WTTS, respectively. Furthermore,
the mean ages of each evolutionary group of SEDs show a progressive trend.
The robust means of the ages of {\bf TNIR},  {\bf TIRAC},  {\bf T24}, 
and diskless objects are 0.6$\pm$0.1, 1.3$\pm$0.3, 1.5$\pm$0.4, 
and 4.1$\pm$0.2~Myr respectively. 
The robust mean ages of on-cloud and off-cloud 
WTTS are 2.3$\pm$0.2 and 4.2$\pm$0.4~Myr respectively. 
%We should take care not to draw too much from the 
Of course, the individual ages have very large uncertainties, 
and depend on which evolutionary model is used to estimate them \citep{2007ApJ...667..308C,2009ApJ...702L..27B}. 
However, the robust mean ages of each
group of stars, especially their relative ages, are clearly less affected by these uncertainties. 

\begin{figure}[ht]
  \centerline{
    \hbox{
      \includegraphics[height=8cm]{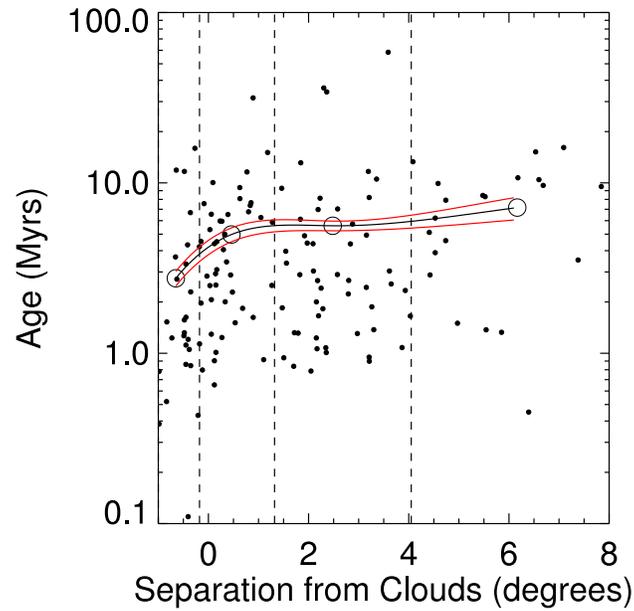}
   }
  }        
  \caption{(Left): The ages obtained from the \citet{2000A&A...358..593S} evolutionary tracks versus the projected separations of the WTTS from their parent cloud. The vertical dashed lines demarcate four cloud-separation bins within which mean ages are calculated. 
The solid black curve is a spline fit through these mean ages. The red curves indicate the 1$\sigma$ errors in the mean ages.
Thus, the mean age of objects increases noticeably with increasing separation from the cloud boundaries.
}
  \label{age_dist}  
\end{figure}

It is difficult to reason that the on-cloud and off-cloud WTTS populations
which are only 2~Myr apart but very well mixed in age, could have such different 
disk fractions (20\% for on-cloud, 5\% for off-cloud). The disk fraction would be much
closer if the ages were really that mixed. We propose that the disk fractions 
are actually very sensitive to age, and the vast majority of disks evolve into the tenuous
end phase of the primordial disk by roughly 3.5~Myr. It is quite plausible
that the ages are actually a strong function of separation from the cloud, but that this relationship
is confused by the uncertainty in the distances.  We investigate this scenario in 
Figure~\ref{age_dist}, where  we plot the measured ages versus the projected separations.
We see that the mean ages in the four bins increase gradually with separation, and the 
errors in the means are small enough to make this trend significant.  At projected
cloud separations of -0.65, 0.46, 2.5 and 6.2 degrees, the mean ages are
 2.75$\pm$0.5, 5$\pm$0.75, 5.6$\pm$0.8  and 7.2$\pm$1.6~Myr. Of course, 
an age-separation relation is not expected to be strict as there many other factors
involved in the spatial distribution of young stars.

\subsection {Connection between Disk Fraction and Multiplicity}

Since even extremely young WTTS have a measurably lower disk fraction than CTTS,
it has been conjectured that unseen stellar companions may be responsible for 
clearing the inner disks of WTTS. The alternative explanations are that WTTS are slightly
older objects or that they resulted from very different initial conditions than CTTS, which 
make them more susceptible to disk dissipation by photo-evaporation 
or tidal interaction, for example.
Nevertheless, we look at published multiplicity studies on our objects, to see if binaries 
show measurably different disk fractions. Also of interest is whether binary 
systems halt disk accretion quickly. In other words, are WTTS preferentially binaries ?
%Of course, for unresolved binaries, the sensitivity to infrared excess is halved, but
%this is a minor issue when we are studying disks with relatively large fractional 
%luminosities.

The most comprehensive study to date on the effect of binarity on the disk excess rate is that of 
Cieza et al. (2009). They found that the IRAC excess fraction for binaries with separation less
40~AU is 38$\pm$6\%, while it was 78$\pm$7\% for systems with larger separation. However, 
their study pertains to inner hole clearing sizes of  $\sim$ 1~AU and somewhat younger objects, since 
their sample was composed of mostly on-cloud objects and they only dealt with the IRAC bands. 

The multiplicity information on our sample comes from several kinds of companion surveys.
Apart from the Cieza et al. 2009 compilation of speckle interferometry, radial velocity, lunar occultation and 
adaptive optics surveys, we found other binarity information in the literature which are listed in Table~\ref{stprops}. To 
summarize the disparate kinds of companion searches, these surveys can find or rule out the 
existence of a companion at two regimes. The spectroscopic surveys are sensitive to systems with
periods of hundreds of days (roughly 1 AU separation) and the imaging surveys are sensitive to companions
with roughly 0.15$''$ separation ($\sim$ 20 AU). We count as binary all objects for which any kind of 
companion was found within 1$''$. All objects which were unsuccessfully searched, we consider single.
Systems with widely separated companions (hundreds of AU) can easily harbor a disk around either
component, as the companion only forces disk truncation to one third the total separation \citep{1977MNRAS.181..441P}. 

In our sample of 33 CTTS, the ratio of the number of CTTS for which companions 
were found versus the total number searched  is 9/19 (47$\pm$16\%). 
The same ratio for the WTTS is 27/82 (33$\pm$6\%). 
%Admittedly, we are dealing with small number statistics here, 
%but we can see that WTTS are not drastically more likely to be binaries.
Despite the small number of sources, there is no evidence in these
data that WTTS are more likely than CTTS to be binaries.

Now, 7 of the 31 (23$\pm$4\%) binary WTTS exhibit IR excess, whereas 
6 of 55 (11$\pm$4\%) purportedly single WTTS show IR excess. 
Since binaries stars have two chances to have a disk,
the disk fraction of an individual star in a binary system is $DF_i = 1-\sqrt{1-DF_s}$,
where $DF_s$ is the disk fraction of the system, which is all we can measure for 
unresolved pair \citep{2009ApJ...696L..84C}. 
Thus, our individual WTTS disk fraction turns out be 12$\pm$3\% virtually the same
as the single star disk fraction. However, our WTTS binaries include systems with 
wide separations ($\sim$ 100 AU) which are expected to be much more conducive to disks.
We also included old WTTS which are likely to have lower disk fractions. A combination
of population contamination and biases is probably masking the effect of binarity on 
disk lifetimes as seen by \citet{2009ApJ...696L..84C}.

%When we consider
%all binary WTTS and CTTS  with projected separation between 0.1$''$ and 1$''$ (roughly 15 to 150 AU) we 
%see an excess rate of 9:25 (36$\pm$14\%). Binaries at these separations should be most inconducive to 
%a circumstellar disk.  

When we try to divide our sample into the two groups chosen by \citet{2009ApJ...696L..84C}, we find that we have very small statistics.
In the first WTTS group we place all the spectroscopic binaries, and the detected binaries with separation less than 0.2$''$ ( roughly $\rho <$ 40~AU), while in the second WTTS group, we put all the wider binaries. 
For the tight binaries, we get an excess fraction of $4/17$ (24$\pm$12\%), while 
the wide binary disk fraction is $3/14$  (21$\pm$12\%). Thus, the results are inconclusive.
If a study complementary to the \citet{2009ApJ...696L..84C} work were to be done for older WTTS and for 
disks detected as MIPS excess, we would need a much larger sample of off-cloud binaries.

Recent studies which have compared disk fractions in single and binary systems 
show how different results are obtained from different samples. \citet{2008ApJ...673..477P} 
found no evidence of the effect of binarity on disk emission in a carefully selected
medium-separation sample ($\rho$~\app 14$-$420 AU) in Taurus (age~\app 1$-$3 years). 
However, for the late-type stars in the 8 Myr old $\eta$ Chamaeleontis cluster, \citet{2006ApJ...653L..57B} 
found that 8 of 9 single stars and 1 of 6 binary systems ($\rho\ <$ 20 AU) had a disk.   
Although, these results seem to show that small separation systems affect disk lifetimes while
large separation systems do not, the two samples have different ages. 
Differing ages, separations, masses and viewing angles of the systems 
under study, which are often quite uncertain, can easily confuse 
the true effect of binarity in small samples. 
\citet{2010ApJ...710..265P} showed that there are no stellar companions in five of the most clearly classifiable 
transition disks, in the region between $\sim$ 0.35 and 4~AU from the stars. Thus, if one has to 
invoke companions as an explanation for the disk clearings, they would have to be substellar.

%such that 
%substellar or planetary mass companions may have to be invoked as an explanation for the 
%inner disk clearings. 
  
%In earlier studies, which had sample sizes in the tens, it was difficult to be sure whether
%biases in the physical separations 
%We conclude that while for the terrestrial planet regions around young stars ($t \sim\ 1$Myr), it is easy to see the clearing 
%effects of a binary companion \citep{2009ApJ...696L..84C},
%such an effect on slightly older stars ($t \sim\ 3$Myr) in regions farther out as detected by 24 \mic\ emission is much harder
%to discern.

\subsection{Physical Interpretation of SEDs}

The objects with SED types {\bf TNIR, TIRAC, T24 and RJ} had mean ages of 
0.6$\pm$0.1, 1.3$\pm$0.3, 1.5$\pm$0.4 and 4.1$\pm$0.2~Myr and 
mean disk fractional luminosities of these groups were
7$\pm$0.7$\times 10^{-2}$, 3.7$\pm$0.7$\times 10^{-2}$, 1.7$\pm$0.5$\times 10^{-3}$, 
3.8$\pm$1$\times 10^{-3}$ and lastly a mean upper limit of 5$\pm$1$\times 10^{-4}$ respectively.
Thus, in terms of age and disk fractional luminosity these statistics strongly
suggest an evolutionary trend through these SED types.

The {\bf T24} objects have comparable $f_{lum}$s to known resolved debris disks, and are probably optically thin 
disks with inner clearings no larger than the terrestrial planet regions. About 6$\pm$4\% of accreting stars (CTTS) have 
SEDs of this type. 
%GM Tau and DM Tau are extreme examples of this type ($M_{disk} \sim 25M_{jup}$) \citep{2007MNRAS.378..369N}.
The {\bf TIRAC} objects
are just a little younger, with smaller inner clearings and larger fractional luminosities which are 
just above the informal debris disk demarcation line at $f_{lum}=1 \times 10^{-2}$. 
About 24$\pm$9\% of accreting stars have SEDs of this type.  The 
{\bf TNIR} disks fall into the category of traditional optically thick disks 
(61$\pm$14\% of CTTS fall into this category).  It should be noted that although we have designated 
the {\bf RJ} objects as diskless they might easily have debris disk of fractional 
luminosity comparable to the $\beta$ Pictoris disk 
in regions analogous to their Kuiper belts.

In Figure~\ref{seds}, we examine the SEDs of these groups of objects. The shapes of the SEDs of most of the
WTTS {\bf TNIR} and {\bf TIRAC} objects indicate a depleted region or a gap in the disks. The excess in the IRAC
bands definitely indicate an inner disk (r\app 1~AU), but the emission coming from the middle disk (r\app 40 AU) 
is obviously much more robust as seen in the MIPS bands. Thus, it seems that the middle disk survives the inner disk.
\citet{2008ApJ...686L.115C} found that objects with low 24~\mic\ 
excesses are not detected in the submillimeter, just as we found that these objects are also not detected at 70~\mic .
They understood this to indicate that the inner disk only dissipates after the outer disk has been significantly depleted
of mass.

\subsection{New Interesting objects}

\begin{figure}[ht]
  \centerline{
    \vbox {
      \hbox {
        \includegraphics[height=3.3cm]{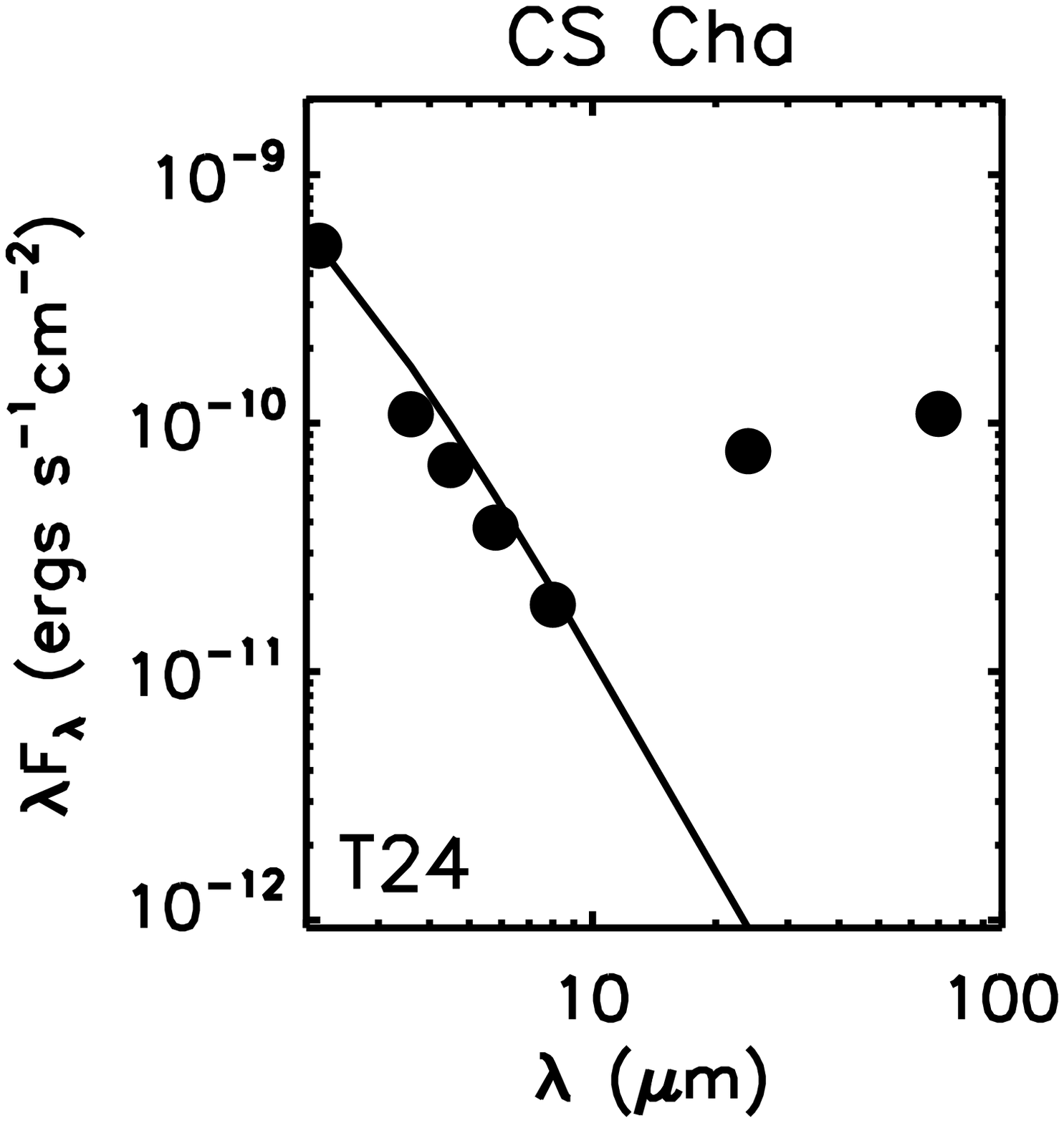}
        \includegraphics[height=3.3cm]{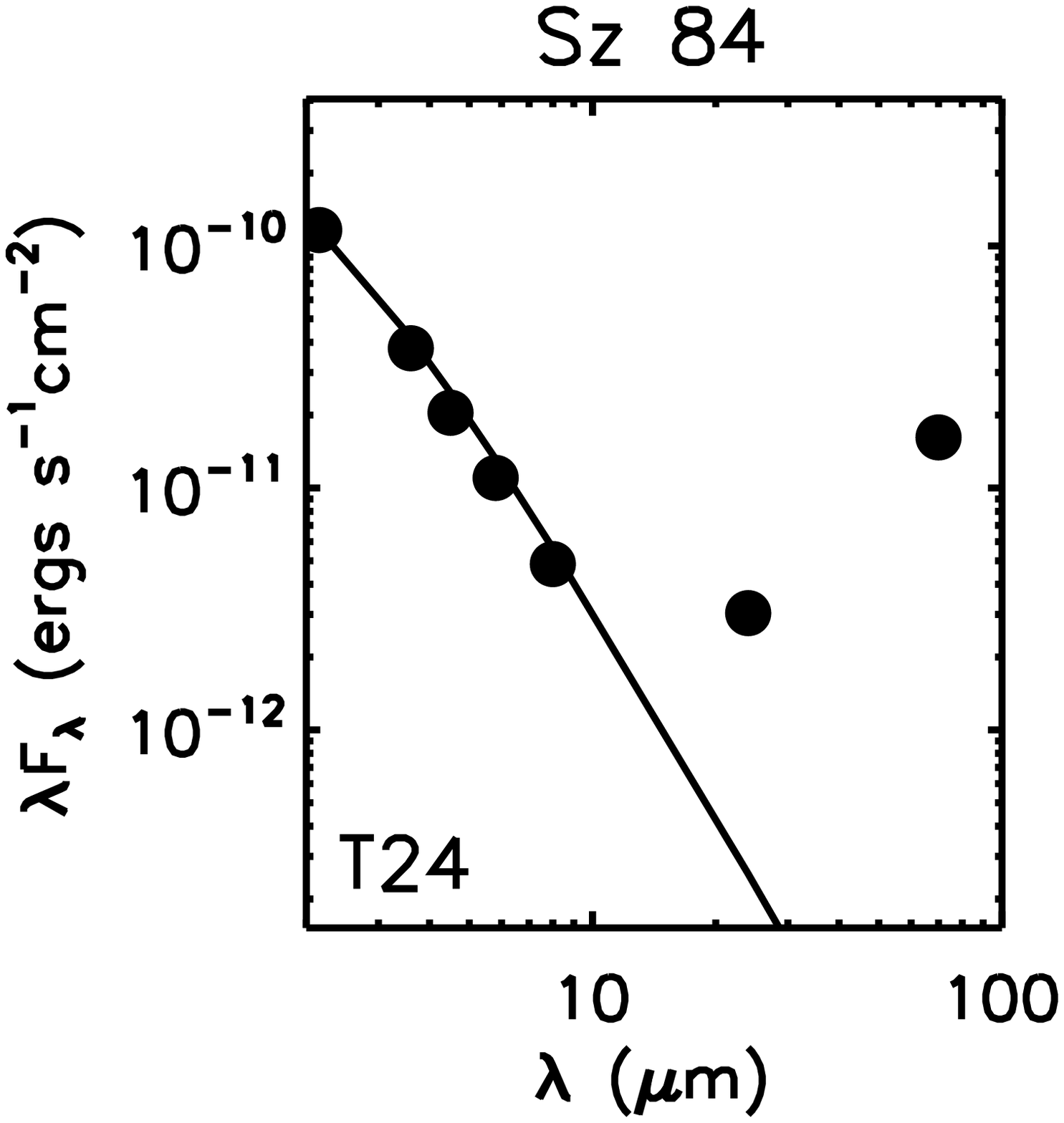}
        \includegraphics[height=3.3cm]{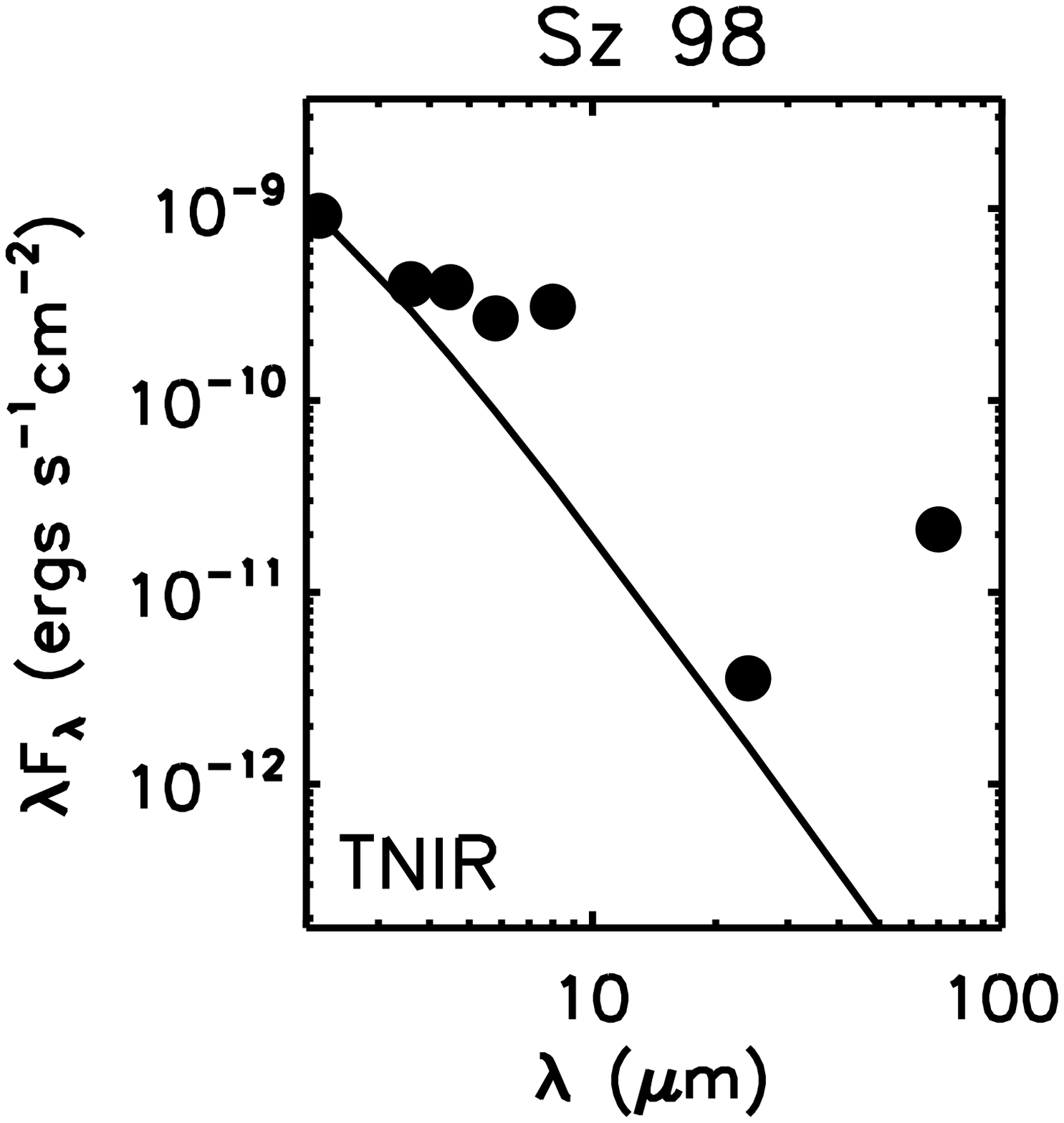}
     }
      \hbox {
        \includegraphics[height=3.3cm]{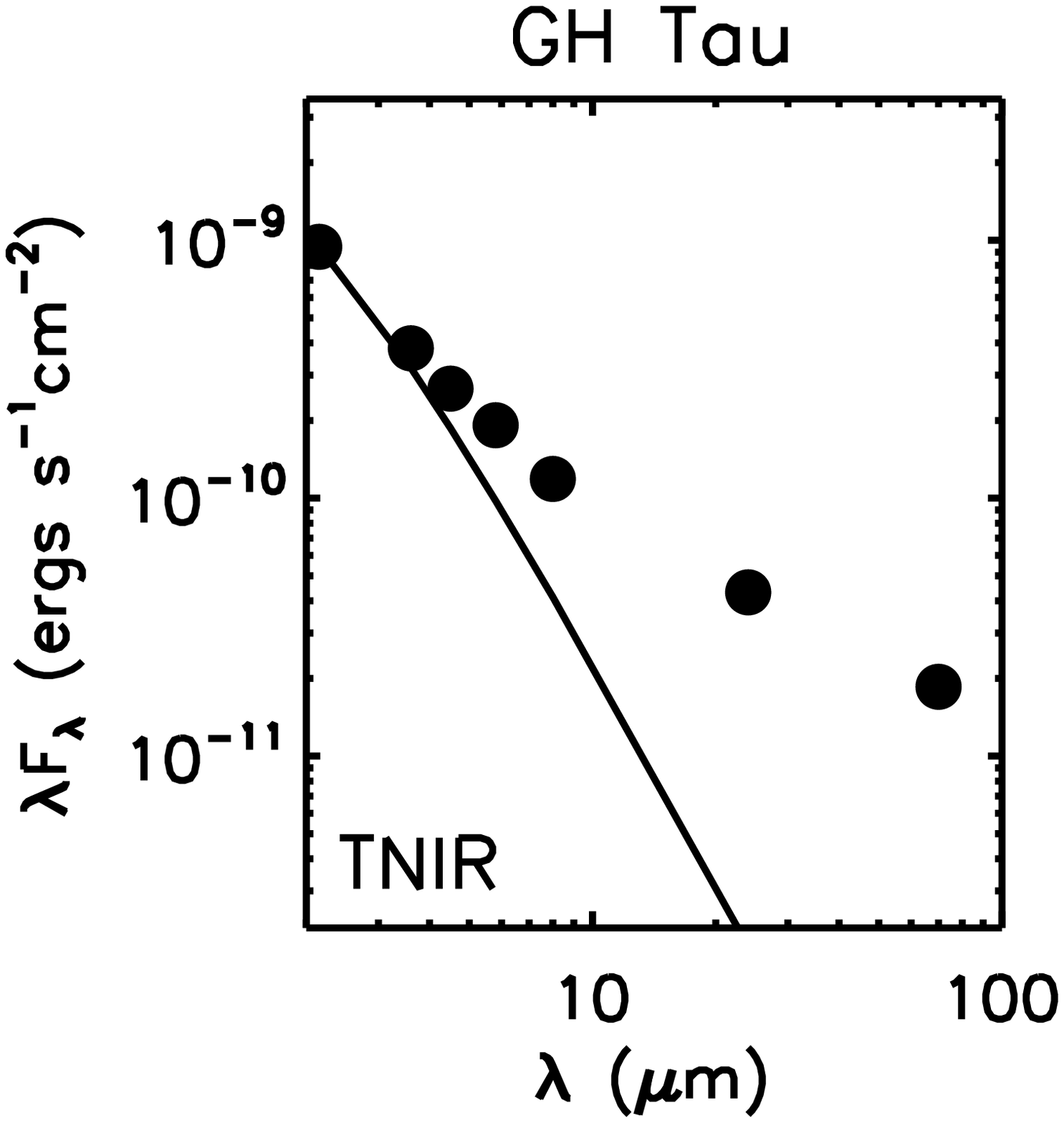}
        \includegraphics[height=3.3cm]{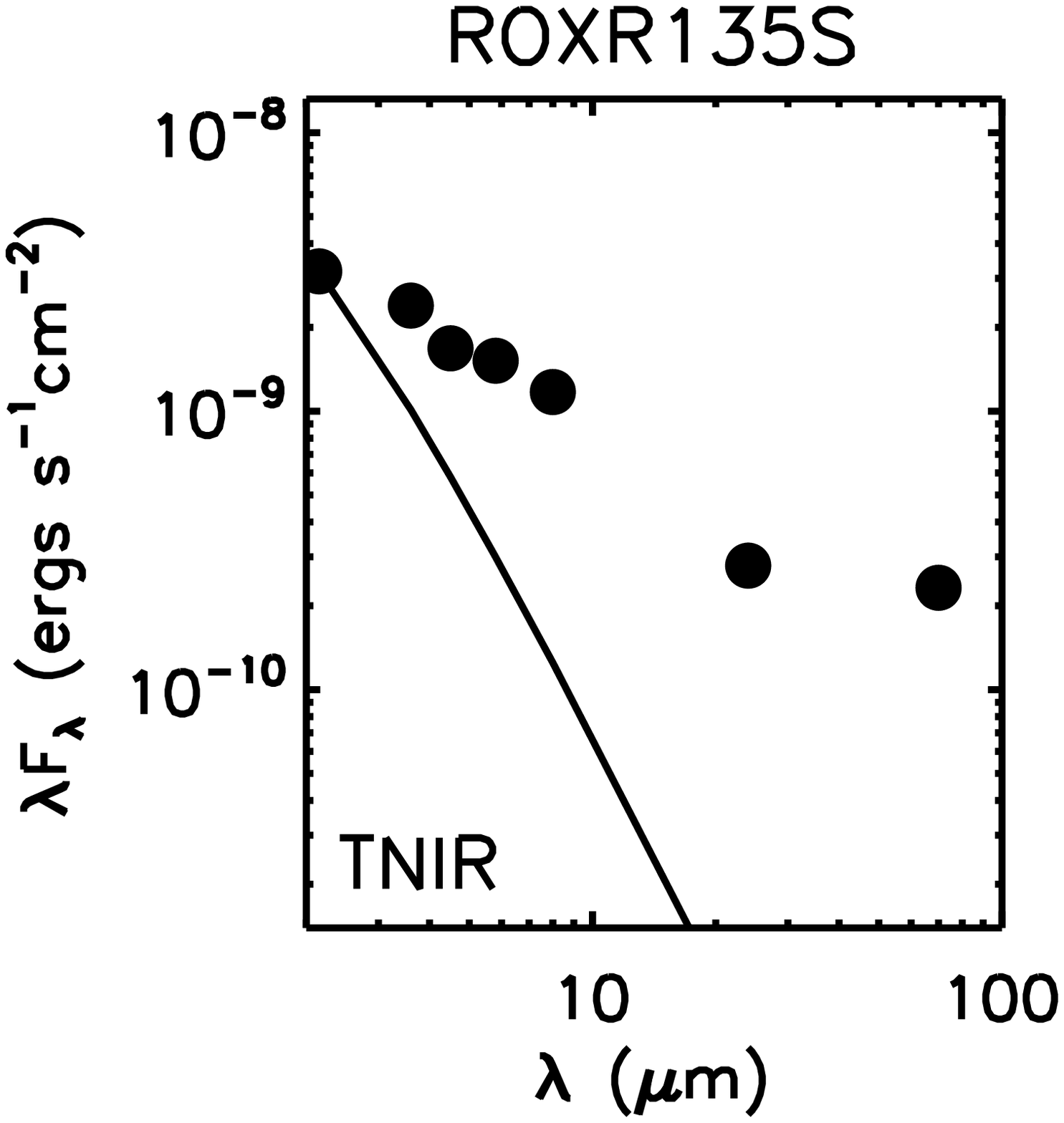}
      }
    }
  }
  \caption{ 
    The SEDs of the interesting CTTS with excess (see Section 4.5). 
    The solid line represents the Planck function of the appropriate
    temperature normalized to the extinction corrected K-band flux. 
    The plots are also labeled with their 
    disk ``turn-on'' classifications which are also given in Table~\ref{stprops}. }
  \label{sedsctts}
\end{figure}

{\bf{\it Accreting Disks with Large Holes.}}
CS Cha and Sz 84 fall into this category of objects.  Was it grain growth that cleared out the dust ?
Are they accreting because there is no inner companion to stop the gas flow or 
are they accreting despite an inner companion?  They show no excess in the IRAC bands 
but show robust excesses at 24 and 70~\mic\ (Figure~\ref{sedsctts}). They both have disks with fractional luminosities of \app 0.1, and they are both solid
accretors with \fwha\ of \app 400 km/s. CS Cha has a companion (Mass~\app 0.1~\msun ) \citep{2007A&A...467.1147G} with a 2435 days 
period ($\rho\sim$~3 AU), while Sz 84 was not targeted in a 
high resolution search. The CS Cha SED from earlier {\it Spitzer} photometry and IRS spectra has been modeled as disk emission in detail to
predict a 43~AU inner hole \citep{2007ApJ...664L.111E}. Given the IRAC and MIPS photometry we obtained, both disks can be modeled
with holes at small as 2~AU.  

\begin{figure}[ht]
  \centerline{
    \vbox {
      \hbox {
        \includegraphics[height=3.3cm]{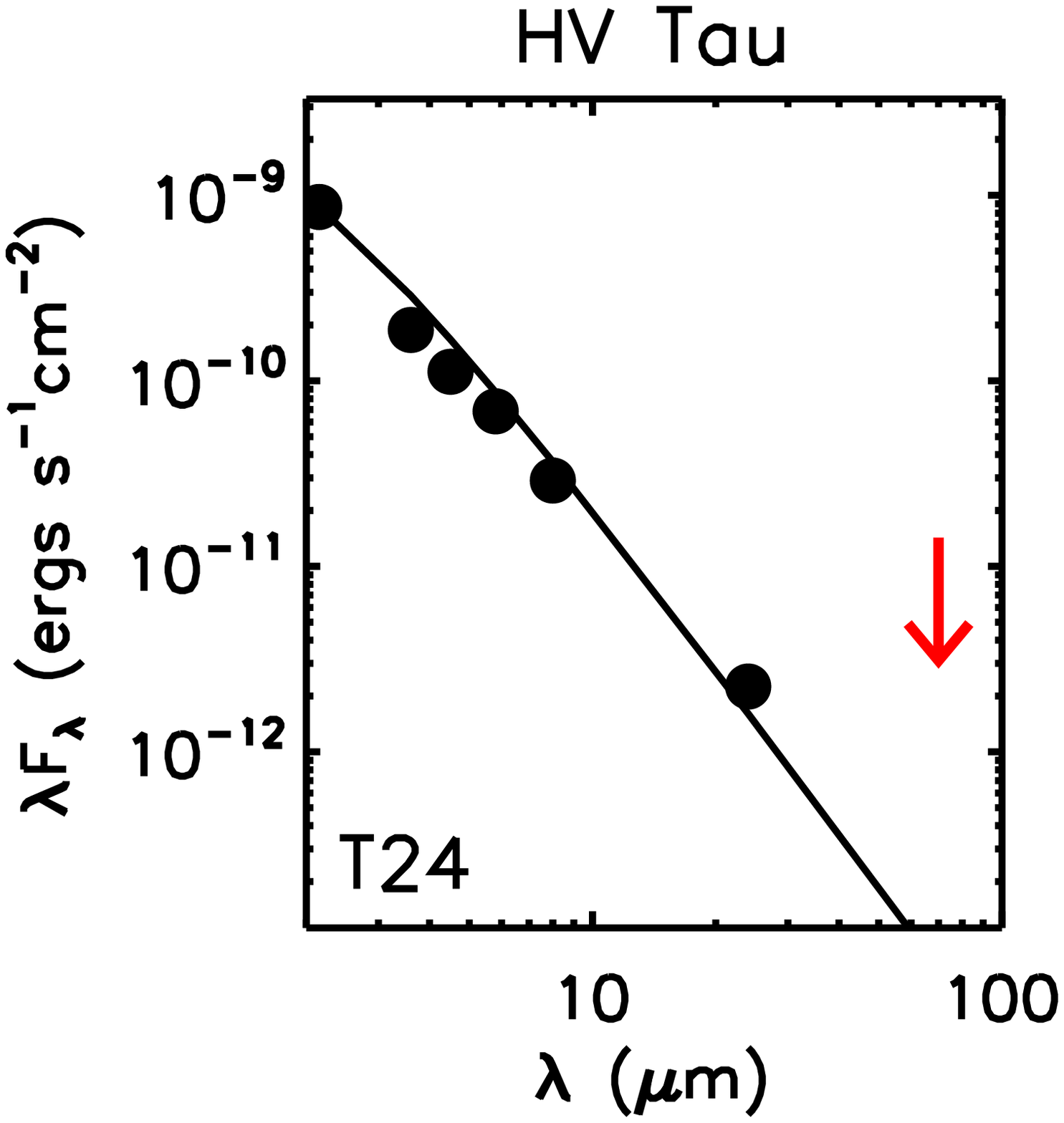}
        \includegraphics[height=3.3cm]{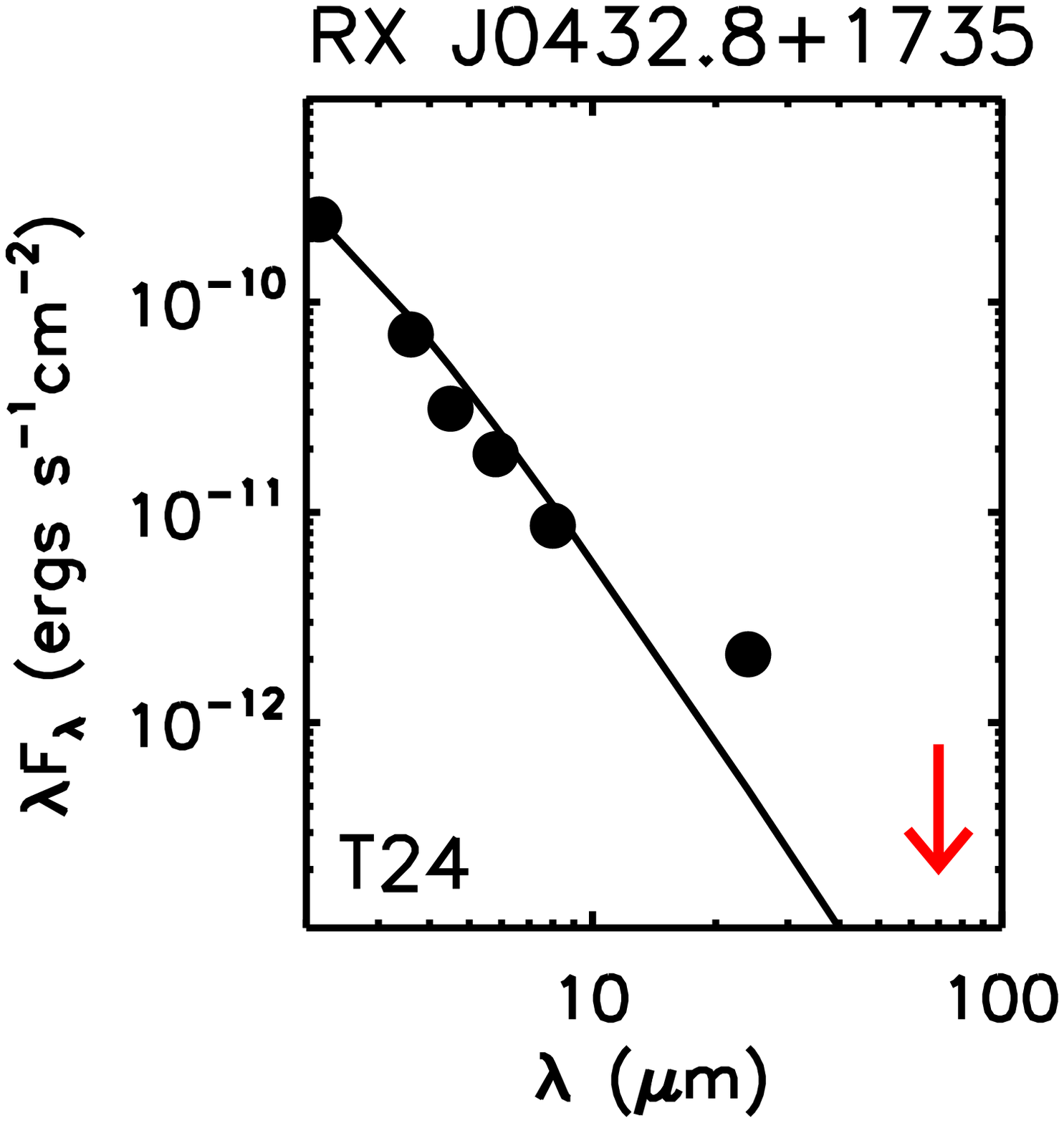}
        \includegraphics[height=3.3cm]{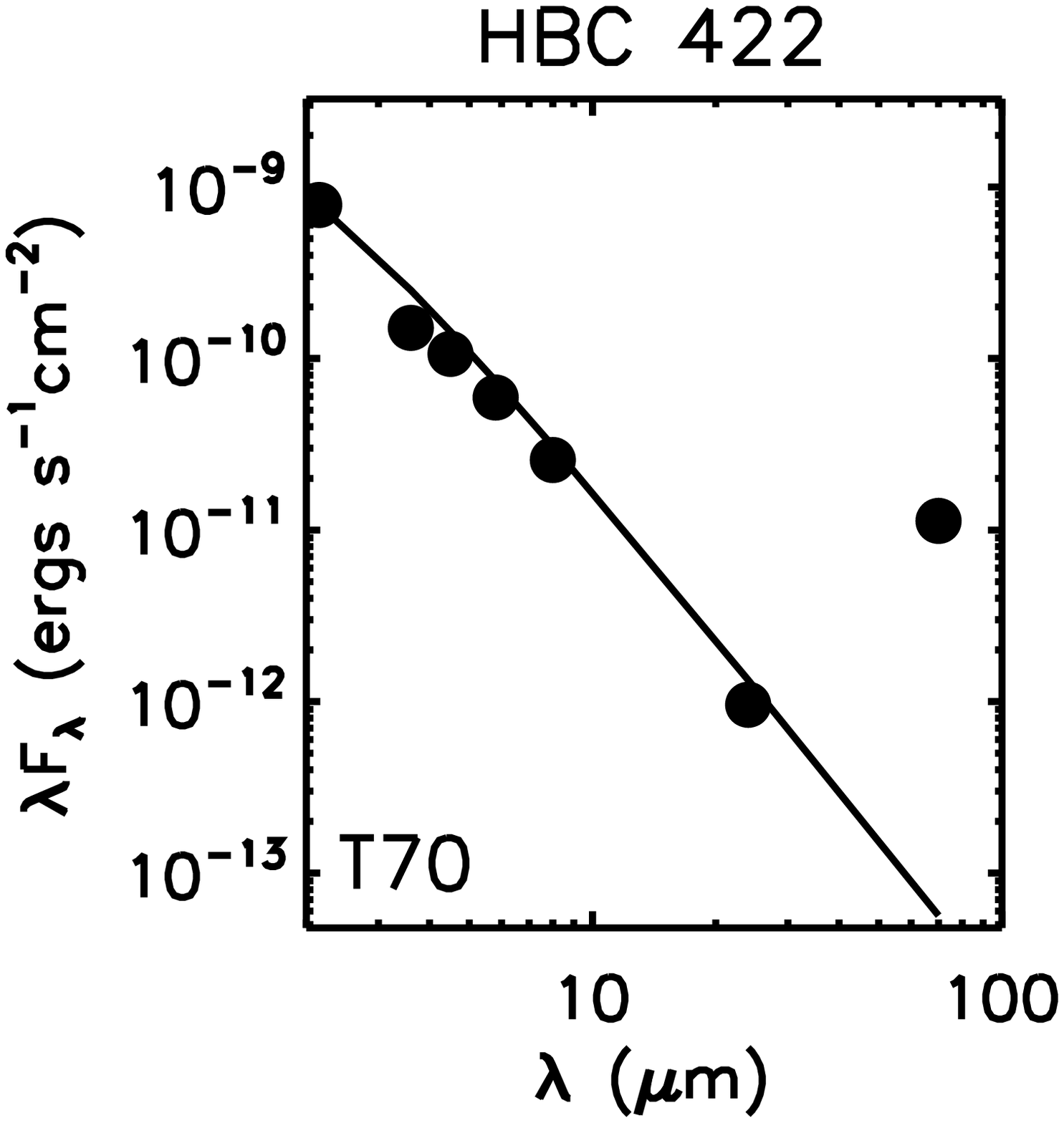}
        \includegraphics[height=3.3cm]{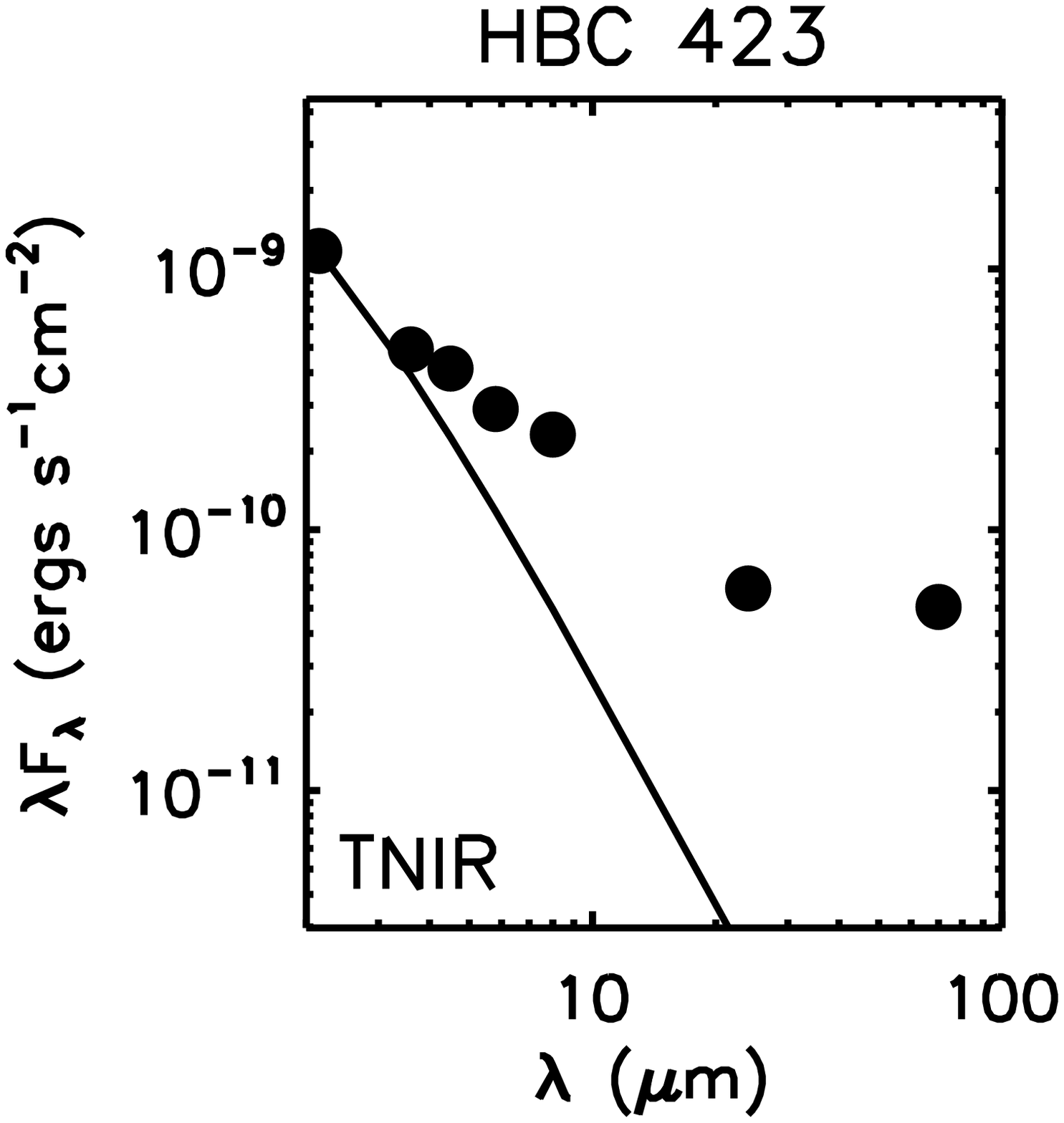}
      }
      \hbox {
        \includegraphics[height=3.3cm]{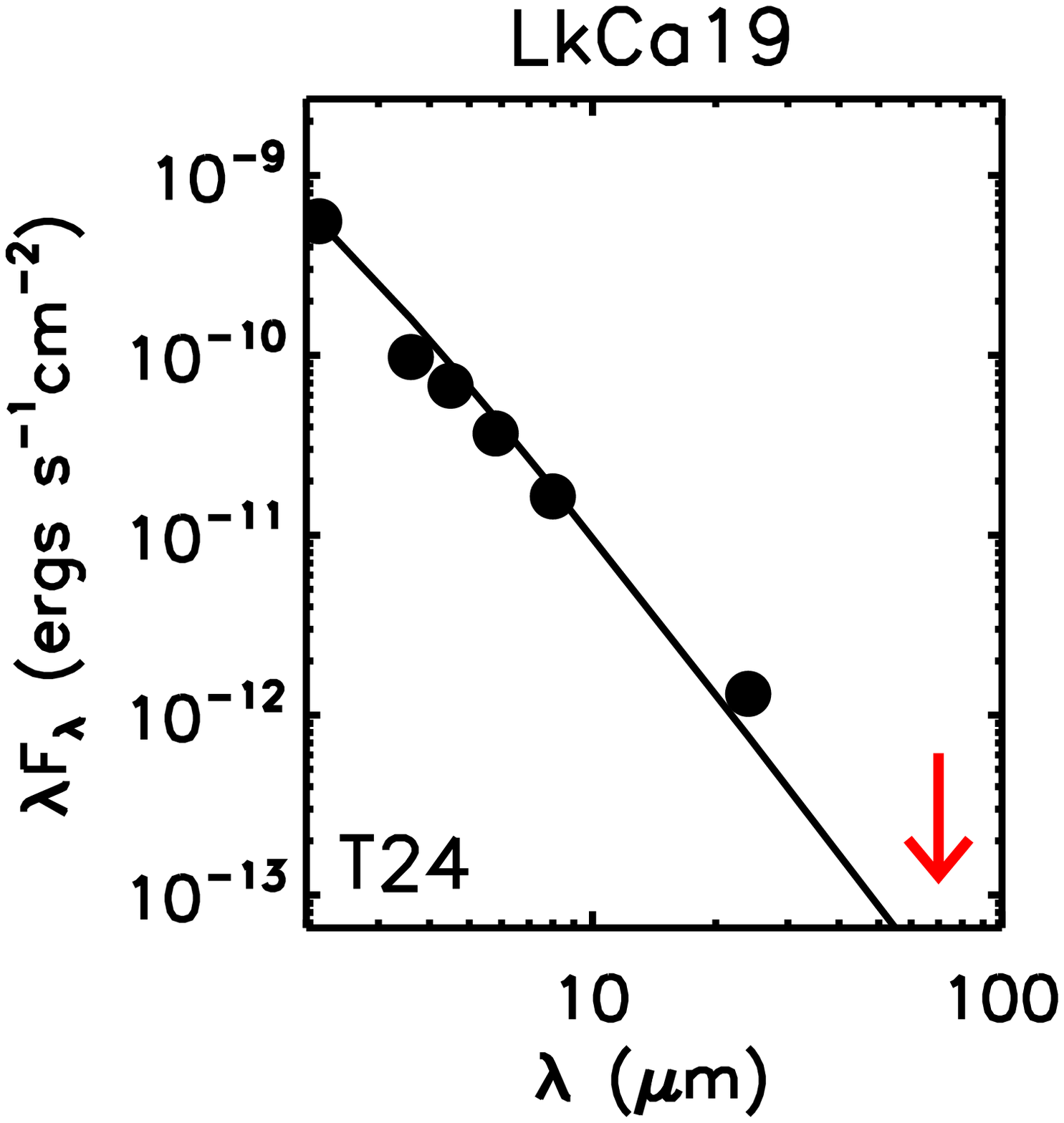}
        \includegraphics[height=3.3cm]{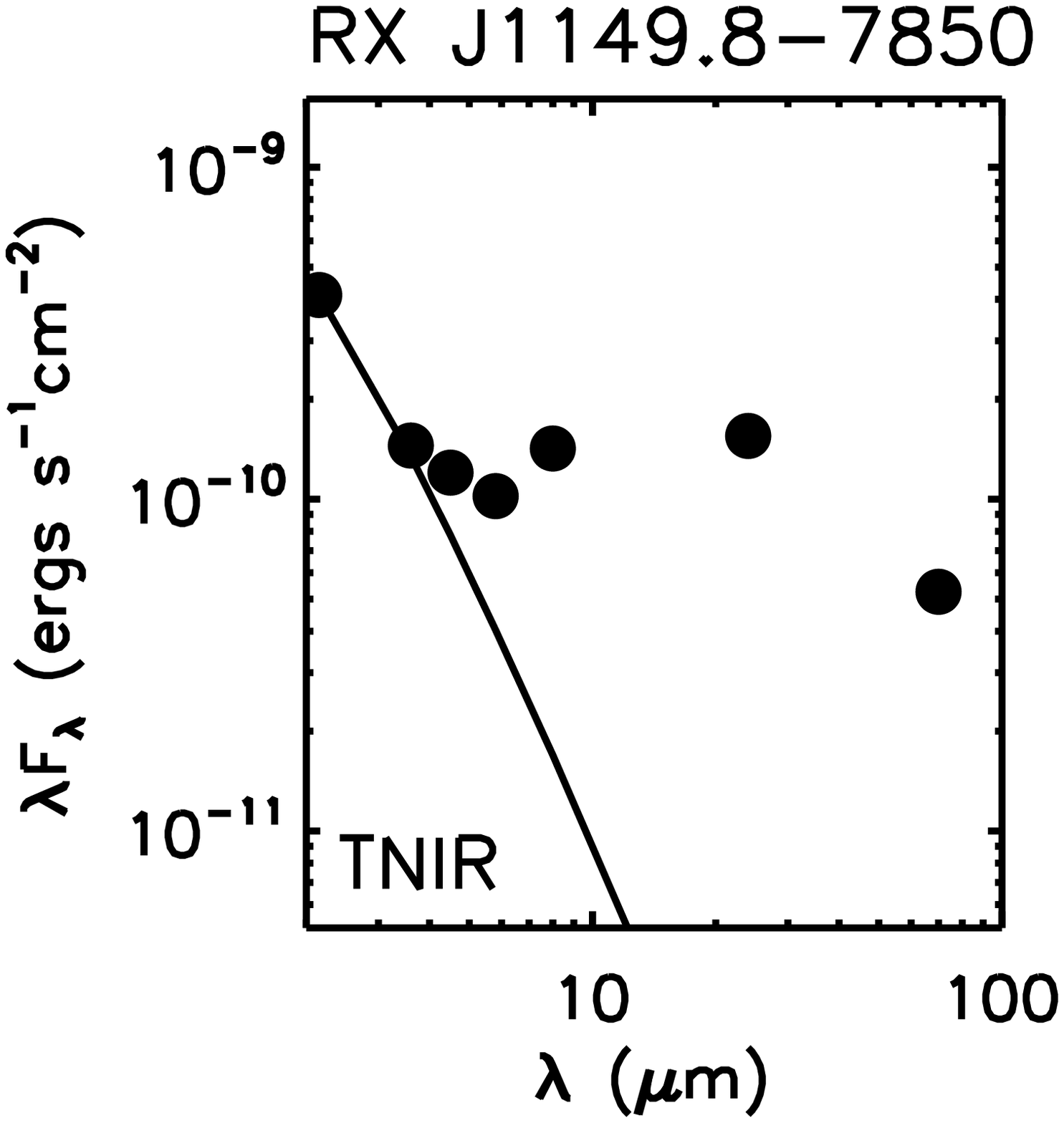}
        \includegraphics[height=3.3cm]{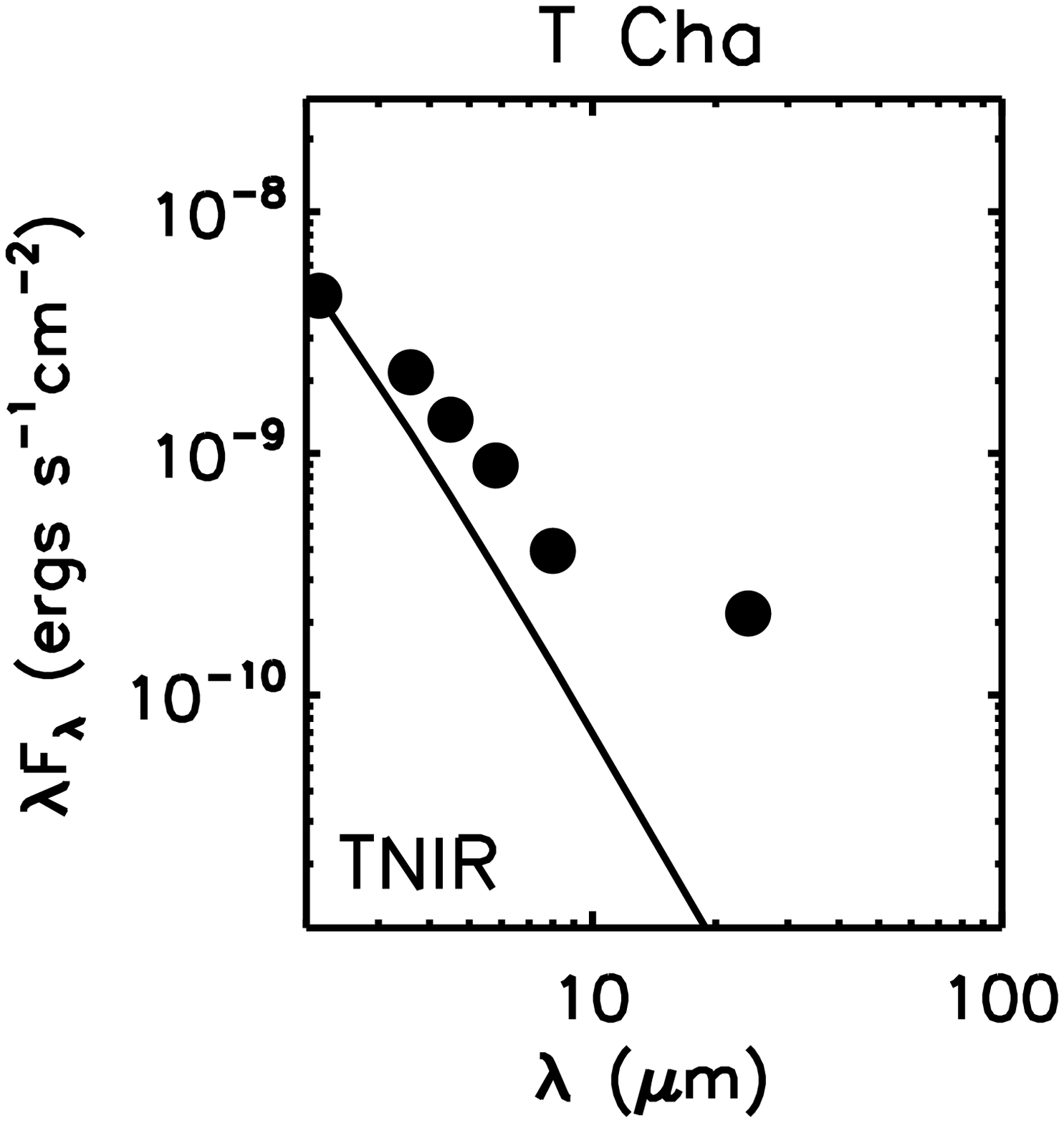}  
        \includegraphics[height=3.3cm]{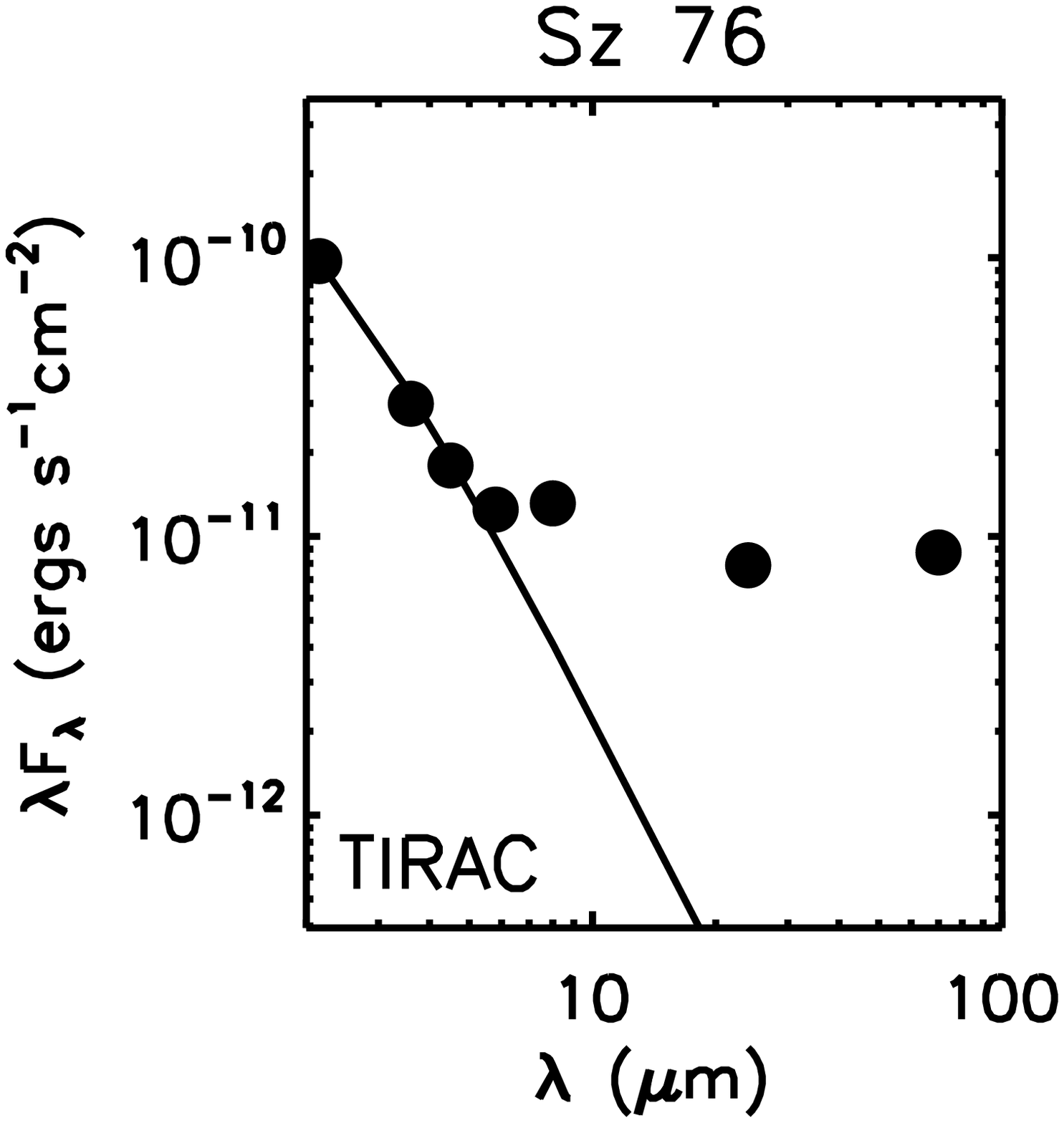}
      }
      \hbox {
        \includegraphics[height=3.3cm]{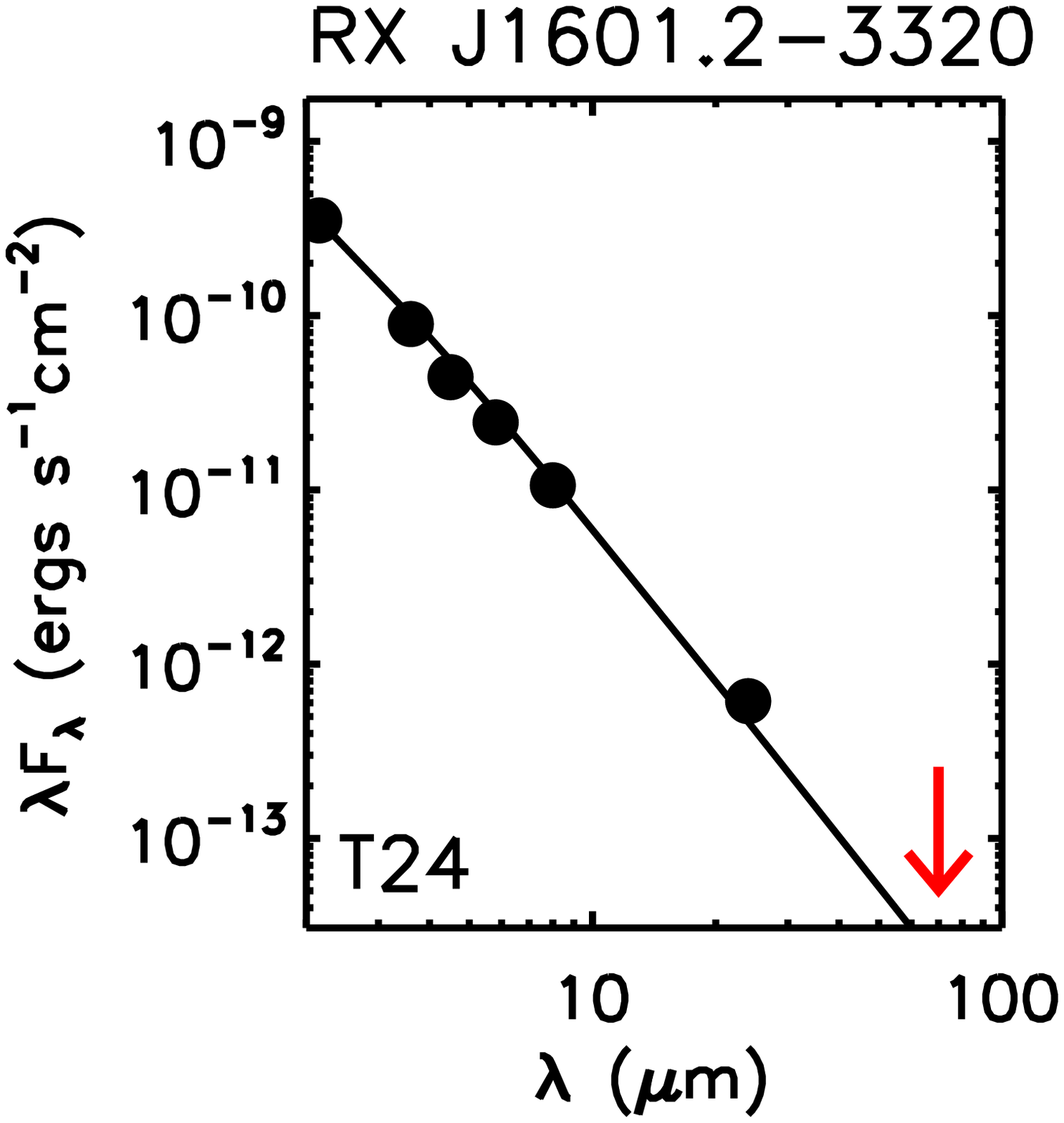}
        \includegraphics[height=3.3cm]{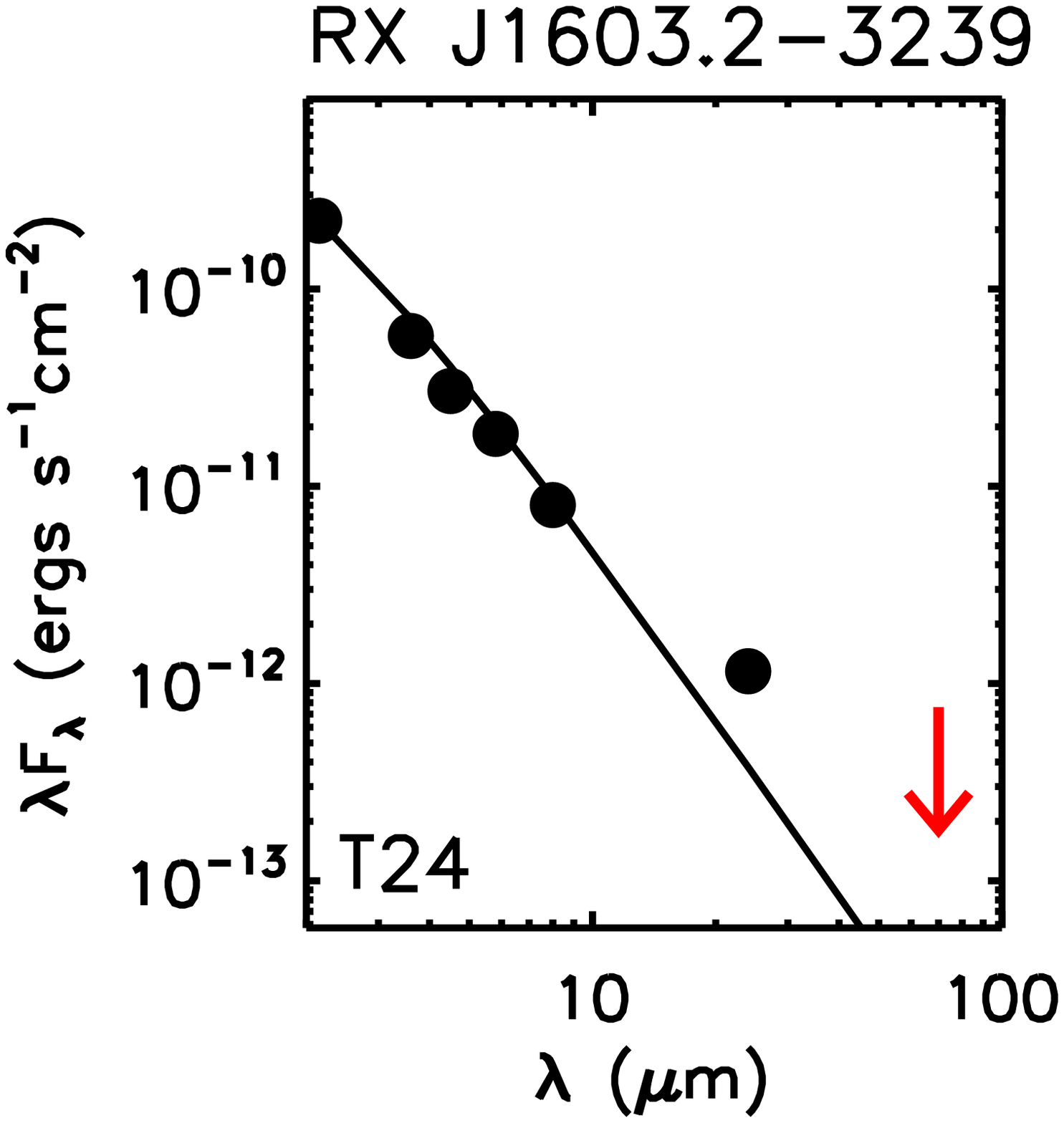}
        \includegraphics[height=3.3cm]{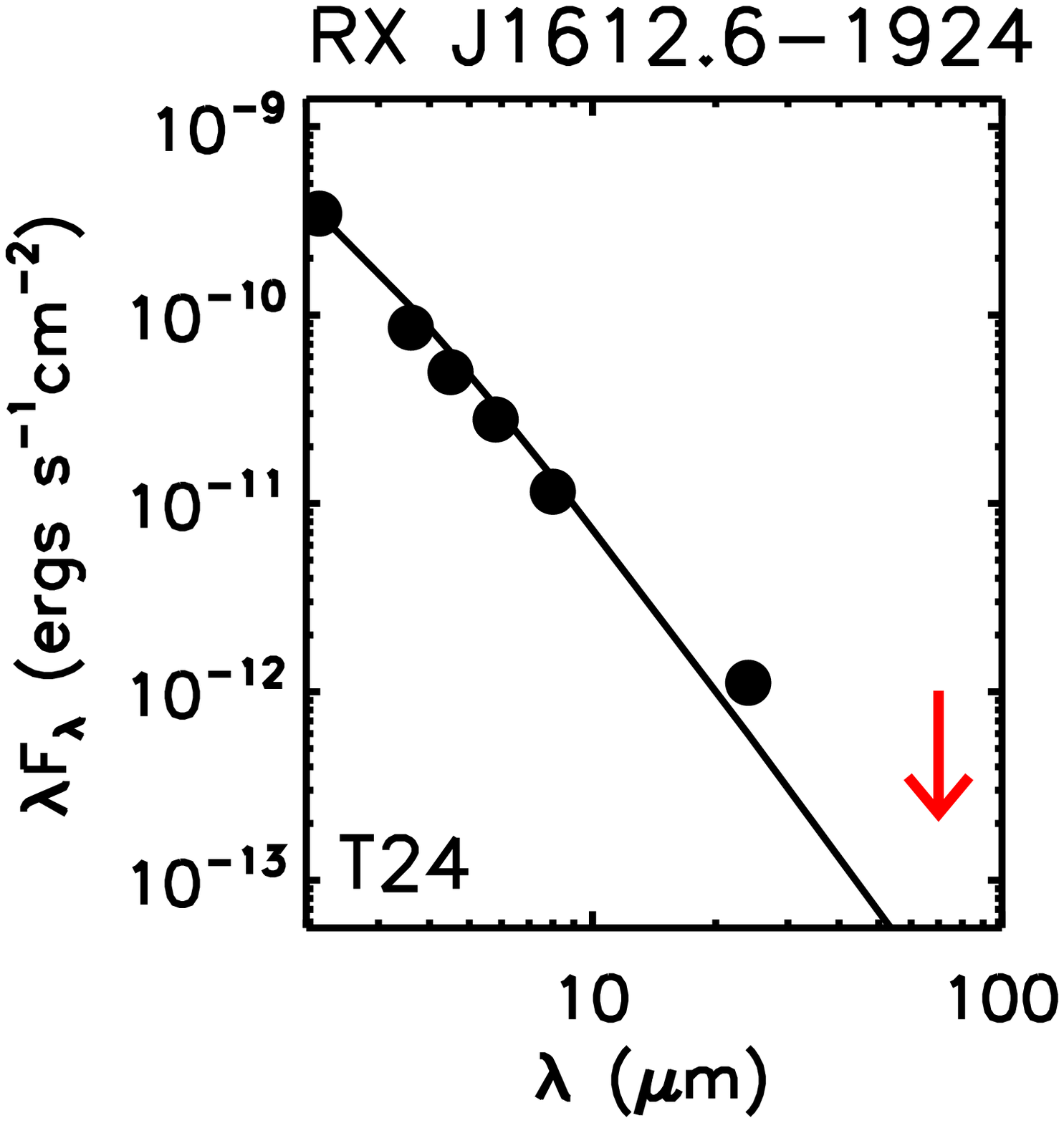}
        \includegraphics[height=3.3cm]{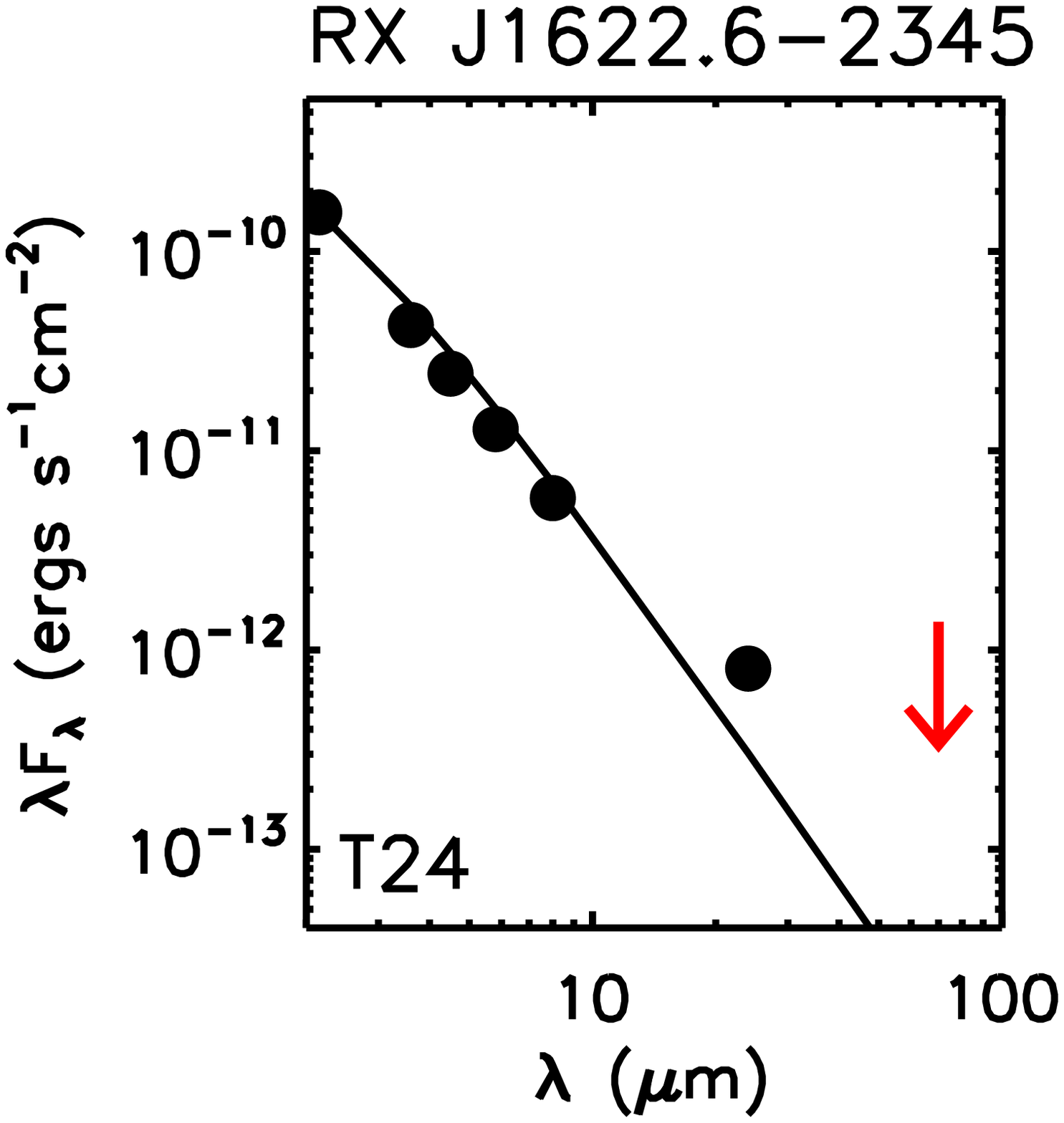}
     }
      \hbox {
        \includegraphics[height=3.3cm]{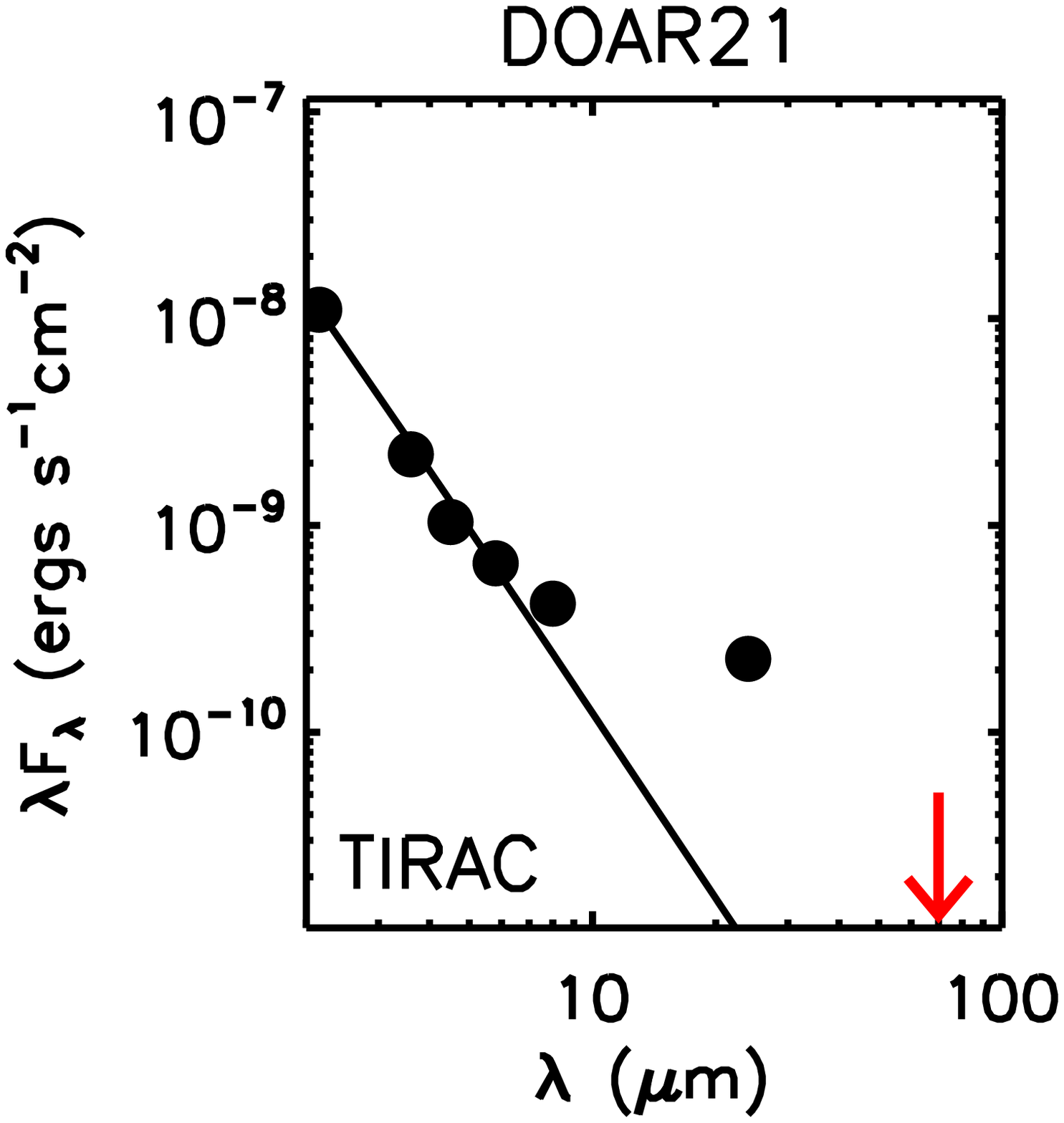}
        \includegraphics[height=3.3cm]{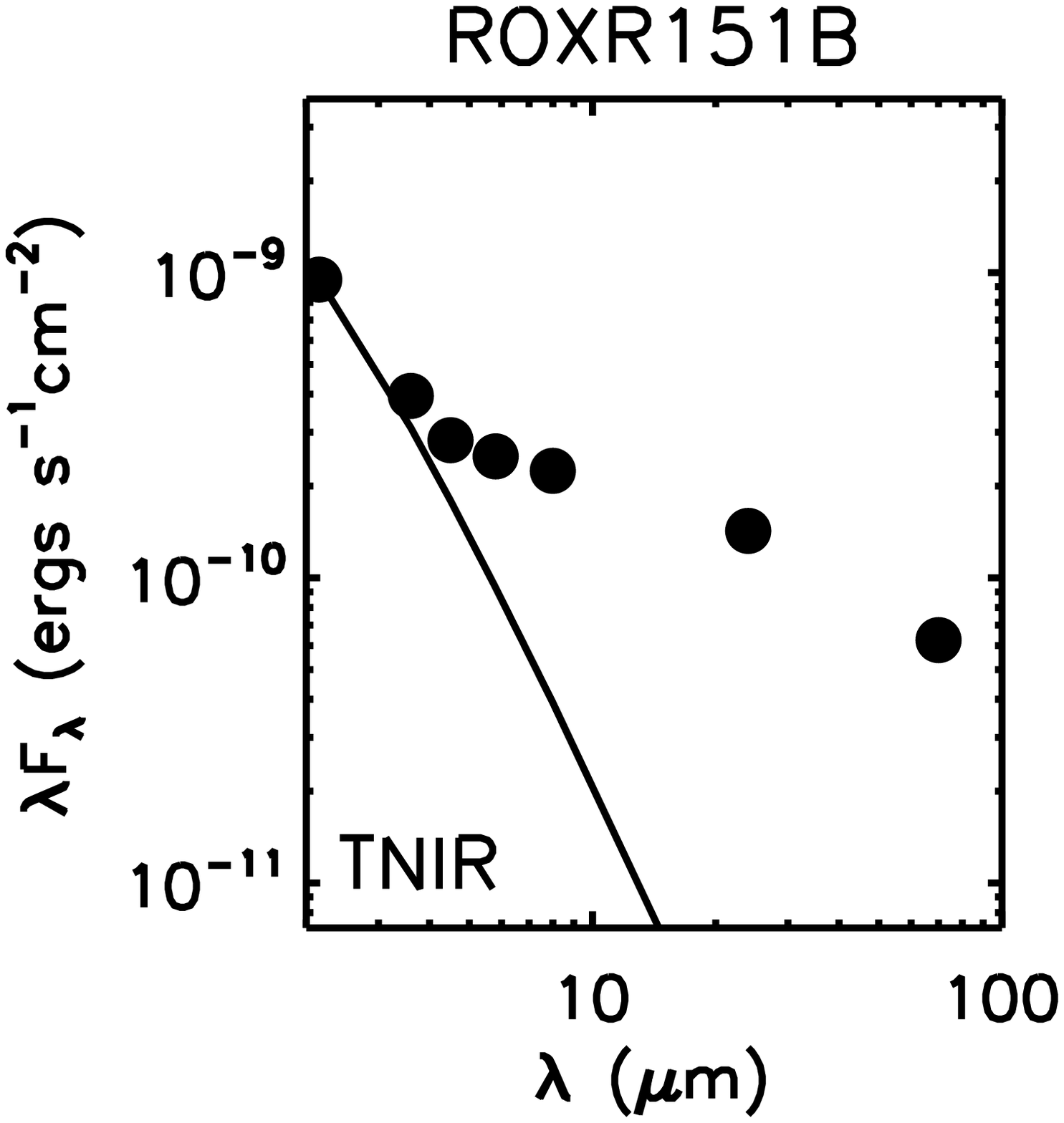}
        \includegraphics[height=3.3cm]{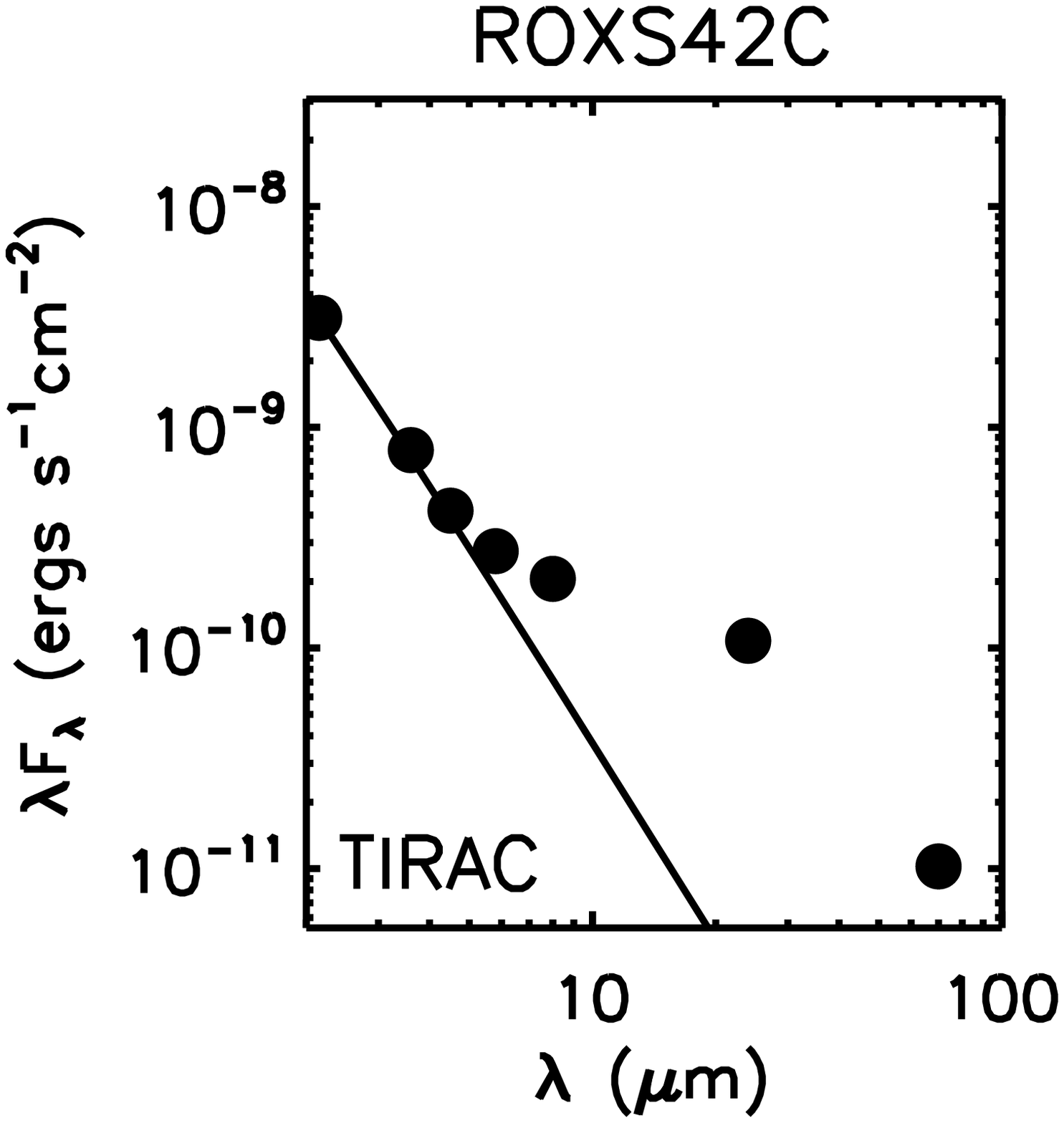}
        \includegraphics[height=3.3cm]{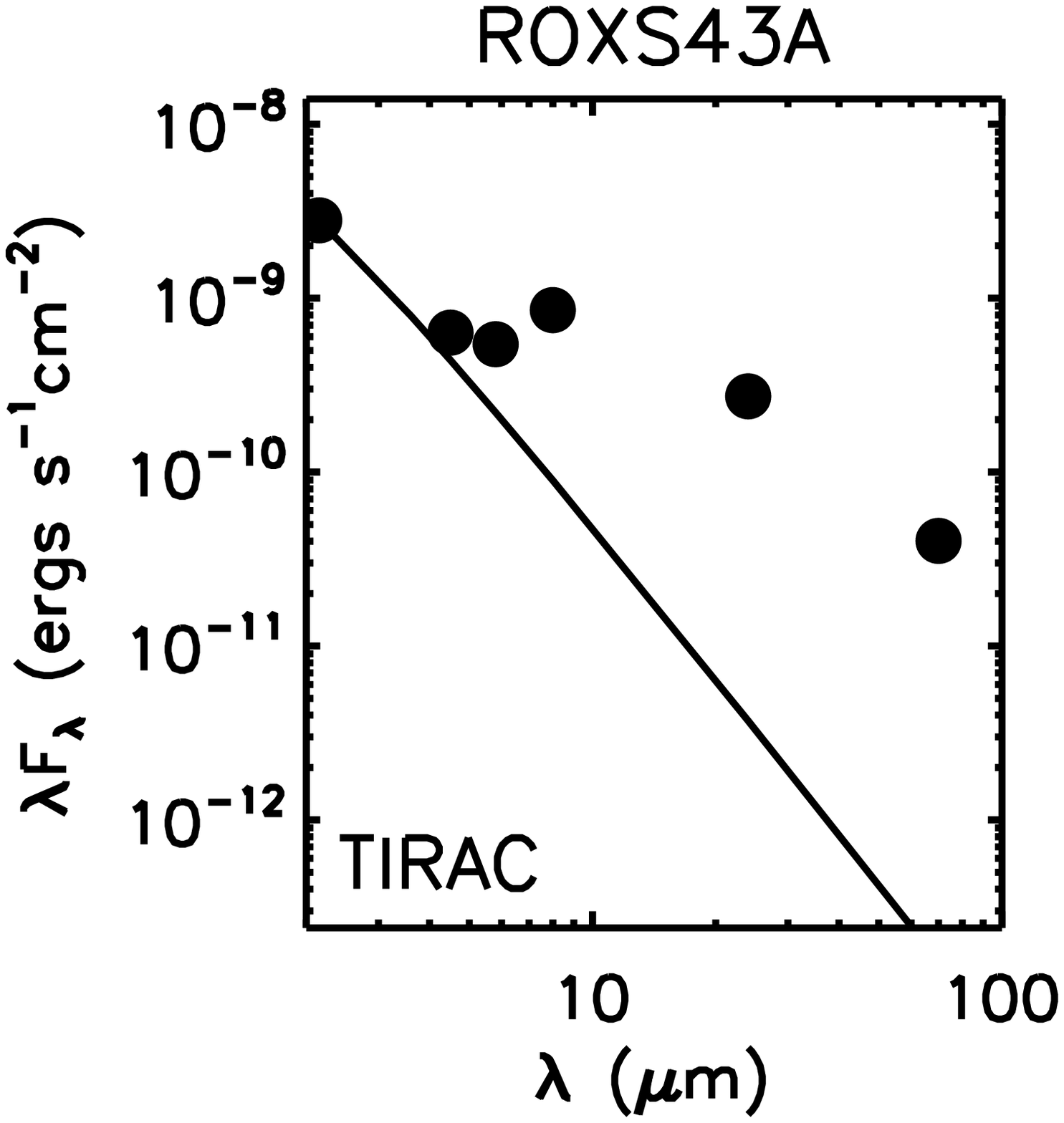}
        \includegraphics[height=3.3cm]{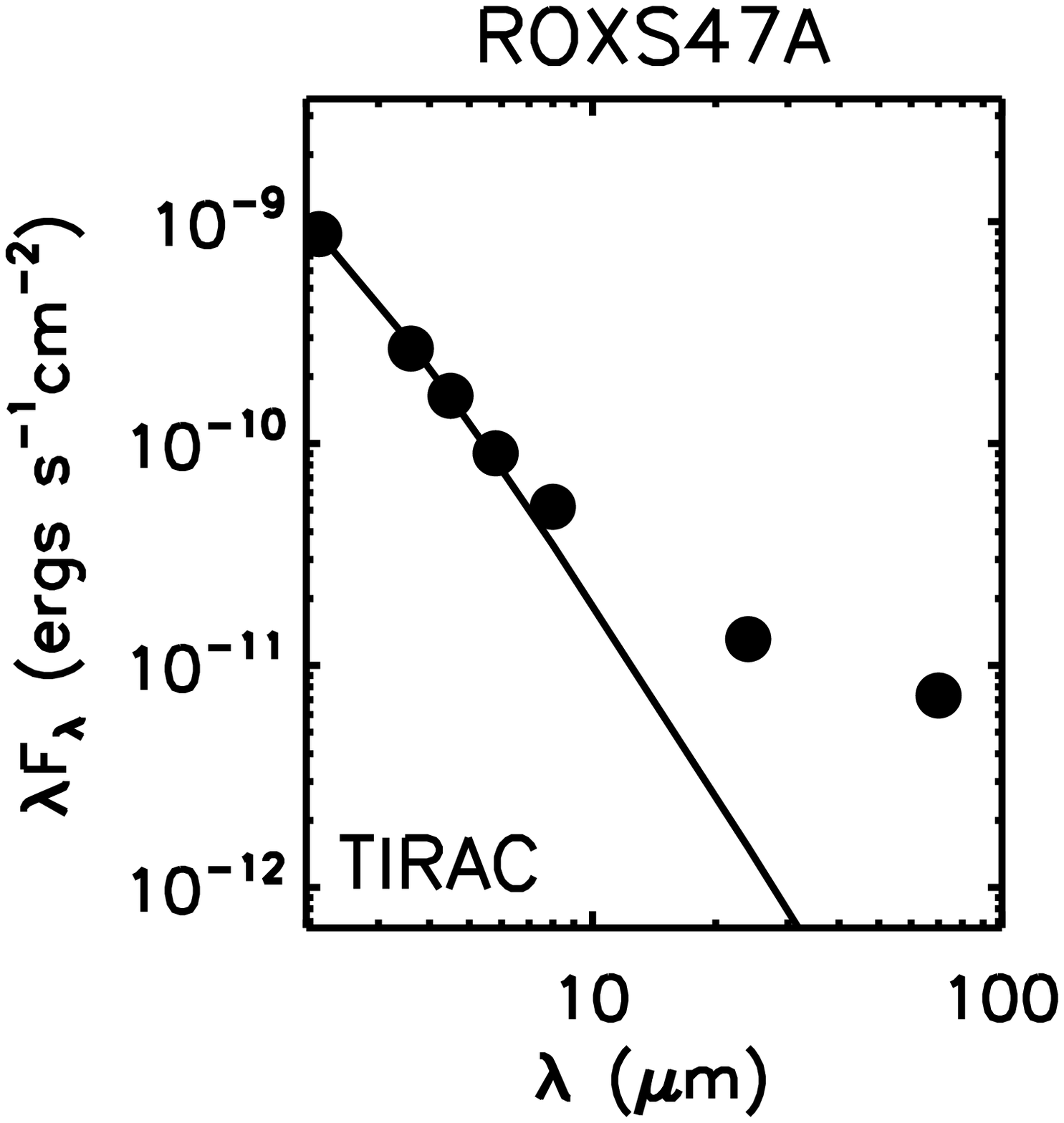}
      }        
    }
  }
  \caption{ 
    The SEDs of the WTTS with excess. The solid line represents the Planck function of the appropriate
    temperature normalized to the extinction corrected K-band flux. The red arrows are 3$\sigma$ upper limits 
    to the 70 \mic\ flux density. The plots are labeled with their 
    disk ``turn-on'' classifications which are also given in Table~\ref{stprops}. }
  \label{seds}
\end{figure}

{\bf{\it WTTS with Robust Disks.}} There are four stars of this type: HBC~423, RX~J1149.8-7850, T~Cha and ROXR1~51b (Figure~\ref{seds}). All these show robust emission at 24 and 70~\mic . The IRAC fluxes of these objects clearly show very depleted inner regions. These objects represent the youngest non-accreting or weakly accreting disks. Why has accretion weakened? How is the disk maintained?

{\bf{\it Zodiacal Dust Disks.}} The CTTS Sz~98 exhibits excess emission which basically
has a Rayleigh-Jeans slope beyond 6~\mic\, and peaks near the IRAC2 band (Figure~\ref{sedsctts}). This indicates a warm contributor to the excess (T \app 600~K), such as would be expected from a dust ring confined to within 0.5~AU of the star. The 8~\mic\ flux is actually anomalously high and 
inconsistent with the 24~\mic\ detection unless it results from some line emission ( e.g. from silicates or Polycyclic Aromatic Hydrocarbons).
Sz 98 is of spectral type K8 and is known to be a spectroscopic single, though it may have a companion at separations between 0.1$''$ and 1$''$.
 
{\bf {\it Intermediate Separation Binaries with Robust Disks.}} Binaries at a few tens of AU separations are supposed to make the least 
conducive environments for massive circumstellar disks. If their physical separations are a few tens of AU, then these objects are interesting because they provide test cases to current 
theories of planet formation in binary systems \citep{2006Icar..185....1Q,2007A&A...467..347M}. Binary systems with projected separations between
0.1$''$ and 0.4$''$ ($\rho$ \app 15 to 60~AU) with fractional disk luminosities 
above 0.01 are WTTS, HBC~423 ({\bf TNIR}) and ROXS~42C ({\bf TIRAC}) (Figure~\ref{seds}) 
and CTTS, GH Tau ({\bf TNIR}) and ROXR1~135S ({\bf TNIR}) (Figure~\ref{sedsctts}). All of these show 
robust 70~\mic\ emission which is either rising or roughly flat with respect 
to  their 24~\mic\ fluxes. The SEDs show no signs of gaps in the disk. These systems may thus be pointing to non-disruptive mechanisms for 
companion and disk interactions that we may not have foreseen.

\section {Conclusions}

We have presented the final results of the first large scale 
far infrared disk survey of weak-line T Tauri stars. The WTTS stars 
in the young star-forming regions within 200~pc (Chamaeleon, Lupus, 
Ophiuchus and Taurus) were probed for 
infrared excess from 3.6 to 70~\mic , in an effort to study 
the evolutionary status of their disks. 
We showed that overall 11\% of WTTS have a disk brighter than
the $\beta$ Pictoris debris disk ($f_{lum} = 2\times10^{-3}$).
However, the disk fraction for WTTS within 1$^o$ of the nearest 
cloud was 19\%, while the disk fraction for WTTS farther away 
was 5\%.  As we move from on-cloud to off-cloud,
stars get gradually older and have progressively lower disk fractions.

Of the fraction detected with disks, objects were classified into 
five groups according to disk ''turn on'' wavelengths {\bf TNIR, TIRAC, T24, T70 and RJ}. 
These groups showed easily identifiable trends in terms of age, 
and fractional luminosity. 
They suggest a sequential  
transition of accreting optically thick disks into passive optically thick disks into optically thin disks
and eventually into apparently diskless systems with $f_{lum} < 5 \times 10^{-4}$
over roughly 4~Myr. Even though the timescales for individual objects
may vary wildly, objects with more evolved SEDs on average tend to be older.  
The incidence rate of $\beta$ Pictoris-like debris disks for 
both the on and off-cloud WTTS may be much higher than the disk rate reported here, 
since fainter disks are just beyond the capability of our survey.

The mean ages of the CTTS, on-cloud and off-cloud WTTS were
also distinguishable from each other because of large statistics in each of these groups.
The on-cloud WTTS seem be measurably older than CTTS (mean age $\sim 0.8\pm0.1$~Myr), but they 
still constitute a very young population (mean age $\sim 2.3\pm0.3$~Myr). One reason why previous studies
have not reported this difference is perhaps because even our on-cloud WTTS include some older stars, due to
 a bias in the ROSAT selection.
However, the mean age of WTTS with disks  ($1.3\pm0.3$~Myr) is basically the same as that of 
the CTTS, which indicates that systems with disks are a younger group.
However, for the off-cloud objects, we may attach a lower age 
bound of  $\sim 3$~Myr,  while the upper bound remains somewhere 
around 100~Myr deduced from X-ray and lithium detections. 
Our analysis, however, suggested that the off-cloud WTTS 
apparent luminosities are consistent with an age of less than 10~Myr.  
Thus, they are indeed predominantly an older population as 
previously surmised \citep{2006ApJ...645.1283P,2007ApJ...667..308C}, and 
the nature of this still arguably young (3$-$10~Myr old) population 
remains a very interesting question. 

The analysis of the effect of binarity on disk fractions was inconclusive because of the small number
of binaries in our sample and the small excess rates we are dealing with.
We are thus not able to complement or test the results of  \citet{2009ApJ...696L..84C}.
An interesting study would be to look at the MIPS disk fractions of a sample of off-cloud 
binary WTTS numbering in the hundreds, to get beyond the diluting effects of the differing 
viewing angles, masses and ages of the systems. Such a study would explain the effect
of binarity on outer disks in relatively older systems.

When looking at the disk fraction as a function of spectral type, we restricted ourselves only
to on-cloud objects, in order to consider only the young objects.  The excess rate for 
spectral types G0 through K5 was 10$\pm$7\% while that for later spectral types was
19$\pm$7\%.  Also it seems that the less evolved SED types are concentrated in the later 
spectral types. 

We also find that, in a large number of cases, nebulosity or confusion with
nearby YSOs can result in false detections of 70~\mic\ excess around WTTS. 
Such problems can be avoided by careful comparisons of the emission centers
at 70~\mic\ and IRAC bands and requiring the shapes of emission to be 
point source-like. However, since false positives were found to be as frequent 
as bona fide detections, it seems that the environs of WTTS within \app 20$''$
are frequently occupied by young objects or cloud material. 
Higher resolution far-infrared and submillimeter imaging of the neighboring cold cloud material may reveal heretofore 
unknown ways in which the WTTS environs affect their disks.

Support for this work, which is part of the Spitzer Legacy Science Program, was provided 
by NASA through contracts 1224608, 1230782, and 1230799 
issued by the Jet Propulsion Laboratory, California Institute of 
Technology under NASA contract 1407. This publication makes use of data 
products from the Two Micron All Sky Survey, which is a joint project 
of the University of Massachusetts and the Infrared Processing 
and Analysis Center funded by NASA and the National Science 
Foundation. We also acknowledge use of the SIMBAD database. 

\vfill
\eject

\bibliographystyle{hapj}
\bibliography{zrefs}

\vfill
\eject

%\begin{tabular}{|l|r|r|r|r|r|r|r|r|}
\begin{table}[h]
%\caption{Relative Extinction in Optical, 2MASS and Spitzer bands.}
\renewcommand{\baselinestretch}{0.2}
\caption{Relative Extinction in Optical, 2MASS and Spitzer bands.}
\renewcommand{\baselinestretch}{1}
%\begin{center}
\begin{tabular}{lrrrrrrrr}
\hline \hline
Band &  $V$ & $J$ & $H$ & $K$ & IRAC1 & IRAC2 & IRAC3 & IRAC4 \\
\hline 
A$_{\lambda}$/A$_k$ & 8.8 & 2.5 & 1.54 & 1.0 & 0.57 & 0.44 & 0.41 & 0.37 \\
\hline
\end{tabular}
%\end{center}
\label{extinc}
\end{table}

\begin{table}[h]
\renewcommand{\baselinestretch}{1}
\caption{Average Properties of Different SED Object Types.}
\renewcommand{\baselinestretch}{1}
%\begin{center}
{\scriptsize
\begin{tabular}{lrrrrrrrr}
\hline \hline
  &  TNIR & TIRAC & T24 & RJ \\
\hline 
N & 24 & 13 & 9 & 134 \\
mean Age (Myr) & 0.6$\pm$0.1 & 1.3$\pm$0.3 & 1.5$\pm$0.4 & 4.1$\pm$0.2 \\
mean \flum & 7$\pm$0.7$\times 10^{-2}$ & 3.7$\pm$0.7$\times 10^{-2}$ & 1.7$\pm$0.5$\times 10^{-3}$  & $<$5$\pm$1$\times 10^{-4}$ \\ 
\hline
\label{props_sedt}
\end{tabular}
}
%\end{center}
\end{table}

\setlength\oddsidemargin{-0.25in}
\begin{center}
{\scriptsize 
\include{table2b}

}
\end{center}
%\setlength\oddsidemargin{0in}
%\setlength\topmargin{0in}

%\pagebreak
%\setlength\oddsidemargin{-0.25in}
\begin{center}
\include{table2c}
\end{center}
\setlength\oddsidemargin{0in}
\setlength\topmargin{0in}

\begin{figure}[ht]
  \centerline{
    \includegraphics[width=10cm, height=10cm]{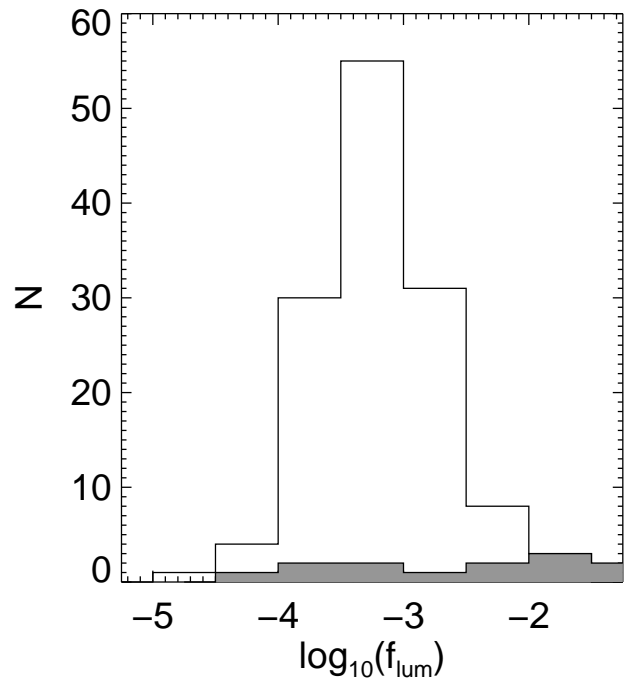}        
  }
  \caption{The filled histogram shows the distribution of the measured fractional luminosities ($f_{lum}$s) for our WTTS. 
The unfilled histogram shows the distribution of the upper limits to $f_{lum}$s for the WTTS without detections of infrared excess.}
  \label{flim}
\end{figure}

\end{document}

%% file: table2b.tex
\begin{table} 
\setlength\oddsidemargin{-0.25in} 
\renewcommand{\baselinestretch}{0.2}
\caption{2MASS AND {\it Spitzer} PHOTOMETRY.
{\scriptsize The absolute calibration uncertainties in the IRAC bands and the MIPS 24 \mic\ band are 5\% and 10\% respectively. These calibration uncertainties are typically more than twice the random errors and thus dominate the total photometric uncertainty.
}}
\renewcommand{\baselinestretch}{1}
{\scriptsize 
\begin{tabular}{lrrrrrrrrrrr} \hline\hline
& RA
& DEC
& $F_J$ 
& $F_H$  
& $F_K$  
& $F_{3.6}$ 
& $F_{4.5}$ 
& $F_{5.8}$ 
& $F_{8}$ 
& $F_{24}$ 
& $F_{70}$ \\
ID
& (hh mm ss.s)
& (dd mm ss.s)
& (mJy)
& (mJy)
& (mJy)
& (mJy)
& (mJy)
& (mJy)
& (mJy)
& (mJy)
& (mJy)
\\
\hline NTTS032641+2420 & 03 29 38.4 & +24 30 37.8 &   119 &   117 &    88 &    38 & 24.2 & 16.0 &  9.9 & 1.1 & $<$  18\\
 NTTS040047+2603 & 04 03 50.9 & +26 10 53.0 &   113 &   129 &   103 &    52 &    32 & 23.8 & 14.2 & 1.6 & $<$  11\\
  RX J0405.3+2009 & 04 05 19.6 & +20 09 25.6 &    534 &    542 &    386 &   144 &    98 &    69 &    43 &  5.1 & $<$  10\\
 NTTS040234+2143 & 04 05 30.9 & +21 51 10.7 &    67 &    78 &    63 & 30.3 & 20.5 & 14.8 &  8.5 & 1.0 & $<$  20\\
  RX J0409.2+1716 & 04 09 17.0 & +17 16 08.2 &   166 &   204 &   160 &    74 &    47 &    33 & 19.2 & 1.8 & $<$  15\\
  RX J0409.8+2446 & 04 09 51.1 & +24 46 20.9 &   146 &   169 &   133 &    57 &    40 & 27.1 & 17.0 & 2.6 & $<$  30\\
  RX J0412.8+1937 & 04 12 50.6 & +19 36 57.9 &   160 &   172 &   134 &    54 &    35 & 22.7 & 13.1 & 2.0 & $<$  24\\
           LkCa 1 & 04 13 14.2 & +28 19 10.7 &   222 &   291 &   237 &    97 &    68 &    50 & 31.1 & 2.7 & $<$  10\\
           LkCa 3 & 04 14 48.0 & +27 52 34.6 &    720 &    913 &    716 &    324 &   208 &   150 &    90 &  9.5 & $<$   9\\
           LkCa 5 & 04 17 39.0 & +28 33 00.4 &   163 &   197 &   159 &    75 &    51 &    35 & 20.6 & 2.5 & $<$   9\\
 NTTS041559+1716 & 04 18 51.7 & +17 23 16.6 &   156 &   175 &   131 &    56 &    36 & 25.0 & 15.7 & 1.6 & $<$  10\\
           LkCa 7 & 04 19 41.3 & +27 49 48.4 &    357 &    455 &    332 &   125 &   108 &    76 &    45 &  5.6 & $<$  19\\
  RX J0420.3+3123 & 04 20 24.1 & +31 23 23.7 &   105 &   115 &    86 &    40 & 25.1 & 16.2 & 10.7 & 1.0 & $<$  19\\
        HD 283572 & 04 21 58.9 & +28 18 06.3 &   1730 &   1610 &   1190 &    485 &    326 &   228 &   133 & 14.7 & $<$  10\\
          LkCa 21 & 04 22 03.2 & +28 25 38.9 &   261 &    348 &   278 &   125 &    91 &    65 &    39 &  4.9 & $<$  29\\
  RX J0424.8+2643 & 04 24 49.0 & +26 43 10.4 &    574 &    650 &    514 &   219 &   157 &   104 &    61 &  6.7 & $<$  34\\
 NTTS042417+1744 & 04 27 10.6 & +17 50 42.6 &    489 &    450 &    320 &   109 &    74 &    57 &    34 &  3.7 & $<$   9\\
           DH Tau & 04 29 41.6 & +26 32 58.2 &   198 &   302 &    357 &   269 &   207 &   172 &   133 &    319 &   384$\pm$ 67\\
           DI Tau & 04 29 42.5 & +26 32 49.2 &   297 &    372 &   293 &   133 &    77 &    64 &    38 &  4.2 & $<$ 346\\
           UX Tau & 04 30 04.0 & +18 13 49.5 &    567 &    670 &    636 &    544 &    391 &    406 &   262 &   1210 &  2643$\pm$407\\
           FX Tau & 04 30 29.6 & +24 26 44.9 &   280 &    448 &    451 &    373 &   299 &   266 &    330 &    414 &   317$\pm$ 57\\
           ZZ Tau & 04 30 51.4 & +24 42 22.3 &   254 &    341 &   280 &   122 &   118 &   109 &   115 &   118 & $<$ 244\\
         V927 Tau & 04 31 23.8 & +24 10 52.8 &   205 &   243 &   207 &    95 &    70 &    53 & 31.0 &  3.6 & $<$   8\\
 NTTS042835+1700 & 04 31 27.2 & +17 06 24.7 &   124 &   134 &   106 &    47 &    32 & 21.2 & 12.6 & 1.3 & $<$  10\\
         V710 Tau & 04 31 57.8 & +18 21 38.2 &   310 &   233 &   230 &   202 &   141 &   169 &   134 &   236 &   343$\pm$ 61\\
 NTTS042916+1751 & 04 32 09.3 & +17 57 22.7 &   210 &   244 &   192 &    80 &    54 &    40 & 23.5 & 2.4 & $<$  10\\
         V928 Tau & 04 32 18.8 & +24 22 27.0 &   244 &    434 &    382 &   152 &   123 &    88 &    59 &  6.7 & $<$  19\\
 NTTS042950+1757 & 04 32 43.7 & +18 02 56.2 &   138 &   168 &   126 &    58 &    37 & 26.5 & 16.2 & 2.0 & $<$  10\\
  RX J0432.8+1735 & 04 32 53.2 & +17 35 33.7 &   159 &   208 &   164 &    80 &    45 &    35 & 22.4 & 16.9 & $<$  28\\
           GH Tau & 04 33 06.2 & +24 09 33.8 &    362 &    521 &    509 &    389 &    352 &    330 &   285 &    345 &   433$\pm$ 75\\
         V807 Tau & 04 33 06.6 & +24 09 54.9 &    879 &   1170 &   1100 &    662 &    568 &    539 &    449 &    420 &   621$\pm$103\\
         V830 Tau & 04 33 10.0 & +24 33 42.9 &   297 &    367 &   285 &    94 &    73 &    53 &    32 &  3.7 & $<$  14\\
           GK Tau & 04 33 34.6 & +24 21 05.8 &    381 &    585 &    687 &    702 &    630 &    551 &    751 &   1510 &  1126$\pm$180\\
          WA Tau1 & 04 34 39.3 & +25 01 01.1 &    682 &    688 &    506 &   159 &   124 &    91 &    48 &  6.3 & $<$   8\\
 NTTS043230+1746 & 04 35 24.5 & +17 51 42.9 &   155 &   191 &   156 &    75 &    51 &    35 & 21.1 & 2.2 & -\\
  RX J0435.9+2352 & 04 35 56.8 & +23 52 05.0 &   224 &   269 &   222 &    90 &    70 &    49 & 30.6 &  3.5 & $<$  19\\
          LkCa 14 & 04 36 19.1 & +25 42 58.9 &   294 &    335 &   247 &   103 &    67 &    42 & 28.7 & 3.0 & $<$  13\\
 RX J0437.4+1851 & 04 37 26.9 & +18 51 26.7 &   271 &    386 &   228 &   104 &    62 &    48 & 29.1 &  3.7 & $<$  22\\
  RX J0438.2+2023 & 04 38 13.0 & +20 22 47.2 &   149 &   157 &   120 &    54 &    35 & 22.3 & 14.4 & 1.3 & $<$  14\\
           HV Tau & 04 38 35.3 & +26 10 38.5 &    325 &    497 &    459 &   190 &   147 &   117 &    69 & 18.0 & $<$ 500\\
  RX J0438.6+1546 & 04 38 39.1 & +15 46 13.6 &    439 &    465 &    338 &   144 &    78 &    60 &    37 &  4.9 & $<$  16\\
 RX J0439.4+3332A & 04 39 25.9 & +33 32 19.4 & 18.1 & 24.7 & 19.1 &  9.1 &  5.8 &  3.9 & 2.5 & 0.1 & $<$  19\\
           IW Tau & 04 41 04.7 & +24 51 06.1 &    320 &    416 &    327 &   132 &   100 &    73 &    44 &  4.9 & $<$  10\\
           ITG33 & 04 41 08.3 & +25 56 07.4 &  5.1 & 14.2 & 24.5 &    32 &    38 &    40 &    52 &    94 &   113$\pm$ 26\\
          HBC422 & 04 42 05.5 & +25 22 56.2 &   194 &    351 &    341 &   136 &   128 &    93 &    57 &  7.7 &   264$\pm$ 50\\
          HBC423 & 04 42 07.8 & +25 23 11.8 &   190 &    371 &    444 &    410 &    470 &    432 &    488 &    476 &  1180$\pm$190\\
 RX J0445.8+1556 & 04 45 51.3 & +15 55 49.7 &   1150 &   1060 &    775 &    329 &   198 &   145 &    88 &  8.1 & $<$ 134\\
  RX J0452.5+1730 & 04 52 30.7 & +17 30 25.8 &   164 &   176 &   133 &    55 &    36 & 25.0 & 15.8 & 1.4 & $<$  10\\
  RX J0452.8+1621 & 04 52 50.1 & +16 22 09.1 &    365 &    413 &    326 &   121 &    97 &    64 &    40 &  4.4 & $<$  10\\
          LkCa 19 & 04 55 37.0 & +30 17 55.0 &    451 &    482 &    367 &   112 &    98 &    69 &    42 & 10.5 & $<$  21\\
 NTTS045251+3016 & 04 56 02.0 & +30 21 03.5 &    416 &    483 &    374 &   183 & - &- &- & 5.1 & $<$  38\\
  RX J0457.2+1524 & 04 57 17.7 & +15 25 09.4 &    708 &    705 &    529 &   250 &   135 &   100 &    60 &  7.6 & $<$  19\\
  RX J0457.5+2014 & 04 57 30.6 & +20 14 29.6 &   309 &   302 &   223 &    95 &    61 &    43 & 26.2 & 2.5 & $<$  15\\
  RX J0458.7+2046 & 04 58 39.7 & +20 46 44.0 &   232 &   267 &   201 &    82 &    56 &    38 & 22.2 & 3.0 & $<$  26\\
  RX J0459.7+1430 & 04 59 46.2 & +14 30 55.4 &   219 &   236 &   176 &    77 &    47 &    34 & 20.4 & 1.8 & $<$  10\\
         V836 Tau & 05 03 06.6 & +25 23 19.9 &   173 &   240 &   243 &   134 &   105 &   110 &   126 &   193 &   220$\pm$ 42\\
  RX J0842.4-8345 & 08 42 22.7 & -83 45 24.5 &   264 &   312 &   243 &    90 &    72 &    48 & 29.6 &  3.4 & $<$  28\\
\hline
\end{tabular}
\label{stphot} 
}
\end{table}
\begin{table} 
{\scriptsize 
\newpage\begin{tabular}{lrrrrrrrrrrr} \hline
   
& RA
& DEC
& $F_J$ 
& $F_H$ 
& $F_k$ 
& $F_{3.6}$ 
& $F_{4.5}$ 
& $F_{5.8}$ 
& $F_{8}$ 
& $F_{24}$ 
& $F_{70}$ \\
ID
& (hh mm ss.s)
& (dd mm ss.s)
& (mJy)
& (mJy)
& (mJy)
& (mJy)
& (mJy)
& (mJy)
& (mJy) 
& (mJy)
& (mJy)
\\
\hline
  RX J0848.0-7854 & 08 47 56.8 & -78 54 53.3 &   297 &    344 &   288 &    74 &    74 &    68 &    40 &  4.7 & $<$   5\\
 RX J0902.9-7759 & 09 02 51.3 & -77 59 34.8 &   145 &   169 &   136 &    64 &    41 & 31.3 & 17.6 & 2.6 & $<$  19\\
 RX J0915.5-7609 & 09 15 29.1 & -76 08 47.1 &   297 &    346 &   268 &    99 &    78 &    56 &    34 &  3.5 & $<$  43\\
 RX J0935.0-7804 & 09 34 56.0 & -78 04 19.4 &   193 &   228 &   185 &    91 &    63 &    42 & 24.3 & 2.6 & $<$   6\\
  RX J0942.7-7726 & 09 42 49.6 & -77 26 40.8 &   114 &   137 &   109 &    49 &    33 & 21.8 & 13.4 & 1.5 & $<$  17\\
  RX J1001.1-7913 & 10 01 08.7 & -79 13 07.6 &   150 &   181 &   137 &    57 &    41 & 29.7 & 16.8 & 2.2 & $<$   6\\
  RX J1005.3-7749 & 10 05 20.0 & -77 48 42.3 &   191 &   239 &   185 &    81 &    58 &    38 & 23.7 & 2.8 & $<$   6\\
           CS Cha & 11 02 24.9 & -77 33 35.6 &    363 &    426 &    350 &   126 &    99 &    72 &    48 &    617 &  2543$\pm$393\\
  RX J1108.8-7519 & 11 08 53.3 & -75 19 37.5 &   203 &   250 &   201 &   104 &    64 &    44 & 27.9 & 3.1 & $<$  20\\
            Sz 30 & 11 09 11.8 & -77 29 12.5 &   170 &   225 &   174 &    92 &    57 &    41 & 26.3 &  3.2 & $<$  20\\
            Sz 41 & 11 12 24.3 & -76 37 06.6 &   311 &    399 &    421 &   243 &   239 &   225 &    317 &   278 &    89$\pm$ 21\\
  RX J1117.0-8028 & 11 16 57.1 & -80 27 52.1 &   186 &   202 &   174 &    97 &    66 &    45 & 26.8 &  3.4 & $<$   5\\
  RX J1123.2-7924 & 11 23 10.6 & -79 24 43.3 & 29.2 & 25.8 & 17.0 &  7.7 &  5.5 &  3.3 & 1.9 & 0.1 & $<$  16\\
  RX J1129.2-7546 & 11 29 12.6 & -75 46 26.4 &   189 &   229 &   187 &    88 &    58 &    37 & 22.6 & 2.5 & $<$  28\\
  RX J1149.8-7850 & 11 49 31.8 & -78 51 00.9 &   265 &    333 &   269 &   165 &   173 &   190 &    366 &   1240 &  1227$\pm$194\\
  RX J1150.4-7704 & 11 50 28.3 & -77 04 38.3 &   207 &   227 &   172 &    81 &    50 &    34 & 20.5 & 2.2 & $<$  12\\
            T Cha & 11 57 13.5 & -79 21 31.5 &    417 &    735 &   1100 &   1410 &   1290 &   1110 &    708 &   1740 & -\\
  RX J1158.5-7754 & 11 58 28.3 & -77 54 29.1 &    822 &    973 &    728 &    325 &   207 &   143 &    87 &  9.3 & $<$  16\\
  RX J1158.5-7913 & 11 58 34.3 & -79 13 17.6 &   202 &   267 &   231 &   101 &    73 &    53 &    32 &  4.9 & $<$  30\\
  RX J1159.7-7601 & 11 59 42.3 & -76 01 26.2 &    352 &    419 &    318 &   125 &    99 &    62 &    40 &  4.7 & -\\
  RX J1202.1-7853 & 12 02 03.7 & -78 53 01.3 &    328 &    425 &    317 &   160 &   106 &    75 &    46 &  5.2 & $<$  12\\
  RX J1204.6-7731 & 12 04 36.1 & -77 31 34.7 &   198 &   229 &   187 &    80 &    64 &    44 & 26.5 &  3.4 & $<$  18\\
  RX J1216.8-7753 & 12 16 45.9 & -77 53 33.4 &   146 &   168 &   135 &    60 &    42 & 28.1 & 17.1 & 2.0 & $<$   7\\
  RX J1219.7-7403 & 12 19 43.7 & -74 03 57.4 &   201 &   246 &   191 &    94 &    58 &    42 & 25.2 & 2.8 & $<$   7\\
  RX J1220.4-7407 & 12 20 21.8 & -74 07 39.6 &   315 &    369 &   300 &   141 &    95 &    66 &    40 &  4.7 & $<$   9\\
  RX J1239.4-7502 & 12 39 21.3 & -75 02 39.3 &    674 &    675 &    517 &   215 &   138 &    95 &    57 &  6.2 & $<$   8\\
  RX J1301.0-7654 & 13 00 56.3 & -76 54 02.2 &    389 &    497 &    385 &   188 &   135 &    91 &    54 &  6.2 & $<$ 102\\
  RX J1507.6-4603 & 15 07 37.8 & -46 03 15.8 &   188 &   212 &   153 &    72 &    46 & 31.3 & 18.5 & 1.9 & $<$  21\\
  RX J1508.6-4423 & 15 08 37.7 & -44 23 17.2 &   288 &   276 &   200 &    85 &    52 &    38 & 23.2 & 2.4 & $<$   7\\
  RX J1511.6-3550 & 15 11 37.0 & -35 50 42.0 &   145 &   159 &   123 &    57 &    38 & 25.3 & 14.8 & 1.6 & $<$   8\\
  RX J1515.8-3331 & 15 15 45.4 & -33 31 59.9 &    407 &    423 &   295 &   116 &    80 &    57 &    32 &  4.4 & $<$  27\\
  RX J1515.9-4418 & 15 15 52.8 & -44 18 17.5 &   135 &   152 &   110 &    48 & 29.8 & 19.8 & 13.0 & 1.4 & $<$   8\\
  RX J1516.6-4406 & 15 16 36.6 & -44 07 20.6 &   175 &   192 &   140 &    54 &    38 & 26.3 & 15.7 & 1.6 & $<$  22\\
  RX J1518.9-4050 & 15 18 52.8 & -40 50 52.9 &    350 &    352 &   254 &    99 &    69 &    49 & 29.6 &  3.3 & $<$  26\\
  RX J1519.3-4056 & 15 19 16.0 & -40 56 07.7 &   241 &   252 &   196 &    88 &    57 &    40 & 23.7 & 2.4 & $<$  11\\
  RX J1522.2-3959 & 15 22 11.6 & -39 59 51.1 &   174 &   196 &   153 &    68 &    44 & 30.0 & 18.2 & 2.0 & $<$   7\\
  RX J1523.4-4055 & 15 23 25.6 & -40 55 46.9 &   166 &   180 &   132 &    60 &    37 & 25.8 & 15.1 & 1.6 & $<$  10\\
  RX J1523.5-3821 & 15 23 30.4 & -38 21 28.9 &   110 &   135 &   106 &    48 &    34 & 23.9 & 14.1 & 1.8 & $<$  10\\
  RX J1524.0-3209 & 15 24 03.1 & -32 09 51.0 &   252 &   304 &   232 &   105 &    69 &    50 & 29.5 &  3.2 & $<$  17\\
  RX J1524.5-3652 & 15 24 32.4 & -36 52 02.9 &   241 &   246 &   179 &    79 &    49 &    33 & 20.0 & 2.2 & $<$   7\\
  RX J1525.5-3613 & 15 25 33.2 & -36 13 46.9 &   238 &   256 &   194 &    78 &    57 &    39 & 24.6 & 2.7 & $<$  15\\
  RX J1525.6-3537 & 15 25 36.7 & -35 37 32.0 &   195 &   227 &   173 &    70 &    50 &    34 & 20.7 & 2.2 & $<$  10\\
  RX J1526.0-4501 & 15 25 59.7 & -45 01 16.0 &   266 &   261 &   183 &    75 &    48 &    33 & 20.4 & 2.0 & $<$  10\\
  RX J1538.0-3807 & 15 38 02.7 & -38 07 23.2 &   144 &   152 &   118 &    49 &    34 & 23.8 & 14.8 & 1.6 & $<$  14\\
  RX J1538.6-3916 & 15 38 38.3 & -39 16 55.5 &   232 &   255 &   192 &    84 &    50 &    35 & 21.2 & 2.2 & $<$  10\\
  RX J1538.7-4411 & 15 38 43.1 & -44 11 47.6 &    479 &    471 &    347 &   140 &    91 &    62 &    36 &  4.4 & $<$  22\\
            Sz 65 & 15 39 27.8 & -34 46 17.4 &    336 &    441 &    428 &   255 &   225 &   231 &   284 &    502 &   533$\pm$ 90\\
  RX J1540.7-3756 & 15 40 41.2 & -37 56 18.7 &   170 &   191 &   141 &    62 &    39 & 26.4 & 16.2 & 1.6 & $<$  10\\
  RX J1543.1-3920 & 15 43 06.3 & -39 20 19.6 &   184 &   211 &   153 &    63 &    46 & 30.8 & 18.0 & 1.5 & $<$  24\\
  RX J1546.7-3618 & 15 46 41.2 & -36 18 47.5 &   255 &   270 &   205 &    95 &    59 &    41 & 25.3 & 2.4 & $<$  11\\
  RX J1547.7-4018 & 15 47 41.8 & -40 18 27.0 &   305 &   306 &   229 &    92 &    62 &    44 & 25.1 & 2.6 & $<$  23\\
 PZ99  J154920.9-260005 & 15 49 21.0 & -26 00 06.5 &    555 &    575 &    456 &   198 &   119 &    84 &    53 &  5.4 & $<$  45\\
            Sz 76 & 15 49 30.7 & -35 49 51.7 &    66 &    79 &    65 &    35 & 26.2 & 23.5 &    34 &    63 &   204$\pm$ 39\\
  RX J1550.0-3629 & 15 49 59.2 & -36 29 57.6 &   239 &   252 &   186 &    80 &    52 &    37 & 21.7 & 2.5 & $<$  12\\
            Sz 77 & 15 51 46.9 & -35 56 44.3 &   266 &    375 &    328 &   218 &   181 &   170 &   199 &    333 &   175$\pm$ 35\\
  RX J1552.3-3819 & 15 52 19.5 & -38 19 31.6 &   115 &   137 &   103 &    46 & 30.6 & 20.4 & 12.7 & 1.2 & $<$   9\\
\hline
\end{tabular}
}
\end{table}
\begin{table} 
{\scriptsize 
\newpage\begin{tabular}{lrrrrrrrrrrr} \hline
   
& RA
& DEC
& $F_J$ 
& $F_H$ 
& $F_k$ 
& $F_{3.6}$ 
& $F_{4.5}$ 
& $F_{5.8}$ 
& $F_{8}$ 
& $F_{24}$ 
& $F_{70}$ \\
ID
& (hh mm ss.s)
& (dd mm ss.s)
& (mJy)
& (mJy)
& (mJy)
& (mJy)
& (mJy)
& (mJy)
& (mJy) 
& (mJy)
& (mJy)
\\
\hline
  RX J1554.9-3827 & 15 54 52.9 & -38 27 56.8 &   109 &   130 &    99 &    46 & 30.7 & 21.4 & 13.2 & 1.4 & $<$   9\\
PZ99  J155506.2-252109 & 15 55 06.3 & -25 21 10.4 &   274 &    366 &   263 &   106 &    78 &    56 &    33 &  3.5 & $<$  37\\
 RX J1555.4-3338 & 15 55 26.2 & -33 38 23.4 &   138 &   156 &   121 &    54 &    32 & 23.6 & 14.5 & 1.9 & $<$  27\\
 RX J1555.6-3709 & 15 55 33.8 & -37 09 41.3 &   165 &   190 &   144 &    63 &    40 & 27.4 & 16.2 & 1.5 & $<$  22\\
           Sz 81 & 15 55 50.3 & -38 01 33.6 &   135 &   157 &   143 &   103 &    96 &    74 &    80 &    95 &   183$\pm$ 36\\
 RX J1556.1-3655 & 15 56 02.1 & -36 55 28.4 &   111 &   149 &   127 &    88 &    72 &    59 &    68 &   205 &   242$\pm$ 45\\
           Sz 82 & 15 56 09.2 & -37 56 06.5 &    489 &    595 &    535 &    324 &   220 &   313 &    370 &    765 &  1458$\pm$229\\
PZ99 J155702.3-195042 & 15 57 02.4 & -19 50 42.2 &    340 &    412 &   299 &   145 &    96 &    63 &    37 &  3.9 & $<$  11\\
            Sz 84 & 15 58 02.5 & -37 36 02.9 &    68 &    85 &    76 &    43 & 29.5 & 20.5 & 12.5 & 24.3 &   377$\pm$ 66\\
  RX J1559.0-3646 & 15 58 59.8 & -36 46 20.9 &   140 &   168 &   129 &    68 &    42 & 30.9 & 18.6 & 1.6 & $<$  10\\
           Sz 129 & 15 59 16.5 & -41 57 10.5 &   170 &   238 &   240 &   178 &   157 &   136 &   184 &    326 &   483$\pm$ 82\\
  RX J1559.8-3628 & 15 59 49.5 & -36 28 28.0 &    494 &    563 &    410 &   175 &   111 &    82 &    48 &  6.0 & $<$  35\\
  RX J1601.2-3320 & 16 01 09.0 & -33 20 14.3 &    391 &    389 &   259 &   109 &    67 &    48 & 28.6 &  4.9 & $<$   9\\
 PZ99  J160151.4-244524 & 16 01 51.5 & -24 45 25.2 &   275 &    352 &   269 &   125 &    75 &    56 &    33 &  3.6 & $<$  11\\
 PZ99  J160158.2-200811 & 16 01 58.2 & -20 08 12.2 &    729 &    771 &    569 &   193 &   142 &   123 &    67 &  7.4 & $<$  12\\
  RX J1602.0-3613 & 16 01 59.2 & -36 12 55.8 &   231 &   264 &   192 &    88 &    58 &    41 & 24.4 & 2.5 & $<$  12\\
 PZ99  J160253.9-202248 & 16 02 54.0 & -20 22 48.2 &    345 &    421 &    353 &   156 &   113 &    88 &    49 &  5.7 & $<$   9\\
  RX J1603.2-3239 & 16 03 11.8 & -32 39 20.4 &   163 &   196 &   149 &    67 &    44 &    35 & 20.9 &  9.2 & $<$  27\\
  RX J1603.8-4355 & 16 03 45.4 & -43 55 49.3 &   1080 &   1110 &    791 &    337 &   213 &   150 &    95 & 10.5 & $<$  40\\
  RX J1603.8-3938 & 16 03 52.5 & -39 39 01.5 &    423 &    466 &    342 &   125 &    94 &    65 &    39 &  4.2 & $<$  49\\
  RX J1604.5-3207 & 16 04 30.6 & -32 07 28.9 &    343 &    342 &   250 &   125 &    73 &    52 & 31.4 &  3.3 & $<$  21\\
  RX J1605.6-3837 & 16 05 33.3 & -38 37 45.4 &    77 &    93 &    73 &    36 & 22.8 & 16.0 &  9.9 & 1.3 & $<$  12\\
 PZ99 J160550.5-253313 & 16 05 50.7 & -25 33 13.8 &    362 &    376 &   275 &   109 &    74 &    53 & 31.2 &  3.5 & $<$  12\\
  RX J1607.2-3839 & 16 07 13.7 & -38 39 24.0 &   212 &   267 &   188 &    87 &    59 &    42 & 25.7 &  3.6 & $<$  16\\
            Sz 96 & 16 08 12.6 & -39 08 33.5 &   142 &   187 &   174 &   168 &   113 &   138 &   173 &   231 &   154$\pm$ 33\\
  RX J1608.3-3843 & 16 08 18.3 & -38 44 05.5 &   238 &   277 &   218 &   102 &    66 &    46 & 26.9 &  3.3 & $<$  13\\
            Sz 98 & 16 08 22.5 & -39 04 46.3 &   246 &    354 &    415 &    373 &    477 &    429 &    693 & 28.5 &   496$\pm$ 85\\
  RX J1608.5-3847 & 16 08 31.6 & -38 47 29.5 &   215 &   275 &   237 &   132 &    94 &    73 &    55 &   162 &    48$\pm$ 16\\
  RX J1608.6-3922 & 16 08 36.2 & -39 23 02.6 &   177 &   247 &   229 &   153 &   138 &   132 &   132 &   113 &   462$\pm$ 80\\
 PZ99 J160843.4-260216 & 16 08 43.4 & -26 02 17.0 &    606 &    617 &    458 &   197 &   112 &    90 &    52 &  5.9 & $<$  19\\
  RX J1609.7-3854 & 16 09 39.5 & -38 55 07.4 &    471 &    534 &    422 &   189 &   112 &    81 &    50 &  7.2 & $<$  14\\
           Sz 117 & 16 09 44.3 & -39 13 30.3 &    86 &   118 &   113 &    65 &    62 &    57 &    55 &    92 &   127$\pm$ 28\\
  RX J1610.1-4016 & 16 10 04.8 & -40 16 12.4 &   293 &   310 &   238 &    97 &    65 &    48 & 29.0 & 2.8 & $<$  20\\
 PZ99 J161019.1-250230 & 16 10 19.2 & -25 02 30.4 &    317 &    394 &   301 &   124 &    86 &    61 &    36 &  3.6 & $<$  28\\
          WA Oph1 & 16 11 08.9 & -19 04 47.1 &    499 &    656 &    558 &   291 &   163 &   121 &    74 &  8.4 & $<$  19\\
 RX J1612.0-1906A & 16 11 59.3 & -19 06 53.5 &    407 &    482 &    386 &   121 &   105 &    74 &    43 &  5.3 & $<$  35\\
  RX J1612.1-1915 & 16 12 05.3 & -19 15 20.0 &   153 &   239 &   190 &    87 &    52 &    40 & 23.3 & 2.1 & $<$  18\\
  RX J1612.3-1909 & 16 12 20.9 & -19 09 04.3 &    84 &   115 &    96 &    48 &    33 & 23.7 & 14.0 & 1.6 & $<$  17\\
  RX J1612.6-1924 & 16 12 41.2 & -19 24 18.5 &   187 &   239 &   207 &    93 &    69 &    50 & 28.8 &  8.9 & $<$  35\\
 RX J1613.1-1904A & 16 13 10.2 & -19 04 13.3 &    89 &   102 &    92 &    50 &    35 & 23.1 & 14.2 & 1.9 & $<$  18\\
  RX J1613.7-1926 & 16 13 43.9 & -19 26 48.7 &   161 &   209 &   183 &    86 &    58 &    43 & 25.0 & 2.9 & $<$  21\\
  RX J1613.8-1835 & 16 13 47.5 & -18 35 00.6 &    57 &    80 &    72 &    38 & 26.6 & 17.7 & 11.0 & 1.3 & $<$  12\\
  RX J1613.9-1848 & 16 13 58.2 & -18 48 29.3 &    73 &    97 &    74 &    37 & 25.3 & 18.7 & 10.1 & 1.1 & $<$  39\\
  RX J1614.2-1938 & 16 14 14.0 & -19 38 28.3 &  6.5 &  7.3 &  5.5 & 2.5 & 1.6 & 1.1 & 0.7 & 0.0 & $<$  14\\
 RX J1614.4-1857A & 16 14 30.0 & -18 57 41.9 & 1.0 & 1.7 & 1.7 & 0.9 & 0.7 & 0.6 & 0.3 & 0.3 & $<$  37\\
 RX J1615.1-1851 & 16 15 08.6 & -18 51 01.2 &    76 &   118 &   102 &    54 &    35 & 24.6 & 15.4 & 0.0 & $<$  14\\
  RX J1615.3-3255 & 16 15 20.2 & -32 55 05.3 &   268 &   316 &   252 &    98 &    85 &    65 &    73 &    322 &  1049$\pm$167\\
 RX J1621.2-2342A & 16 21 14.5 & -23 42 20.0 &   145 &   200 &   169 &    92 &    56 &    39 & 23.4 & 2.1 & $<$  78\\
 RX J1621.2-2342B & 16 21 15.4 & -23 42 26.5 &    56 &   145 &   159 &    97 &    59 &    45 & 27.3 &  3.4 & $<$ 115\\
  RX J1621.4-2312 & 16 21 28.4 & -23 12 11.0 &   169 &   250 &   213 &   110 &    71 &    51 & 30.7 & 2.8 & $<$  25\\
  RX J1622.6-2345 & 16 22 37.6 & -23 45 50.8 &    60 &    97 &    85 &    44 &    32 & 22.1 & 13.9 &  6.5 & $<$  48\\
 RX J1622.7-2325A & 16 22 46.8 & -23 25 33.0 &   190 &    354 &    339 &   208 &   144 &    92 &    55 &  6.7 & $<$ 208\\
 RX J1622.7-2325B & 16 22 47.2 & -23 25 44.9 &    54 &    98 &    93 &    54 &    36 & 25.8 & 16.0 & 1.6 & $<$ 261\\
  RX J1622.8-2333 & 16 22 53.3 & -23 33 10.3 &    65 &   115 &   117 &    61 &    43 & 29.7 & 17.9 & 1.8 & $<$  54\\
  RX J1623.5-3958 & 16 23 29.6 & -39 58 01.0 &   261 &   236 &   170 &    70 &    49 &    33 & 19.6 & 2.2 & $<$  14\\
  RX J1623.8-2341 & 16 23 49.4 & -23 41 27.1 &   158 &   254 &   242 &   129 &    87 &    62 &    38 &  4.8 & $<$ 424\\
\hline
\end{tabular}
}
\end{table}
\begin{table} 
{\scriptsize 
%\newpage
\begin{tabular}{lrrrrrrrrrrr} \hline
   
& RA
& DEC
& $F_J$ 
& $F_H$ 
& $F_k$ 
& $F_{3.6}$ 
& $F_{4.5}$ 
& $F_{5.8}$ 
& $F_{8}$ 
& $F_{24}$ 
& $F_{70}$ \\
ID
& (hh mm ss.s)
& (dd mm ss.s)
& (mJy)
& (mJy)
& (mJy)
& (mJy)
& (mJy)
& (mJy)
& (mJy) 
& (mJy)
& (mJy)
\\
\hline
  RX J1624.0-2456 & 16 24 06.3 & -24 56 47.0 &   266 &    373 &    325 &   152 &   109 &    74 &    47 &  7.7 & $<$  33\\
 RX J1624.8-2359 & 16 24 48.4 & -23 59 16.0 &   269 &    502 &    480 &   266 &   171 &   120 &    72 &  7.4 & $<$ 118\\
 RX J1625.2-2455 & 16 25 14.7 & -24 56 07.0 &   233 &    379 &    328 &   171 &   112 &    81 &    49 &  5.5 & $<$  47\\
         EM*SR8 & 16 25 26.9 & -24 43 09.0 &   186 &   263 &   229 &   107 &    69 &    49 & 30.0 & 2.5 & $<$ 157\\
         DOAR21 & 16 26 03.0 & -24 23 36.0 &    926 &   1840 &   2150 &   1260 &    878 &    743 &    689 &   1810 & $<$1786\\
        ROXR123 & 16 26 23.7 & -24 43 14.0 &   279 &    448 &    484 &    367 &   292 &   299 &   258 &    399 &   956$\pm$173\\
          ROX16 & 16 26 46.4 & -24 11 59.9 &   214 &    487 &    676 &    603 &    355 &    495 &    386 &   278 & $<$ 311\\
       ROXR135S & 16 26 58.4 & -24 45 31.8 &   114 &    360 &    637 &   1390 &   1440 &   1740 &   1950 &   2230 &  5416$\pm$829\\
          ROX21 & 16 27 19.5 & -24 41 40.2 &   271 &    361 &   289 &   140 &    95 &    68 &    40 &  5.2 & $<$2063\\
       ROXR151A & 16 27 39.4 & -24 39 15.3 &    80 &   211 &   274 &   282 &   284 &    333 &    427 &   1140 &  1458$\pm$231\\
         EM*SR9 & 16 27 40.3 & -24 22 04.0 &    671 &    878 &    873 &    784 &    575 &    465 &    484 &   1040 & $<$ 860\\
NTTS162649-2145 & 16 29 48.7 & -21 52 12.0 &    539 &    644 &    527 &   238 &   148 &   105 &    64 &  7.3 & $<$  23\\
           ROX39 & 16 30 35.6 & -24 34 18.7 &    367 &    500 &    411 &   197 &   138 &    98 &    58 &  6.3 & $<$  52\\
         ROXS42C & 16 31 15.7 & -24 34 02.0 &    724 &   1010 &    938 &    575 &    428 &    371 &    397 &    862 &   239$\pm$ 53\\
         ROXS43A & 16 31 20.1 & -24 30 04.9 &   1100 &   1390 &   1360 &  4.6 &    800 &    895 &   1970 &   2180 &   938$\pm$152\\
         ROXS47A & 16 32 11.8 & -24 40 21.5 &    320 &    468 &    449 &   265 &   212 &   152 &   122 &   105 &   171$\pm$ 41\\
          WA Oph6 & 16 48 45.6 & -14 16 35.9 &    519 &    957 &   1200 &   1090 &    814 &    966 &   1010 &   1270 &  1298$\pm$205\\
\hline
\end{tabular}
}
\end{table}

%% file: table2c.tex
\begin{table}
\renewcommand{\baselinestretch}{0.2}
\caption{  Stellar Properties.
{\scriptsize FW.1H: Notes of the full at 10\% of H alpha line. (Keller 2004, 0* indicates non-detections). Negative velocities indicate absorption.
Bin. Notes: The separation of binary component is arcseconds or their period in days. Limits on non detection are also given, e.g. $<$0.13 indicate a companion
is only possible for $\rho < 0.13''$.
Binary References: 1. \citet{1993A&A...278..129L}, 2. \citet{1995ApJ...443..625S}, 3. \citet{2008ApJ...683..844L}, 4. \citet{2007ApJ...657..338P}, 
5. \citet{2005A&A...437..611R}, 6. \citet{2001AJ....122.3325K}, 7. \citet{2007A&A...467.1147G}, 8. \citet{1998A&A...331..977K}, 9. \citet{1998A&A...334..592S}, 
10. \citet{2001AJ....122..997S}, 11. \citet{2003A&A...410..269M}, 12. \citet{1997ApJ...481..378G}, 13. \citet{2003ApJ...591.1064B} .
Dedge : Distance from cloud edge in degrees. Negative numbers indicate that objects are within cloud boundary as defined in the text.}}
\renewcommand{\baselinestretch}{1}
{\tiny
\begin{tabular}{lrrrrrrrrrrrrr} \hline\hline
 
& Type
& EW($H_{\alpha}$)
& FW.1H
& SpT
& SED
& Bin.
& Bin.
& $ T_{eff}$
& \lstar  
& Age 
& $ A_{v}$ 
& $ d_{edge} $
& $ L_d/L_*$ \\
ID
& 
& \AA
& km/s
&
& type
& notes
& ref.
& K
& \lsun
& years
& mags
& deg.
& 
\\
\hline\hline     NTTS032641+2420 &  WTTS & 0.00 & 0 &  K1 &  RJ & $<$0.2$”$ & 9 & 5050 & 0.49 & 4.80E+07 & 0.1 & 3.6 &   $<$6.7E-04
\\
     NTTS040047+2603 &  WTTS & -10.00 & 155 &  M2 &  RJ &  &  & 3550 & 0.28 & 2.50E+06 & 0.1 & 2.2 &   $<$1.1E-03
\\
      RX J0405.3+2009 &  WTTS & -1.80 & -114 &  K1 &  RJ & $<$0.13$”$ & 8 & 5050 & 2.14 & 7.70E+06 & 0.1 & 5.5 &   $<$1.5E-04
\\
     NTTS040234+2143 &  WTTS & -6.60 & - &  M2 &  RJ &  &  & 3550 & 0.18 & 4.50E+06 & 0.2 & 4.7 &   $<$2.1E-03
\\
      RX J0409.2+1716 &  WTTS & -3.70 & 220 &  M1 &  RJ & $<$0.13$”$ & 8 & 3700 & 0.49 & 1.80E+06 & 0.4 & 3.3 &   $<$6.6E-04
\\
      RX J0409.8+2446 &  WTTS & -1.80 & 84 &  M1.5 &  RJ & $<$0.13$”$ & 8 & 3700 & 0.39 & 2.40E+06 & 0.0 & 1.3 &   $<$1.0E-03
\\
      RX J0412.8+1937 &  WTTS & -0.60 & 84 &  K6 &  RJ & $<$0.13$”$ & 8 & 4200 & 0.50 & 7.70E+06 & 0.2 & 3.2 &   $<$1.1E-03
\\
               LkCa 1 &  WTTS & -2.80 & 159 &  M4V &  RJ & $<$0.13$”$ & 1 & 3350 & 0.62 & 1.10E+06 & 0.7 & -0.2 &   $<$6.1E-04
\\
               LkCa 3 &  WTTS & -0.14 & 176 &  M1V &  RJ & 0.47$”$ & 1 & 3700 & 2.22 & 4.20E+05 & 0.6 & -0.2 &   $<$1.1E-04
\\
               LkCa 5 &  WTTS & -2.50 & 160 &  M2V &  RJ & $<$0.13$”$ & 1 & 3550 & 0.45 & 1.60E+06 & 0.4 & -0.4 &   $<$6.7E-04
\\
     NTTS041559+1716 &  WTTS & -1.90 & - &  K7 &  RJ &  &  & 4050 & 0.45 & 5.80E+06 & 0.1 & 1.8 &   $<$8.7E-04
\\
               LkCa 7 &  WTTS & -0.53 & 133 &  K7V &  RJ & 1.05$”$ & 1 & 4050 & 1.23 & 1.20E+06 & 0.7 & -0.5 &   $<$2.8E-04
\\
      RX J0420.3+3123 &  WTTS & -0.50 & 111 &  K4 &  RJ & $<$0.13$”$ & 8 & 4550 & 0.39 & 2.80E+07 & 0.3 & 2.3 &   $<$1.2E-03
\\
            HD 283572 &  WTTS & -0.63 & -187 &  G2III &  RJ & $<$0.13$”$ & 1 & 5850 & 9.75 & 6.30E+06 & 0.8 & 0.1 &   $<$1.7E-05
\\
              LkCa 21 &  WTTS & -5.50 & 155 &  M3... &  RJ & $<$0.13$”$ & 1 & 3450 & 0.77 & 8.60E+05 & 0.8 & 0.1 &   $<$6.3E-03
\\
      RX J0424.8+2643 &  WTTS & -2.10 & -125 &  K0 &  RJ &  &  & 5250 & 3.51 & 6.30E+06 & 1.4 & -0.4 &   $<$1.8E-04
\\
     NTTS042417+1744 &  WTTS & 1.60 & -217 &  K1 &  RJ &  &  & 5050 & 1.89 & 8.70E+06 & 0.7 & 0.8 &   $<$1.5E-04
\\
               DH Tau &  CTTS & -38.50 & - &  M0.5V:e &  TNIR & $<$0.005$”$ & 1;2 & 3850 & 1.77 & 6.70E+05 & 4.5 & -0.4 & 3.30E-02
\\
               DI Tau &  WTTS & -2.00 & 128 &  M0.5V:e &  RJ & 0.12$”$ & 1;2 & 3850 & 0.97 & 1.10E+06 & 0.6 & -0.4 & $<$1.2E-02
\\
               UX Tau &  CTTS & -0.67 & 503 &  G5V:e... &  TNIR & $<$0.13$”$ & 1 & 5750 & 6.69 & 7.90E+06 & 3.5 & 0.2 & 4.70E-02
\\
               FX Tau &  CTTS & -14.50 & 302 &  M4e &  TNIR & 0.91$”$ & 1;2 & 3350 & 1.55 & 6.60E+04 & 3.3 & -0.6 & 6.40E-02
\\
               ZZ Tau &  CTTS & -14.00 & 309 &  M3 &  TIRAC & 0.029$”$ & 1 & 3450 & 0.80 & 8.00E+05 & 1.0 & -0.3 & 3.60E-02
\\
             V927 Tau &  WTTS & -10.00 & - &  M4 &  RJ & 0.3$”$ & 1;2 & 3350 & 0.52 & 1.30E+06 & 0.3 & -0.5 &   $<$7.9E-04
\\
     NTTS042835+1700 &  WTTS & -1.10 & 98 &  K5 &  RJ &  &  & 4350 & 0.44 & 1.40E+07 & 0.4 & 0.9 &   $<$8.1E-04
\\
             V710 Tau &  CTTS & -3.87 & 313 &  K7 &  TNIR &  &  & 4050 & 0.85 & 2.10E+06 & 0.7 & 0.3 & 1.00E-01
\\
     NTTS042916+1751 &  WTTS & -0.56 & 114 &  K7 &  RJ &  &  & 4050 & 0.70 & 2.80E+06 & 0.6 & 0.1 &   $<$4.9E-04
\\
             V928 Tau &  WTTS & -1.80 & 125 &  M... &  RJ & 0.18$”$ & 1;2 & 3200 & 2.36 & 2.80E+04 & 9.5 & -0.6 &   $<$1.2E-04
\\
     NTTS042950+1757 &  WTTS & -1.07 & 114 &  K7 &  RJ &  &  & 4050 & 0.46 & 5.60E+06 & 0.6 & 0.2 &   $<$7.4E-04
\\
      RX J0432.8+1735 &  WTTS & -1.90 & 138 &  M2 &  T24 & $<$0.13$”$ & 8 & 3550 & 0.48 & 1.50E+06 & 0.7 & 0.5 & 3.70E-03
\\
               GH Tau &  CTTS & -27.50 & 480 &  M2e &  TNIR & 0.35$”$ & 1 & 3550 & 1.83 & 1.10E+05 & 2.7 & -0.3 & 5.50E-02
\\
             V807 Tau &  CTTS & -13.50 & 405 &  M... &  TIRAC & 0.41$”$ & 2 & 3200 & 5.85 & 2.80E+04 & 8.1 & -0.3 & 1.10E-02
\\
             V830 Tau &  WTTS & -1.80 & - &  K7 &  RJ & $<$0.13$”$ & 1 & 4050 & 1.08 & 1.50E+06 & 0.9 & -0.5 &   $<$3.6E-04
\\
               GK Tau &  CTTS & -30.50 & 344 &  K7 &  TNIR & 12.2$”$ & 1;2 & 4050 & 3.95 & 4.30E+05 & 5.0 & -0.4 & 9.20E-02
\\
              WA Tau1 &  WTTS & -0.60 & 122 &  K0IV &  RJ & $<$0.13$”$ & 1 & 5250 & 3.04 & 7.30E+06 & 0.2 & -0.1 &   $<$1.3E-04
\\
     NTTS043230+1746 &  WTTS & -9.00 & -193 &  M2 &  RJ &  &  & 3550 & 0.45 & 1.60E+06 & 0.6 & 0.9 &   $<$6.0E-04
\\
      RX J0435.9+2352 &  WTTS & -6.27 & 143 &  M1.5 &  RJ & 0.069$”$ & 8 & 3700 & 0.68 & 1.30E+06 & 0.5 & 0.1 &   $<$6.6E-04
\\
              LkCa 14 &  WTTS & -0.90 & 133 &  M0:V &  RJ & $<$0.13$”$ & 1 & 3850 & 0.79 & 1.30E+06 & 0.3 & -0.8 &   $<$7.6E-04
\\
RX J0437.4+1851 &  WTTS & -1.10 & 111 &  K6 &  RJ &  &  & 4200 & 0.86 & 3.20E+06 & 0.2 & 1.6 &   $<$5.3E-04
\\
      RX J0438.2+2023 &  WTTS & -1.20 & - &  K2 &  RJ & 0.464$”$ & 8 & 4900 & 0.64 & 2.20E+07 & 0.5 & 2.4 &   $<$9.3E-04
\\
               HV Tau &  WTTS & -8.50 & 194 &  M1 &  T24 & 0.035$”$ & 2 & 3700 & 1.80 & 5.10E+05 & 2.9 & -0.8 & 3.8E-04
\\
      RX J0438.6+1546 &  WTTS & -0.07 & -65 &  K2 &  RJ &  &  & 4900 & 1.74 & 6.80E+06 & 0.2 & 0.8 &   $<$2.3E-04
\\
RX J0439.4+3332A &  WTTS & -0.07 & 100 &  K5 &  RJ & $<$0.13$”$ & 8 & 4350 & 0.09 & 1.20E+08 & 1.8 & 0.0 &   $<$6.0E-03
\\
               IW Tau &  WTTS & -4.00 & 133 &  K7V &  RJ & 0.27$”$ & 1;2 & 4050 & 1.29 & 1.20E+06 & 1.3 & -0.7 &   $<$3.4E-04
\\
               ITG33 &  CTTS & -53.00 & - &  M3 &  TNIR &  &  & 3450 & 0.19 & 3.30E+06 & 10.4 & -0.7 & 1.20E-01
\\
              HBC422 &  WTTS & -1.92 & 94 &  K7 &  T70 & 0.3$”$ & 1 & 4050 & 1.93 & 7.50E+05 & 4.8 & -1.0 & 6.9E-03
\\
              HBC423 &  WTTS & -1.38 & 77 &  M1 &  TNIR & 0.33$”$ & 1 & 3700 & 2.43 & 3.70E+05 & 6.1 & -1.0 &  8.2E-02
\\
      RX J0445.8+1556 &  WTTS & 0.93 & -320 &  G5 &  RJ &  &  & 5750 & 5.80 & 9.00E+06 & 0.2 & 0.6 & $<$7.7E-04
\\
      RX J0452.5+1730 &  WTTS & -0.05 & -23 &  K4 &  RJ & $<$0.13$”$ & 8 & 4550 & 0.60 & 1.30E+07 & 0.2 & 1.2 &   $<$4.9E-04
\\
      RX J0452.8+1621 &  WTTS & -0.95 & 144 &  K6 &  RJ & 0.478$”$ & 8 & 4200 & 1.28 & 1.70E+06 & 0.6 & 1.5 &   $<$3.3E-04
\\
              LkCa 19 &  WTTS & -1.20 & 110 &  K0V &  T24 &  &  & 5250 & 2.35 & 9.50E+06 & 0.8 & 0.1 & 2.50E-04
\\
     NTTS045251+3016 &  WTTS & -1.40 & 77 &  K5 &  RJ & 0.034$”$ & 10 & 4350 & 1.60 & 2.00E+06 & 0.8 & 0.1 &   $<$3.0E-04
\\
      RX J0457.2+1524 &  WTTS & -0.13 & -114 &  K1 &  RJ & 0.57$”$ & 8 & 5050 & 2.93 & 5.50E+06 & 0.1 & 2.9 &   $<$1.4E-04
\\
      RX J0457.5+2014 &  WTTS & -0.18 & -96 &  K3 &  RJ & 6.87$”$ & 8 & 4700 & 1.09 & 8.90E+06 & 0.4 & 4.6 &   $<$4.8E-04
\\
      RX J0458.7+2046 &  WTTS & 0.00 & -53 &  K7 &  RJ & 6.11$”$ & 8 & 4050 & 0.71 & 2.70E+06 & 0.3 & 4.4 &   $<$7.4E-04
\\
      RX J0459.7+1430 &  WTTS & -0.01 & 57 &  K4 &  RJ & $<$0.13$”$ & 8 & 4550 & 0.79 & 1.10E+07 & 0.2 & 4.1 &   $<$4.3E-04
\\
             V836 Tau &  CTTS & -7.70 & 240 &  K7V &  TIRAC &  &  & 4050 & 1.18 & 1.30E+06 & 3.4 & -0.1 & 3.50E-02
\\
      RX J0842.4-8345 &  WTTS & -1.00 & 171 &  K4 &  RJ & $<$0.13$”$ & 6 & 4550 & 1.80 & 3.30E+06 & 1.1 & 7.4 & $<$7.8E-04
\\
      RX J0848.0-7854 &  WTTS & -5.57 & 210 &  M3.2Ve &  RJ & $<$0.13$”$ & 6 & 3450 & 1.13 & 4.50E+05 & 0.2 & 6.4 &   $<$2.2E-04
\\
      RX J0902.9-7759 &  WTTS & -1.70 & 179 &  M3 &  RJ & $<$0.13$”$ & 6 & 3450 & 0.52 & 1.30E+06 & 0.0 & 5.9 &   $<$9.8E-04
\\
      RX J0915.5-7609 &  WTTS & -1.15 & 120 &  K6 &  RJ & 0.111$”$ & 6 & 4200 & 1.59 & 1.30E+06 & 0.7 & 5.5 & $<$1.3E-03
\\
      RX J0935.0-7804 &  WTTS & -4.90 & 117 &  M1 &  RJ & 0.36$”$ & 6 & 3700 & 0.84 & 1.10E+06 & 0.4 & 3.9 &   $<$4.4E-04
\\
      RX J0942.7-7726 &  WTTS & -2.16 & 109 &  K8 &  RJ & $<$0.13$”$ & 6 & 3950 & 0.59 & 2.90E+06 & 0.8 & 3.6 &   $<$7.4E-04
\\
      RX J1001.1-7913 &  WTTS & -1.80 & 113 &  K8 &  RJ & $<$0.13$”$ & 6 & 3950 & 0.72 & 2.10E+06 & 0.5 & 2.8 &   $<$5.4E-04
\\
      RX J1005.3-7749 &  WTTS & -2.80 & 127 &  M1 &  RJ & $<$0.13$”$ & 6 & 3700 & 0.85 & 1.10E+06 & 0.5 & 2.3 &   $<$4.3E-04
\\
               CS Cha &  CTTS & -54.30 & 395 &  M0 &  T24 & 2435 days & 7 & 3850 & 1.73 & 6.70E+05 & 0.6 & -0.5 & 1.40E-01
\\
      RX J1108.8-7519 &  WTTS & -1.70 & 218 &  M2 &  RJ & 0.15$”$ & 6 & 3550 & 0.86 & 9.00E+05 & 0.5 & 1.1 &   $<$5.2E-04
\\
                Sz 30 &  WTTS & -4.90 & 176 &  M0 &  RJ & 1.24$”$ & 3 & 3850 & 0.90 & 1.20E+06 & 0.9 & -0.4 &   $<$7.0E-04
\\
\hline
\label{stprops}
\end{tabular}
}
\end{table}
\newpage\begin{table}
{\tiny
\begin{tabular}{lrrrrrrrrrrrrr} \hline\hline
 
& Type
& EW($H_{\alpha}$)
& FW.1H
& SpT
& SED
& Bin.
& Bin.
& $ T_{eff}$
& \lstar  
& Age 
& $ A_{v}$ 
& $ d_{edge} $
& $ L_d/L_*$ \\
ID
& 
& \AA
& km/s
&
& type
& notes
& ref.
& K
& \lsun
& years
& mags
& deg
& 
\\
\hline
                Sz 41 &  CTTS & 0.00 & 378 &  K0 &  TIRAC & 1.97$”$ & 3 & 5250 & 5.70 & 4.00E+06 & 4.0 & 0.0 & 2.10E-02
\\
     RX J1117.0-8028 &  WTTS & -11.77 & 159 &  M2 &  RJ & $<$0.13$”$ & 6 & 3550 & 0.72 & 1.00E+06 & 0.1 & 2.2 &   $<$3.3E-04
\\
     RX J1123.2-7924 &  WTTS & -3.00 & 173 &  K8 &  RJ & $<$0.13$”$ & 6 & 3950 & 0.11 & 4.20E+07 & 2.4 & 1.4 &   $<$3.9E-03
\\
     RX J1129.2-7546 &  WTTS & -0.23 & 49 &  K3 &  RJ & 0.534$”$ & 6 & 4700 & 1.57 & 5.50E+06 & 1.6 & 1.3 & $<$9.0E-04
\\
      RX J1149.8-7850 &  WTTS & -32.00 & 120 &  M0 &  TNIR & $<$0.13$”$ & 6 & 3850 & 1.38 & 7.70E+05 & 0.9 & 2.0 & 2.80E-01
\\
      RX J1150.4-7704 &  CTTS & -1.70 & 388 &  K2 &  RJ & $<$0.13$”$ & 6 & 4900 & 1.41 & 9.00E+06 & 0.7 & 1.9 &   $<$3.0E-04
\\
                T Cha &  WTTS & -2.70 & -293 &  F5 &  TNIR &  &  & 6400 & 45.88 & 9.20E+05 & 10.3 & 1.5 & 2.20E-02
\\
      RX J1158.5-7754 &  WTTS & -0.50 & 120 &  K2 &  RJ & 0.073$”$ & 6 & 4900 & 6.21 & 1.70E+06 & 1.1 & 2.3 &   $<$6.3E-05
\\
      RX J1158.5-7913 &  WTTS & -3.34 & 194 &  K3 &  RJ & $<$0.13$”$ & 6 & 4700 & 2.14 & 3.70E+06 & 2.6 & 1.5 &   $<$2.5E-04
\\
      RX J1159.7-7601 &  WTTS & -0.39 & 91 &  K2 &  RJ & $<$0.13$”$ & 6 & 4900 & 2.75 & 4.10E+06 & 1.2 & 2.8 &   $<$4.3E-04
\\
      RX J1202.1-7853 &  WTTS & -2.48 & 170 &  K7 &  RJ & $<$0.13$”$ & 6 & 4050 & 1.81 & 8.00E+05 & 1.0 & 1.7 &   $<$2.1E-04
\\
      RX J1204.6-7731 &  WTTS & -4.20 & 99 &  M3 &  RJ & $<$0.13$”$ & 6 & 3450 & 0.72 & 9.80E+05 & 0.0 & 2.4 &   $<$6.0E-04
\\
      RX J1216.8-7753 &  WTTS & -4.00 & 117 &  M4 &  RJ & $<$0.13$”$ & 6 & 3350 & 0.51 & 1.30E+06 & 0.2 & 1.7 &   $<$7.8E-04
\\
      RX J1219.7-7403 &  WTTS & -3.30 & 117 &  K8 &  RJ & $<$0.13$”$ & 6 & 3950 & 1.02 & 1.30E+06 & 0.7 & 3.3 &   $<$3.5E-04
\\
      RX J1220.4-7407 &  WTTS & -2.24 & 155 &  K7m... &  RJ & 0.296$”$ & 6 & 4050 & 1.70 & 8.60E+05 & 0.9 & 3.2 &   $<$2.1E-04
\\
      RX J1239.4-7502 &  WTTS & -0.07 & 137 &  K2 &  RJ & $<$0.13$”$ & 6 & 4900 & 4.00 & 2.70E+06 & 0.1 & 1.8 &   $<$7.3E-05
\\
      RX J1301.0-7654 &  WTTS & -3.90 & -350 &  M0.5 &  RJ & $<$0.13$”$ & 6 & 3850 & 1.92 & 6.40E+05 & 0.7 & 0.1 & $<$2.7E-03
\\
      RX J1507.6-4603 &  WTTS & -0.36 & 131 &  K2 &  RJ &  &  & 4900 & 1.56 & 8.00E+06 & 0.5 & 5.5 &   $<$4.6E-04
\\
      RX J1508.6-4423 &  WTTS & -0.55 & -229 &  G8 &  RJ & not SB & 7 & 5500 & 2.52 & 1.50E+07 & 0.0 & 7.1 &   $<$2.2E-04
\\
      RX J1511.6-3550 &  WTTS & -0.14 & 143 &  K5 &  RJ &  &  & 4350 & 0.96 & 4.20E+06 & 0.4 & 2.0 &   $<$5.6E-04
\\
      RX J1515.8-3331 &  WTTS & -1.57 & 85 &  K0 &  RJ &  &  & 5250 & 3.32 & 6.70E+06 & 0.0 & 2.2 &   $<$2.8E-04
\\
      RX J1515.9-4418 &  WTTS & -1.03 & 117 &  K1 &  RJ & not SB & 7 & 5050 & 1.23 & 1.50E+07 & 0.7 & 6.5 &   $<$4.9E-04
\\
      RX J1516.6-4406 &  WTTS & 0.00 & -137 &  K2 &  RJ & not SB & 7 & 4900 & 1.41 & 8.90E+06 & 0.4 & 6.7 &   $<$5.6E-04
\\
      RX J1518.9-4050 &  CTTS & -1.37 & 914 &  G8 &  RJ &  &  & 5500 & 3.30 & 1.10E+07 & 0.3 & 7.2 &   $<$2.3E-04
\\
      RX J1519.3-4056 &  WTTS & -0.28 & -504 &  K0 &  RJ &  &  & 5250 & 2.38 & 9.70E+06 & 0.8 & 7.8 &   $<$3.0E-04
\\
      RX J1522.2-3959 &  WTTS & -1.80 & 166 &  K3 &  RJ &  &  & 4700 & 1.50 & 5.90E+06 & 0.9 & 4.5 &   $<$3.8E-04
\\
      RX J1523.4-4055 &  WTTS & 0.00 & 81 &  K2 &  RJ & not SB & 7 & 4900 & 1.32 & 9.80E+06 & 0.4 & 6.6 &   $<$5.4E-04
\\
      RX J1523.5-3821 &  WTTS & -5.93 & 302 &  M2 &  RJ & not SB & 7 & 3550 & 0.56 & 1.30E+06 & 0.3 & 3.0 &   $<$9.1E-04
\\
      RX J1524.0-3209 &  WTTS & -2.26 & 127 &  K7 &  RJ & 3000 days & 7 & 4050 & 1.62 & 9.10E+05 & 0.7 & 3.2 &   $<$3.3E-04
\\
      RX J1524.5-3652 &  WTTS & -0.07 & -50 &  K1 &  RJ & not SB & 7 & 5050 & 1.88 & 8.70E+06 & 0.1 & 1.5 &   $<$3.2E-04
\\
      RX J1525.5-3613 &  WTTS & -0.29 & 166 &  K2 &  RJ & not SB & 7 & 4900 & 1.98 & 5.90E+06 & 0.5 & 1.0 &   $<$4.6E-04
\\
      RX J1525.6-3537 &  WTTS & -1.56 & 170 &  K6 &  RJ & not SB & 7 & 4200 & 1.28 & 1.70E+06 & 0.6 & 0.7 &   $<$4.5E-04
\\
      RX J1526.0-4501 &  WTTS & -2.40 & -159 &  G5 &  RJ & not SB & 7 & 5750 & 2.64 & 1.50E+07 & 0.4 & 4.7 &   $<$2.5E-04
\\
      RX J1538.0-3807 &  WTTS & -0.46 & 113 &  K5 &  RJ & not SB & 7 & 4350 & 0.90 & 4.60E+06 & 0.2 & 3.2 &   $<$6.7E-04
\\
      RX J1538.6-3916 &  WTTS & -0.12 & 80 &  K4 &  RJ & not SB & 7 & 4550 & 1.67 & 3.60E+06 & 0.4 & 4.5 &   $<$3.8E-04
\\
      RX J1538.7-4411 &  WTTS & 0.00 & -56 &  G5 &  RJ &  &  & 5750 & 5.19 & 1.00E+07 & 0.7 & 3.4 &   $<$1.1E-04
\\
                Sz 65 &  CTTS & -3.30 & 336 &  K8 &  TIRAC &  &  & 3950 & 3.53 & 4.50E+05 & 2.6 & -0.2 & 6.20E-02
\\
      RX J1540.7-3756 &  WTTS & -0.49 & 145 &  K6 &  RJ & not SB & 7 & 4200 & 1.00 & 2.50E+06 & 0.1 & 2.8 &   $<$6.5E-04
\\
      RX J1543.1-3920 &  WTTS & -0.13 & 127 &  K6 &  RJ &  &  & 4200 & 1.09 & 2.20E+06 & 0.1 & 3.9 &   $<$1.6E-03
\\
      RX J1546.7-3618 &  WTTS & -0.57 & 127 &  K1 &  RJ &  &  & 5050 & 2.27 & 7.10E+06 & 0.6 & 0.9 &   $<$2.8E-04
\\
      RX J1547.7-4018 &  WTTS & -0.63 & -99 &  K1 &  RJ &  &  & 5050 & 2.42 & 6.70E+06 & 0.1 & 2.6 &   $<$3.4E-04
\\
PZ99 J154920.9-2600 &  WTTS & 0.00 & -123 &  K0 &  RJ &  &  & 5250 & 2.18 & 1.10E+07 & 0.8 & 6.2 &   $<$8.8E-05
\\
                Sz 76 &  WTTS & -10.30 & 227 &  M1 &  TIRAC &  &  & 3700 & 0.38 & 2.40E+06 & 0.6 & 0.1 & 8.60E-02
\\
      RX J1550.0-3629 &  WTTS & -0.06 & -46 &  K2 &  RJ &  &  & 4900 & 1.84 & 6.40E+06 & 0.2 & 0.8 &   $<$3.3E-04
\\
                Sz 77 &  CTTS & -17.10 & 304 &  M0 &  TNIR & not SB & 7 & 3850 & 2.41 & 5.30E+05 & 2.1 & 0.5 & 5.60E-02
\\
      RX J1552.3-3819 &  WTTS & -0.70 & 174 &  K7 &  RJ &  &  & 4050 & 0.71 & 2.70E+06 & 0.5 & 2.6 &   $<$8.2E-04
\\
      RX J1554.9-3827 &  WTTS & -1.94 & 91 &  K7 &  RJ &  &  & 4050 & 0.69 & 2.80E+06 & 0.6 & 2.1 &   $<$8.9E-04
\\
PZ99 J155506.2-2521 &  WTTS & -0.76 & 152 &  M1 &  RJ &  &  & 3700 & 0.59 & 1.50E+06 & 0.4 & 5.0 &   $<$5.9E-04
\\
      RX J1555.4-3338 &  WTTS & -1.20 & 95 &  K5 &  RJ & not SB & 7 & 4350 & 0.97 & 4.20E+06 & 0.6 & 2.1 &   $<$9.4E-04
\\
      RX J1555.6-3709 &  WTTS & -0.61 & 146 &  K6 &  RJ &  &  & 4200 & 1.06 & 2.30E+06 & 0.4 & 2.2 &   $<$1.0E-03
\\
                Sz 81 &  CTTS & -35.80 & 330 &  M5.5 &  TNIR &  &  & 3200 & 0.65 & 7.90E+05 & 0.4 & 2.5 & 9.80E-02
\\
      RX J1556.1-3655 &  CTTS & -82.60 & 416 &  M1 &  TNIR &  &  & 3700 & 0.82 & 1.10E+06 & 1.5 & 1.9 & 9.70E-02
\\
                Sz 82 &  CTTS & -39.00 & - &  M0 &  TNIR &  &  & 3850 & 3.63 & 2.90E+05 & 1.4 & 2.2 & 1.00E-01
\\
PZ99 J155702.3-1950 &  WTTS & -0.81 & 159 &  K7 &  RJ &  &  & 4050 & 0.79 & 2.30E+06 & 0.4 & 3.2 &   $<$2.8E-04
\\
                Sz 84 &  CTTS & -43.70 & 422 &  M5.5 &  T24 &  &  & 3200 & 0.36 & 1.10E+06 & 0.8 & 2.3 & 7.70E-02
\\
      RX J1559.0-3646 &  WTTS & -3.05 & 174 &  M1.5 &  RJ &  &  & 3700 & 0.72 & 1.20E+06 & 0.1 & 2.2 &   $<$9.4E-04
\\
               Sz 129 &  CTTS & -43.90 & 346 &  K8 &  TNIR &  &  & 3950 & 2.12 & 6.20E+05 & 3.3 & 0.3 & 7.20E-02
\\
      RX J1559.8-3628 &  WTTS & -0.45 & 103 &  K3 &  RJ &  &  & 4700 & 3.86 & 1.90E+06 & 0.5 & 2.2 &   $<$2.4E-04
\\
      RX J1601.2-3320 &  WTTS & -1.92 & -137 &  G8 &  T24 & not SB & 7 & 5500 & 3.34 & 1.10E+07 & 0.3 & 3.2 & 7.20E-05
\\
PZ99 J160151.4-2445 &  WTTS & -0.07 & 141 &  K7 &  RJ &  &  & 4050 & 0.76 & 2.40E+06 & 1.0 & 3.6 &   $<$3.4E-04
\\
PZ99 J160158.2-2008 &  WTTS & -1.50 & -117 &  G5 &  RJ &  &  & 5250 & 2.82 & 7.70E+06 & 1.2 & 2.2 &   $<$6.8E-05
\\
      RX J1602.0-3613 &  WTTS & -0.85 & - &  K3 &  RJ &  &  & 4700 & 1.81 & 4.60E+06 & 0.5 & 1.9 &   $<$3.4E-04
\\
PZ99 J160253.9-2022 &  WTTS & -2.77 & 180 &  K7 &  RJ &  &  & 4050 & 1.03 & 1.60E+06 & 1.3 & 2.2 &   $<$2.2E-04
\\
      RX J1603.2-3239 &  WTTS & -2.45 & 132 &  K7 &  T24 &  &  & 4050 & 1.04 & 1.60E+06 & 0.6 & 4.0 & 1.50E-03
\\
      RX J1603.8-4355 &  WTTS & 0.00 & -212 &  G8V &  RJ &  &  & 5500 & 10.33 & 3.90E+06 & 0.4 & 0.3 &   $<$1.9E-04
\\
      RX J1603.8-3938 &  WTTS & -0.03 & 153 &  K3 &  RJ &  &  & 4700 & 3.17 & 2.40E+06 & 0.4 & 0.0 & $<$9.7E-04
\\
\hline
\end{tabular}
}
\end{table}
\newpage\begin{table}
{\tiny
\begin{tabular}{lrrrrrrrrrrrrr} \hline\hline
 
& Type
& EW($H_{\alpha}$)
& FW.1H
& SpT
& SED
& Bin.
& Bin.
& $ T_{eff}$
& \lstar  
& Age 
& $ A_{v}$ 
& $ d_{edge} $
& $ L_d/L_*$ \\
ID
& 
& \AA
& km/s
&
& type
& sep.
& ref.
& K
& \lsun
& years
& mags
& deg
& 
\\
\hline
      RX J1604.5-3207 &  WTTS & -0.19 & -74 &  K2 &  RJ &  &  & 4900 & 2.45 & 4.70E+06 & 0.2 & 4.4 &   $<$2.3E-04
\\
     RX J1605.6-3837 &  WTTS & -2.63 & 124 &  M1 &  RJ &  &  & 3700 & 0.41 & 2.20E+06 & 0.3 & 0.5 &   $<$2.3E-03
\\
PZ99 J160550.5-2533 &  WTTS & -0.21 & 85 &  G7 &  RJ &  &  & 5600 & 1.52 & 2.30E+04 & 0.7 & 2.6 &   $<$2.5E-04
\\
     RX J1607.2-3839 &  WTTS & -2.39 & 181 &  K7 &  RJ &  &  & 4050 & 1.28 & 1.20E+06 & 0.4 & 0.3 &  $<$7.7E-04
\\
               Sz 96 &  CTTS & -6.10 & 369 &  M1.5 &  TNIR & not SB & 11 & 3700 & 1.17 & 8.30E+05 & 1.9 & -0.2 & 1.20E-01
\\
     RX J1608.3-3843 &  WTTS & -1.04 & 131 &  K7 &  RJ & not SB & 7 & 4050 & 1.52 & 9.80E+05 & 0.6 & 0.1 & $<$5.3E-04
\\
               Sz 98 &  CTTS & -29.10 & 382 &  K8 &  TNIR & not SB & 7 & 3950 & 4.12 & 3.80E+05 & 4.4 & -0.2 & 6.60E-02
\\
     RX J1608.5-3847 &  CTTS & -6.17 & 438 &  M2 &  TIRAC &  &  & 3550 & 1.38 & 6.40E+05 & 1.2 & 0.1 & 3.10E-02
\\
      RX J1608.6-3922 &  CTTS & -14.20 & 656 &  K6 &  TNIR & not SB & 7 & 4200 & 2.18 & 8.80E+05 & 3.0 & 0.0 & 3.90E-02
\\
PZ99 J160843.4-2602 &  WTTS & -0.32 & -120 &  G7 &  RJ & not SB & 7 & 5600 & 2.53 & 1.50E+07 & 0.7 & 1.8 &   $<$1.5E-04
\\
      RX J1609.7-3854 &  WTTS & -0.39 & 137 &  K5 &  RJ &  &  & 4350 & 3.42 & 7.50E+05 & 0.7 & -0.1 &  $<$2.4E-04
\\
               Sz 117 &  CTTS & -20.50 & 183 &  M2 &  TIRAC & $<$0.13$”$ & 12 & 3550 & 0.74 & 1.00E+06 & 2.3 & -0.2 & 5.30E-02
\\
      RX J1610.1-4016 &  WTTS & -0.27 & 192 &  K2 &  RJ &  &  & 4900 & 2.42 & 4.80E+06 & 0.5 & 0.3 &   $<$3.8E-04
\\
PZ99 J161019.1-2502 &  WTTS & -0.75 & 132 &  M1 &  RJ &  &  & 3700 & 0.67 & 1.30E+06 & 0.3 & 1.8 &   $<$5.3E-04
\\
              WA Oph1 &  WTTS & -1.70 & 137 &  K2 &  RJ & 145 days & 7 & 4900 & 2.74 & 4.10E+06 & 2.6 & 0.1 &   $<$7.0E-04
\\
RX J1612.0-1906A &  WTTS & -0.50 & -156 &  K3 &  RJ &  &  & 4700 & 1.55 & 5.60E+06 & 1.4 & 0.3 &   $<$3.9E-04
\\
      RX J1612.1-1915 &  WTTS & -1.80 & 139 &  K5 &  RJ &  &  & 4350 & 0.75 & 6.10E+06 & 2.8 & 0.4 &   $<$6.6E-04
\\
      RX J1612.3-1909 &  WTTS & -18.00 & 246 &  M2.5 &  RJ & SB & 4 & 3550 & 0.22 & 3.30E+06 & 1.3 & 0.4 &   $<$2.6E-03
\\
      RX J1612.6-1924 &  WTTS & -2.40 & 131 &  K8 &  T24 & SB & 4 & 3950 & 0.61 & 2.70E+06 & 1.7 & 0.4 & 6.80E-04
\\
RX J1613.1-1904A &  WTTS & -2.70 & 117 &  M4 &  RJ & 0.5$”$ & 4 & 3350 & 0.18 & 3.00E+06 & 0.5 & 0.2 &   $<$2.5E-03
\\
      RX J1613.7-1926 &  WTTS & -2.90 & 153 &  M1 &  RJ & 0.7$”$ & 4 & 3700 & 0.46 & 1.90E+06 & 1.5 & 0.3 &   $<$9.2E-04
\\
      RX J1613.8-1835 &  WTTS & -7.60 & - &  M2e &  RJ & $<$0.13$”$ & 4 & 3550 & 0.18 & 4.40E+06 & 2.1 & 0.2 &   $<$3.2E-03
\\
      RX J1613.9-1848 &  WTTS & -1.50 & 128 &  M2 &  RJ & $<$0.13$”$ & 4 & 3550 & 0.16 & 5.00E+06 & 0.7 & 0.0 &   $<$4.3E-03
\\
      RX J1614.2-1938 &  WTTS & -0.60 & -91 &  K2 &  RJ &  &  & 4900 & 0.02 & 2.90E+08 & 0.7 & 0.4 &   $<$3.0E-02
\\
RX J1614.4-1857A &  WTTS & 0.08 & 153 &  M2 &  T24 & $<$0.13$”$ & 4 & 3550 & 0.01 & 4.50E+04 & 4.2 & -0.2 & 3.90E-03
\\
      RX J1615.1-1851 &  WTTS & -2.60 & 121 &  K7-M0 &  RJ & $<$0.13$”$ & 4 & 3900 & 0.63 & 2.00E+06 & 9.2 & 0.0 &   $<$7.0E-04
\\
      RX J1615.3-3255 &  CTTS & -18.90 & 397 &  K5 &  TIRAC &  &  & 4350 & 2.11 & 1.40E+06 & 1.1 & 2.5 & 7.00E-02
\\
RX J1621.2-2342A &  WTTS & -0.80 & 142 &  K7 &  RJ &  &  & 4050 & 0.54 & 4.20E+06 & 2.2 & -0.1 & $<$3.6E-03 
\\
RX J1621.2-2342B &  WTTS & -0.80 & - &  K7 &  RJ &  &  & 4050 & 0.92 & 1.90E+06 & 7.8 & -0.1 & $<$3.1E-03 
\\
      RX J1621.4-2312 &  WTTS & -1.60 & 181 &  K7 &  RJ &  &  & 4050 & 0.72 & 2.70E+06 & 2.7 & 0.0 &   $<$1.1E-03
\\
      RX J1622.6-2345 &  WTTS & -3.60 & 128 &  M2.5 &  T24 & $<$0.13$”$ & 4 & 3550 & 0.23 & 3.30E+06 & 2.7 & -0.4 & 1.80E-03
\\
RX J1622.7-2325A &  WTTS & -1.70 & 124 &  M1 &  RJ & SB & 4 & 3700 & 1.15 & 8.40E+05 & 4.4 & -0.4 &   $<$5.0E-03
\\
RX J1622.7-2325B &  WTTS & -4.10 & - &  M3 &  RJ & SB & 4 & 3450 & 0.27 & 2.30E+06 & 3.9 & -0.4 &   $<$1.6E-02
\\
      RX J1622.8-2333 &  WTTS & -1.70 & 91 &  K8 &  RJ & SB & 4 & 3950 & 0.47 & 4.00E+06 & 4.8 & -0.4 &   $<$6.3E-03
\\
      RX J1623.5-3958 &  WTTS & -2.53 & -313 &  G0 &  RJ &  &  & 6000 & 2.81 & 1.90E+07 & 0.7 & -0.3 &   $<$3.8E-04
\\
      RX J1623.8-2341 &  WTTS & -0.40 & 197 &  K5 &  RJ &  &  & 4350 & 1.10 & 3.40E+06 & 4.2 & -0.7 &   $<$3.2E-03
\\
      RX J1624.0-2456 &  WTTS & -0.92 & 153 &  K0 &  RJ &  &  & 5250 & 2.03 & 1.10E+07 & 3.4 & -0.5 & 2.60E-05
\\
      RX J1624.8-2359 &  WTTS & -0.06 & 0 &  K3 &  RJ &  &  & 4700 & 2.95 & 2.60E+06 & 5.4 & -0.6 &   $<$1.1E-03
\\
      RX J1625.2-2455 &  WTTS & -2.90 & 206 &  M0 &  RJ & $<$0.13$”$ &  & 3850 & 1.03 & 1.00E+06 & 3.0 & -0.4 &   $<$1.1E-03
\\
              EM*SR8 &  WTTS & -1.30 & 129 &  K2 &  RJ & $<$0.13$”$ & 5 & 4900 & 1.20 & 1.10E+07 & 3.2 & -0.6 &   $<$3.0E-03
\\
              DOAR21 &  WTTS & -0.70 & -197 &  B2V &  TIRAC & $<$0.005$”$ & 2;5 & 22000 & 1477.07 & 1.00E+05 & 12.4 & -0.7 & 3.50E-04
\\
             ROXR123 &  CTTS & -16.10 & 489 &  K7 &  TNIR & $<$0.13$”$ & 5 & 4050 & 2.02 & 7.20E+05 & 4.7 & -0.6 & 4.50E-02
\\
               ROX16 &  CTTS & -10.00 & - &  B5 &  TNIR & 0.577$”$ & 5 & 15400 & 201.20 & 1.00E+05 & 14.0 & -0.6 & 2.20E-04
\\
            ROXR135S &  CTTS & -73.00 & - &  K7 &  TNIR & 0.197$”$ & 2;5 & 4050 & 5.80 & 3.30E+05 & 12.2 & -0.6 & 1.20E-01
\\
               ROX21 &  CTTS & -3.00 & - &  Me &  RJ & 0.3$”$ & 2;5 & 3200 & 1.03 & 1.60E+04 & 7.1 & -0.5 &   $<$3.4E-03
\\
            ROXR151B &  WTTS & -3.70 & - &  M0 &  TNIR & $<$0.13$”$ & 4 & 3350 & 1.23 & 1.20E+05 & 8.7 & -0.4 & 1.50E-01
\\
              EM*SR9 &  CTTS & -6.40 & 394 &  K3.5e &  TNIR & 0.638$”$ & 5 & 4700 & 4.31 & 1.70E+06 & 3.3 & -0.4 & 4.60E-02
\\
     NTTS162649-2145 &  WTTS & -0.20 & 234 &  K0 &  RJ &  &  & 5250 & 2.83 & 7.60E+06 & 2.0 & 0.6 &   $<$1.9E-04
\\
               ROX39 &  CTTS & -3.60 & - &  K5 &  RJ & $<$0.13$”$ & 5 & 4350 & 1.51 & 2.10E+06 & 2.2 & -0.1 &   $<$3.7E-04
\\
             ROXS42C &  WTTS & -1.60 & 183 &  M &  TIRAC & 0.277$”$ & 5 & 3200 & 3.80 & 1.30E+04 & 8.3 & -0.2 & 2.10E-02
\\
             ROXS43A &  WTTS & -1.80 & - &  G8+... &  TIRAC & 89 days & 7 & 5500 & 9.84 & 4.10E+06 & 3.7 & -0.2 & 4.50E-02
\\
             ROXS47A &  WTTS & -9.20 & - &  K8 &  TIRAC & 0.046$”$ & 13 & 3950 & 1.54 & 8.30E+05 & 3.2 & -0.4 & 1.10E-02
\\
              WA Oph6 &  CTTS & -17.05 & 442 &  K7 &  TNIR &  &  & 4050 & 6.06 & 3.10E+05 & 6.5 & -0.3 & 3.90E-02\\
\hline
\end{tabular}
}
\end{table}